\documentclass[3p,number,sort&compress]{elsarticle}

\journal{Annals of Physics}

\usepackage{amsmath}
\usepackage{amsfonts}
\usepackage{graphicx}
\usepackage{subfigure}
\usepackage{amssymb,mathrsfs}
\usepackage{multirow}

\begin{document}
\renewcommand{\arraystretch}{1.5}

\begin{frontmatter}

\title{Spinful fermionic ladders at incommensurate filling: Phase
  diagram, local perturbations, and ionic potentials}

\author[kent,kit]{Sam T. Carr}
\author[kit]{Boris N. Narozhny}
\author[trieste]{Alexander A. Nersesyan}

\address[kent]{School of Physical Sciences, University of Kent, Canterbury CT2 7NH, UK}
\address[kit]{Institut f\"ur Theorie der
  Kondensierten Materie and DFG Center for Functional Nanostructures,
  Karlsruher Institut f\"ur Technologie, 76128 Karlsruhe, Germany}
 \address[trieste]{The Abdus Salam
  International Centre for Theoretical Physics, 34100, Trieste, Italy}

\date{\today}

\begin{abstract}

We study the effect of external potential on transport properties of
the fermionic two-leg ladder model. The response of the system to a
local perturbation is strongly dependent on the ground state
properties of the system and especially on the dominant
correlations. We categorise all phases and transitions in the model
(for incommensurate filling) and introduce ``hopping-driven
transitions'' that the system undergoes as the inter-chain hopping is
increased from zero. We also describe the response of the system to an
ionic potential. The physics of this effect is similar to that of the
single impurity, except that the ionic potential can affect the bulk
properties of the system and in particular induce true long range
order.

\end{abstract}


\end{frontmatter}

\section{Introduction}

The wonderful world of electrons confined to move in one dimension has
fascinated physicists for more than fifty years. Since the pioneering
work of Tomonaga \cite{tom} and Luttinger \cite{lut}, powerful
theoretical methods have been devised to treat interacting
one-dimensional (1D) systems \cite{mal,lap,dal,eal}, culminating in
establishing the concept of the Tomonaga-Luttinger (TL) liquid
\cite{eal,hal} (see for a review \cite{gnt,gb}). The distinctive
feature of the TL liquid is that the elementary excitations have
nothing to do with free electrons, but rather consist of plasmon-like
collective modes.  Moreover, various perturbations, such as
backscattering or Umklapp processes, can lead to the development of a
strong-coupling regime where spectral gaps are
dynamically generated \cite{dl72, lae, elp} without spontaneous
breakdown of any continuous symmetry \cite{maw}.

For a long time, beautiful one-dimensional models mainly remained in
the theoretical domain. However, due to recent technological
advances, quantum wires have become experimentally realizable
\cite{naw,dbgy}, and one-dimensional physics is undergoing a
renaissance. TL-liquid effects have been observed in carbon nanotubes
\cite{nanotube_experiment1,nanotube_experiment2,nanotube_experiment3,nanotube_experiment4},
and cleaved edge
\cite{cleaved_edge_experiment1,cleaved_edge_experiment2} and V-groove
\cite{V_groove_experiment} semiconductor quantum wires.

Though much experimental effort goes into making quantum wires as
clean as possible, any real system inevitably contains some degree of
disorder. The traditional approach to disorder in solids builds on
the approximation of nearly free electrons. At a certain
concentration of impurities the system undergoes an Anderson
localization transition \cite{and,gof}. This one-particle description
of disorder has been realized in all possible dimensions -- in
particular, it was established that in one-dimensional disordered
systems electrons are always localized \cite{mot}.

Going beyond the free electron limit one must study the interplay of
disorder and electron-electron interactions -- one of central topics
in modern condensed matter physics. If disorder strength is
insufficient to reach the Anderson transition, the electrons remain
mobile and one finds interaction corrections to physical observables
\cite{aar}. On the other hand, when free electrons are localized by
disorder, then the electron-electron interaction is not expected to
change the insulating nature of the ground state \cite{lrr}.  This can
not be the whole story in 1D systems, however; here interactions may
dramatically change the nature of the ground state and therefore
should never be considered as a small perturbation.

In a seminal paper \cite{kaf}, Kane and Fisher have shown that, due to
strong correlation effects, transport properties of the TL liquid can
dramatically change even in the presence of a single impurity. While a
clean TL liquid is an ideal conductor, a single impurity, no matter
how weak, completely reflects the charge carriers driving the
zero-temperature conductance to zero in the physical case of repulsive
bulk interactions.  This may also be understood \cite{ymg} as an
extreme limit of the Altshuler-Aronov corrections \cite{aar}; the
reduced dimensionality enhances the effect such that it may no longer
be considered a correction.

The result of Kane and Fisher is specific to purely 1D models where a
strongly renormalized impurity potential effectively splits the chain
into two disconnected pieces. Conversely, if a single impurity is
added to a two- or three-dimensional system, then, no matter how
strong, such perturbation will have no effect whatsoever on global
observable properties. Exactly how one can interpolate between one-
and two-dimensional (2D) behavior (i.e. between the TL and Fermi
liquids) is still an open question, despite considerable theoretical
effort \cite{wen,csa,arr}. Instead, one can consider a simpler
situation where only a small number of 1D systems are coupled to form
a ladder-like lattice.

The simplest ladder model is that of a two-leg ladder. The extensive
research in this field
\cite{lrk,mae,fab,kar,baf,lbf,kal,wlf,cts,us1,nlk,fpt,fal,nws,khv,nsw,sch,scfmt,fms,tdsg,moh,tas,tsvelik,snt}
is motivated in part by purely theoretical reasons (e.g. the crossover
between the 1D and 2D), but also by a plethora of experimental
realizations. Many solids are structurally made up of weakly coupled
ladders \cite{dar}, which leaves a wide temperature range within which
the properties are dominated by one-dimensional physics.  Of
particular relevance are the metallic ladder compounds which include
the ``telephone number" compound Sr$_{14-x}$Ca$_{x}$Cu$_{24}$O$_{41}$
\cite{tnc1,tnc2} and PrBa$_2$Cu$_4$O$_8$ \cite{pbc} as well as members
of the cuprate family Sr$_{n-1}$Cu$_{n+1}$O$_{2n}$ after hole doping
\cite{lac}; for a review of such compounds see Ref.~\cite{dag}.  It was
also suggested that such physics may be seen in the (fluctuating)
stripe ordered phase in certain cuprate high-temperature
superconductors \cite{kiv}. More recently, the development of
nanotechnology has reached the point where systems manufactured in the
laboratory can be ``tailored'' to resemble microscopic models of
interest. Double-chain nanostructures can be manufactured
\cite{cleaved_edge_experiment1} while multi-sub-band quantum wires
\cite{tns,ysw} have a theoretical low-energy description equivalent to
that of ladders \cite{smh}.  Similarly, metallic single wall carbon
nanotubes \cite{ebb} have a low energy description equivalent to that
of a two-leg ladder
\cite{nanotube_theory1,nanotube_theory2,Carr-2008}, the two channels
(legs) originating from the valley degeneracy of graphene. The ladder
geometry can also be created in optical lattices in cold atoms
experiments \cite{cae1,cae2,cae3}.

Taking the viewpoint that ladder models may serve as an intermediate
step between purely 1D and 2D systems, we ask a natural question: does
the dramatic response of interacting 1D systems to a local potential
extend onto the ladder structures as well? In a recent publication
\cite{us2}, we addressed this question in the context of the spinless
ladder (e.g. in the model where the particles hopping on the ladder
are spin polarized electrons). We found that in the case where the
bulk interaction is repulsive, the effect of a generic local potential
is described by the Kane-Fisher scenario leading to vanishing
conductance, $G\rightarrow 0$ at $T\rightarrow 0$. Physically, this
result follows from the fact that the external potential couples to
the local density-wave order parameter which determines leading
dynamical correlations in the ground state of the system.

While this result might have been expected, it has a rather surprising
corollary: if the impurity potential is tuned to the form that {\it
  does not} couple to the dominant order parameter, then the density
wave does not get pinned by the impurity and the ladder exhibits ideal
conductance (at $T=0$). Thus, contrary to a naive expectation, a
potential applied symmetrically at the two opposite sites of a given
rung of the ladder remains transparent \cite{us2}.

In this paper we extend our analysis to the more realistic case where
the charge carriers in the system are real electrons. Our strategy
remains the same as above: to describe transport properties of the
system, we need to (i) identify the nature of the ground state of the
model, (ii) determine which local operator acquires a non-zero
expectation value in a given ground state, and, finally, (iii)
establish whether the external perturbation couples to the dominant
order parameter. Throughout the paper we consider a generic, {\it
  incommensurate} filling.

The problem of a single impurity is essentially that of $2k_F$
backscattering at a single point in the one-dimensional structure.
One can extend this to the case where such a backscattering occurs
uniformly throughout the entire ladder, a perturbation known as an
{\it ionic potential}.  The theoretical attraction of this problem is
apparent: the structure of the gaps and correlations in the ground
state of the unperturbed ladder and the ionic crystal are not
necessarily the same -- and therefore the path from one to the other
may show rich physics with one or more quantum phase transition as
happens in the interplay between the Mott insulator and band insulator
in single chains \cite{fgn}.  Beyond the theoretical interest, there
are also a number of natural experimental realizations of such an
ionic potential.  The telephone number compound \cite{tnc1,tnc2}
consists of both ladders and chains with incommensurate lattice
spacings.  Hence the chains provide an ionic potential for the ladders
within the same system.  By varying the doping, the periodicity of
this potential may be tuned to the Fermi level.  Similarly, ordered
monolayers of atoms adsorbed onto the surface of carbon nanotubes
\cite{ipo} act as an ionic potential on the electrons within the
nanotube.  Furthermore, such ionic potentials may be seen as
intermediate between single impurities and true disordered systems
\cite{rbm}.  In fact, within cold-atom systems in an optical lattice,
the addition of a second incommensurate optical lattice is often used
to mimic disorder \cite{bichromatic1,bichromatic2}, the combined
bichromatic lattice having only quasi-periodicity.

The first step in the above program, i.e.  finding the ground state
correlations of the ladder model, is well studied
\cite{lrk,mae,fab,kar,baf,lbf,kal,wlf,cts,us1,nlk,fpt,fal,nws,khv,nsw,sch,scfmt,fms,tdsg,moh,tas,tsvelik,snt},
with all possible ground states \cite{wlf} and transitions between
them \cite{cts} discussed in the literature. However, the scattered
nature and sheer number of relevant publications in the field presents
a certain challenge when one wants to find the answer to a seemingly
straightforward question: {\it given a set of values of the
  microscopic parameters of the model Hamiltonian, what is the ground
  state of the system and what is the nature of dominant
  correlations?} Having been unable to identify a single source, where
this question could be answered for arbitrary values of the model
parameters, we decided that this paper would be incomplete without an
overview of the phase diagram of the ladder model in the absence of
the perturbation.  In particular, previous works have mostly
concentrated on the limits when inter-chain hopping is either large or
small; to our knowledge, how one of these limits evolves into the
other has not yet been studied.  We therefore spend some time on this
question before studying the effects of the perturbations.

The structure of the paper is as follows: in Sec.~\ref{sec:model} we
introduce the model, review previous literature and summarize our main
results.  The main body of the paper then provides the formalism from
which these results are obtained.  In Section~\ref{lowtperp} we
present the phase diagram for the model of two capacitively coupled
chains, followed by the discussion in Section~\ref{hightperp} of the
phase diagram of the two-leg ladder, where the inter-chain hopping is
sufficiently strong.  Comparing the phase diagrams in
Sections~\ref{lowtperp} and \ref{hightperp}, we notice significant
differences between the two. Therefore, in Section~\ref{pts} we
introduce a concept of a ``hopping-driven phase transition'' and
describe the evolution of the ground state of the model as the
inter-chain hopping parameter $t_\perp$ is increased from zero.

Having identified the necessary properties of the model in the absence
of external perturbations, we turn to the central issue of this paper,
namely the effect of a local perturbation, which we discuss in
Section~\ref{impurity}. In Section~\ref{ionic} we argue that the
analysis of Section~\ref{impurity} can be extended to the case of the
ionic potential. We conclude the paper with a summary and outlook. Technical details are relegated
to Appendices.

\section{The model and summary of results}
\label{sec:model}

In this section, we define the ladder model and external perturbations
that we will consider in this paper. We then present a historic
summary of what is already known about the ladder model before
summarizing the results that will be derived in the remainder of this
paper.

\subsection{The Hamiltonian of the ladder model}

\begin{figure}
\centering
\subfigure[]{\includegraphics[width=6cm]{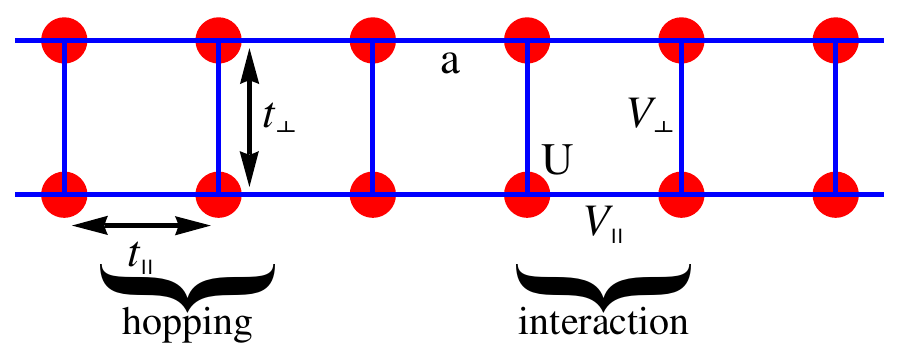}\label{fig:Ladder}}
\hspace{2cm}
\subfigure[]{\includegraphics[width=6cm]{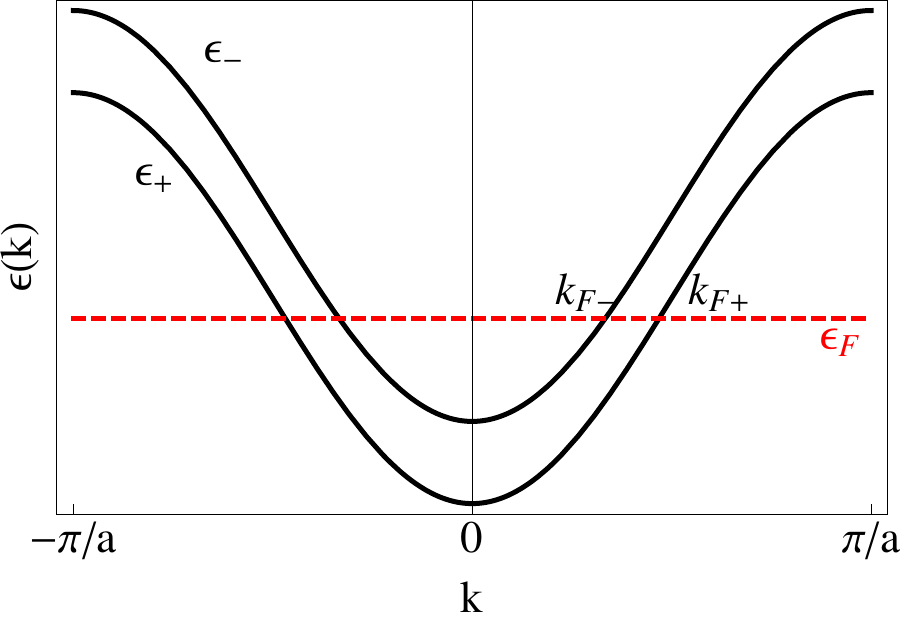}\label{fig:Dispersion}}
\caption{(a) The geometry and interactions of the two-leg ladder.  The
  legs consist of tight binding chains with hopping integral
  $t_\parallel$, and on-site ($U$) and nearest-neighbor ($V_\|$)
  interaction. The legs are then coupled via inter-chain hopping
  $t_\perp$ and interaction $V_\perp$. (b) Dispersion of the two-leg
  ladder. We consider the case of an incommensurate filling such that
  the Fermi level goes through four Fermi points $\pm k_{F+}$ and $\pm
  k_{F-}$.}
\end{figure}

We consider a model Hamiltonian consisting of the single-particle part
${\cal H}_0$ and the interaction terms ${\cal H}_{\text{int}}$
\begin{subequations}
\label{h}
\begin{equation}
{\cal H} = {\cal H}_0 + {\cal H}_{\text{int}}.
\end{equation}
The single-particle part of the Hamiltonian ${\cal H}_0$ represents a
nearest-neighbor tight-binding model
\begin{equation} 
{\cal H}_0 = -\frac{t_{\parallel}}{2} 
\sum_{im\sigma}c_{i,m,\sigma}^\dagger
c_{i+1,m,\sigma}  - t_{\perp} \sum_{i\sigma}
c_{i,1,\sigma}^\dagger c_{i,2,\sigma} + H.c,
\label{h0}
\end{equation}
\noindent
while ${\cal H}_{\text{int}}$ contains an on-site Hubbard term as well
as in-chain and inter-chain nearest-neighbor density-density
interactions 
\footnote{\label{foot-exch}While it is possible (see e.g. Ref~\cite{wlf}) to
  include also an inter-chain exchange interaction $J_\perp \sum_{i}
  \vec{S}_{i,1} \vec{S}_{i,2}$ (where
  $\vec{S}_{i,m}=c^\dagger_{i,m,\sigma}\vec{\sigma}_{\sigma\sigma'}c_{i,m,\sigma'}$
  and $\vec{\sigma}$ is the vector of Pauli matrices), the model
  (\ref{hint}) is quite representative: it is general enough to
  encompass all possible ground states of the two-leg ladder with
  short range interactions respecting the SU(2) spin
  symmetry. Moreover, the inter-chain exchange interaction is
  dynamically generated, see Section~\ref{sec:tperp_chain}.}:
\begin{equation}
{\cal H}_{\text{int}} = U\sum_{i,m}  n_{i,m,\uparrow} n_{i,m,\downarrow} 
+ V_\| \sum_{i,m,\sigma,\sigma'} n_{i,m,\sigma}n_{i+1,m,\sigma'} 
+ V_\perp \sum_{i,\sigma,\sigma'} n_{i,1,\sigma} n_{i,2,\sigma'}.
\label{hint}
\end{equation}
\end{subequations}
In above formulas $c_{i,m,\sigma}$ is the annihilation operator of an
electron with the spin projection $\sigma=\uparrow(\downarrow)$
localized at the site $i$ of the chain $m=1,2$, (hereafter this set of
fermionic operators will be referred to as the {\it chain basis}),
$n_{im\sigma} = c^{\dagger}_{i,m,\sigma}c_{i,m,\sigma}$ are the
occupation number operators, and $t_\|$ and $t_{\perp}$ are the intra-
and inter-chain hopping amplitudes, respectively. This Hamiltonian is
illustrated schematically in Fig.~\ref{fig:Ladder}.

The kinetic part of the Hamiltonian has a spectrum consisting of two cosine bands
\begin{equation}
\epsilon_{\pm}(k) = -t_\| \cos(ka) \mp t_\perp.
\end{equation}
where $a$ is the longitudinal lattice spacing.  In this paper, we
analyze main features of this model at {\it incommensurate} filling
factors such that the Fermi level goes through four Fermi points, as
illustrated in Fig.~\ref{fig:Dispersion}. Consequently Umklapp
processes play no role in our analysis and therefore the unperturbed
model always remains in the metallic regime: while certain collective
modes may acquire a gap, the total charge mode always remains gapless.

\subsection{External perturbations}

The main goal of this paper is to describe the effect of external
perturbations on the ground state and transport properties of the
ladder model (\ref{h}). We will restrict our discussion to
perturbations that couple to the carrier density in the system,
either locally, in which case the perturbation potential
describes a single impurity, or globally which is the case of 
an ionic potential.

\subsubsection{Local perturbation (single impurity)}

A local external perturbation respecting the SU(2) spin symmetry
can be described by the general expression
\begin{equation}
{\cal H}_{\rm imp} = \sum_b \lambda_b \sum_{m,m',\sigma} 
c^\dagger_{i=0,m,\sigma} \tau^b_{mm'} c_{i=0,m',\sigma},
\label{imp}
\end{equation}
where $b=0,x,y,z$ and $\tau^b$ are the Pauli matrices acting in chain space (with $\tau^0$
being the identity). Here the $b=0$ and $b=z$ terms are local analogs
of the two types of charge density waves in the ladder, CDW$^+$ and
CDW$^-$, illustrated in Fig.~\ref{cdwp} below (more details on the
local operators in the model are given in \ref{op}). The $b=x$ term
describes a local fluctuation in the inter-chain hopping amplitude
$t_\perp$ and $b=y$ corresponds to a local (orbital) magnetic field.

A usual scattering center on the chain $m=1$ is described by
Eq.~(\ref{imp}) with $\lambda_0=\lambda_z$, while an impurity on the
chain $m=2$ is described by $\lambda_0=-\lambda_z$; in both cases the
off-diagonal terms $\lambda_x=\lambda_y=0$. For clarity, we eliminate
these off-diagonal terms in \eqref{imp} for the bulk of the
manuscript; generalization of our results to the case
$\lambda_x,\lambda_y\ne 0$ is straightforward.

\subsubsection{Ionic potential}

We now extend the local potential \eqref{imp} periodically to the
whole ladder to obtain a bulk perturbation:
\begin{equation}
{\cal H}_{\rm ion} = \sum_b \lambda_b \sum_{i,m,m',\sigma}  \cos(K a i)
c^\dagger_{i,m,\sigma} \tau^b_{mm'} c_{i,m',\sigma}.
\label{ip0}
\end{equation}
If the modulation wave vector $K$ is commensurate with the density of
particles
\footnote{Of course, at a finite $t_\perp$ there are two separate
  bands in the non-interacting picture, and therefore two separate
  Fermi-points $k_{F\pm}$ (see Fig.~\ref{fig:Dispersion}.  Here we
  define the single $k_F=\bar{n}/2\pi a=(k_{F+}+k_{F-})/2$ simply in
  terms of the filling fraction $\bar{n}$; which also coincides with
  $k_{F\pm}$ at $t_\perp=0$.  },
$K=2k_F$, then the $b=0$ and $b=z$ terms are proportional to the local
operators that serve as order parameters for CDW$^+$ and CDW$^-$,
respectively. Such a perturbation is known as an ionic potential.

If the energy scale associated with the ionic potential is the largest
in the system, then the particles will arrange themselves at the
minima of the cosine potential, with the energy levels forming bands
as in the usual Bloch theory.  The case of commensurability $K=2k_F$
then corresponds to the lowest band being completely filled.  The
system in this limit is a band insulator.  In the absence of ${\cal
  H}_{\rm ion}$, the ground state of the ladder model \eqref{h} is a
conducting state, but typically with a spin gap.  The question then is
whether one can go smoothly between these two limits, or whether there
is necessarily a quantum phase transition at some finite value of
$\lambda_b$.

While at first sight the physics of the ionic potential and that of
the local impurity may seem completely different, we will show they
are related.  The analysis of the local impurity is dominated by the
$2k_F$ backscattering spawned from the perturbation \eqref{imp}. The
physical effect this term has is then determined by how it is affected
by the correlated ground state of the unperturbed system.  Similarly,
the important contribution of the ionic-potential \eqref{ip0} is the
$2k_F$ backscattering -- this is affected by the ground state in the
same way as the local impurity, however because it occurs in the bulk
rather than just at a point it may now have a back effect on this
ground state, potentially giving rise to the aforementioned quantum
phase transitions.

\subsection{Summary of previous work on Ladder models}
\label{el}

To set the scene for the present work, we briefly review the existing
literature on ladder models. Initially, the studies were motivated by
the question: what happens when TL liquids are coupled?  Earlier works
were mainly focused on density-wave structures in arrays of chains
coupled by interaction only (see e.g. \cite{bk, gruner} and references
therein). In the beginning of the nineties, however, the main
attention was shifted towards the possible role of inter-chain
hopping. Most notably, the involved issues were the relevance of the
single-particle hopping \cite{braz,schulz,and-confinement,csa} which may
lead to a confinement phase, along with the importance of the
generated pair-hopping \cite{bac}. This was studied in detail for the
case without spin \cite{nlk}, where it was seen that the confinement
phase manifests itself in a commensurate-incommensurate (C-IC)
transition at which the Fermi-points belonging to the two bands become
split. Furthermore, the generated pair-hopping terms gave rise to
various phases with quasi-long-range order, including the previously
elusive \textit{orbital antiferromagnetic} phase, also known in the
literature as the \textit{staggered flux} or \textit{$d$-density wave}
phase.

When a similar study was conducted for the spinful ladder
\cite{fpt,fal,fab,nws}, something quite different was discovered. It
appeared that, for generic interactions, there was a spin gap in the
spectrum, in strong contrast to the single chain case. Furthermore, it
was demonstrated that even for repulsive interactions dominant
correlations in the system may be of the superconducting type. These
exciting properties led to a renewed interest in the ladder models.
It was soon understood that the spin gap is present for all ladders
with an even number of legs, while those with an odd number of legs,
including the single chain, show gapless spin excitations
\cite{khv,dar}. This is closely related to the existence of a gap
in antiferromagnetic spin chains with an integer spin and its absence
for chains with a half-integer spin \cite{hal_spin}. This implies that
the ground state of the system may be a Haldane spin-liquid
\cite{snt}, a topological phase of matter showing Majorana-fermion
edge states \cite{ntm}.  On top of this, in half-filled ladders with a
ground state of a spin-gapped Mott insulator, already weak doping
causes dominance of superconducting fluctuations -- the fact going in
a remarkable parallel with the two-dimensional cuprate
high-temperature superconductors \cite{dag}. It was therefore believed
that one could gain insight into the properties of high-$T_c$
materials by studying ladder toy models. Indeed, by coupling many
ladders together into a 3D structure, one finds a true phase
transition to a strongly correlated superconductor
\cite{eal2,coke,cat,ram}.

A lot of effort was therefore expended establishing the complete phase
diagram of the two-leg ladder model. Initially, the work concentrated
on the Hubbard model \cite{nsw,sch,baf}; later studies included more
generic interactions, and, in addition to superconducting
correlations, density wave and orbital antiferromagnetic phases were
found \cite{wlf,scfmt,fms}. This gave a complete picture of the phase
diagram \cite{wlf} in the regime where the Fermi-points are split by
the inter-chain hopping term. Studies of the confinement regime and
consequent C-IC transitions were limited to densities close to
half-filling, where the charge (Mott) gap was the catalyst for
confinement \cite{tdsg}.

The analytic studies of the ladders are mostly based on a weak
coupling RG approach and bosonization. This is backed up by numerics
which is particularly useful for tracing the phase diagram at
intermediate and large bare couplings. However, the presence of gaps
means that even weak bare couplings flow to a strong coupling fixed
point which cannot be reliably described using perturbative RG
methods.  The quest to describe the strong-coupling phases led to
various parallel developments in the theory of ladders. One of the
most important is the phenomenon of \textit{dynamical symmetry
  enhancement} \cite{lbf}. This occurs when the low-energy fixed point
of the theory has a higher symmetry than that of the underlying
lattice model. In Ref.~\cite{lbf}, it was shown that, in the case of a
weakly interacting two-leg ladder, the low-energy effective theory has
an O(6)$\times$ U(1) symmetry for the generic incommensurate case; at
half filling it is further enhanced to O(8). Technically this means
that the weak-coupling RG flow converges towards high-symmetry rays
\cite{klls} (perturbations that break the high symmetry down to the
bare lattice symmetry are irrelevant in the RG sense). This
high-symmetry fixed point is often integrable \cite{kal} implying that
the strong-coupling regime may be described non-perturbatively at the
solvable point, while treating irrelevant symmetry-breaking operators
as weak perturbations.

Another concept that is important for undersrtanding of the
strong-coupling phases of the model is \textit{duality}
\cite{bal}. There is not one but many high-symmetry rays in the
parameter space. These correspond to different ground states of the
system and are related to each other by some non-local transformation.
Some of the dualities may even be exact on the lattice \cite{moh}, but
in the most general form they are a property emerging in the
low-energy limit. Now we can interpret the phase transitions between
different ground states as located at the separatrices between the RG
basins of attraction of two of these high-symmetry rays in parameter
space. These represent self-dual points of the corresponding duality
transformations \cite{bal}.

The final important tool developed to understand the strong-coupling
phases is the refermionization approach \cite{snt}. The advantage of
the refermionized theory over abelian bosonization is that the
underlying symmetry of the model remains explicit. This fact
simplifies identification of various criticalities \cite{tas}. 

Such an approach was used in Ref.~\cite{cts} in order to describe a
complete picture of the phase diagram previously propounded in
\cite{wlf}. The phase transitions were all classified into the
universality classes Z$_2$ (Ising), U(1)
(Berezinskii-Kosterlitz-Thouless) and that of the SU(2)$_2$
Wess-Zumino-Novikov-Witten model. The elementary excitations and
excitation spectrum throughout the phase diagram were also calculated.
However, the consideration was limited to the region where the Fermi
points in the two bands were split, i.e. the system was not in a
confined phase. Further works have extended this study to the case
when spin and charge velocities may be very different \cite{tsvelik}.
Experimentally relevant properties such as the NMR relaxation rate
have also been calculated
\cite{ts_nmr,Dora-Gulasci-Simon-Kuzmany-2007,Dora-Gulasci-Simon-Wzietek-Kuzmany-2008}.

Important further developments address the role of
disorder\cite{og1,og2,oas}. One of the main conclusions reached is
that for any repulsive interactions (or weak attractive ones), the
disorder is always relevant driving the system to a localized phase.
Some other works worthy of note are Ref.~\cite{smh}, where disorder
and some transport properties of a two-subband quantum wire were also
studied; Ref.~\cite{scgm} where weak localization in the two-leg
ladder was studied; and the recent publication \cite{ret} which looks
at some aspects of the two-leg ladder in an ionic
potential.

\subsection{Summary of present results}
\label{gen}

For the benefit of the reader not interested in technical details, we
summarize our results here. Our principal results are
three-fold. Firstly, we introduce the notion of hopping-driven phase
transitions, in-particular those relating to confinement. Secondly, we
establish the effect of the single impurity on the ground state and
transport properties of the system. Finally, we extend our results on
the single-impurity problem to the effect of the ionic potential.

These results rely on the precise knowledge of the phase diagram of
the model including the nature of the phase transitions. Although all
of the phases of the ladder model are already known, we have included
a detailed discussion of the phase diagram in order to make the paper
self-contained.

\subsubsection{Hopping-driven phase transitions}
\label{subsubhopping}

Microscopic Hamiltonians describing electronic systems, including
strongly correlated ones, often contain separate single-particle
(e.g. tight-binding or kinetic) and interaction terms (our Hamiltonian
(\ref{h}) is no exception). In a typical experiment, one may vary the
carrier density (e.g. by applying external electro-static potentials),
temperature, pressure, etc., and try to modify the effective strength
of electron-electron interaction. At the same time, the parameters
describing the bare single-particle spectrum are determined by the
crystal properties of the material and are assumed to remain constant
in the course of the measurements (a notable exception is the case of
pressure-induced structural phase transitions \cite{spt}). 

Recently there has been renewed interest in the Hubbard-type models
due to the rapid advances in the experimental techniques related to
optically trapped cold atoms \cite{coa}. In such measurements, almost
any parameter of the optical lattice can be controlled by tuning the
laser fileds that create the optical traps. Thus it is conceivable to
study the evolution of the observable properties of the system with
the change in tunneling amplitudes, and in particular, the interchain
hopping parameter $t_\perp$.

The fact that such evolution should go through critical regions
associated with quantum phase transitions can be seen by comparing the
phase diagrams of the ladder model (\ref{h}) in the two cases of
vanishing and sufficiently large inter-chain hopping shown in
Figs.~\ref{pd1} and \ref{Phase-Diagram}, respectively. The most
obvious differences between the two is the absence of the CDW$^+$ and
TL liquid phases in Fig.~\ref{Phase-Diagram}. At the same time, the
superconducting (SC) phases and the orbital anti-ferromagnet (OAF) can
only appear in the presence of $t_\perp$.

Our primary technical tool is the perturbative (``one-loop'')
renormalization group (RG). The RG equations are formulated on the
basis of the effective low-energy field theory, which in turn relies
upon the details of the single-particle spectrum \cite{gnt,gb}. The
latter undergoes significant changes as the inter-chain hopping is
introduced. Consequently, the RG equations for $t_\perp=0$ and large
$t_\perp$ (given by Eqs.~(\ref{RG1}) and (\ref{RGfive}), respectively)
are rather different. However in the representation used in
Eq.~(\ref{RGfive}), known as the band-basis, the inter-chain hopping
parameter $t_\perp$ itself is not part of the RG equations, as it
couples to a topological current of the theory. As soon as $t_\perp$
is changed, the RG flow has to be re-calculated. Thus, in order to
describe the evolution of the system as $t_\perp$ increases from zero,
we use a two-cut-off scaling procedure, explained in
Section~\ref{rg2}. One may use an alternative approach and derive the
RG equations in the chain basis. Then the operator that couples to
$t_{\perp}$ is a vertex operator with a nonzero conformal spin and
subject to renormalization. We outline how this looks in section
\ref{sec:tperp_chain}, but do not develop this further, since this
alternative procedure does not allow us to follow the phase diagram to
the large $t_\perp$ limit.

Numerical integration of the RG equations within the two-cut-off
scaling procedure results in the $t_\perp-V_\perp$ phase diagram (see
Fig.~\ref{hdpt}), that illustrates the hopping-driven phase
transitions. Note, that the TL liquid phase that exists for
$t_\perp=0$ is unstable with respect to an infinitesimal addition of
inter-chain hopping -- as soon as $t_\perp$ is switched on, the system
forms the $OAF$ phase.

In the most physical case of repulsive interaction and under an
additional assumption $U\gg V_\|$ the system undergoes the first-order
transition between the CDW$^-$ phase at low $t_\perp$ and SC$^d$ phase
at higher $t_\perp$. However, if $U<V_\|$, then the system remains in
the CDW$^-$ phase and there is no phase transition. This result agrees
with the naive expectation of the CDW$^-$ phase being compatible with
the inter-chain hopping, based on the cartoon description of this
phase shown in Fig.~\ref{cdwp}.

On the contrary, the CDW$^+$ phase (also illustrated in
Fig.~\ref{cdwp}) appears to be incompatible with $t_\perp$. As
$t_\perp$ is increased past some low but non-zero value, the system
undergoes an Ising-type (Z$_2$) transition to one of the SC phases
(depending on the value of $V_\perp$). Moreover in a narrow range of
$V_\perp$, the system exhibits a second transition from the SC$^s$
phase to the CDW$^-$ phase. This transition nominally belongs to the
SU(2)$_2$ universality class and corresponds to the change of sign of
the dynamically generated mass of the spin-triplet sector in the
re-fermionised formulation of the effective low energy field theory,
however marginal interaction terms are expected to drive it weakly
first order.

\subsubsection{The effect of a local perturbation}

As in the Kane-Fisher problem \cite{kaf}, there are generically two
possibilities for the response of the system to the local impurity.
Either the impurity is a relevant perturbation completely reflecting
the electrons and driving the system towards insulating behavior
\begin{equation}
\label{insulator}
G(T) \propto T^{\gamma'},
\end{equation}
or the perturbation is irrelevant and one sees only a small correction
to the perfect conductance
\begin{equation}
\label{conductor}
G(T) = \frac{4e^2}{h} - C T^\gamma.
\end{equation}
Here the positive exponents $\gamma$ and $\gamma'$ are nonuniversal
and depend on the interactions in the system.

For a single chain, realization of a particular scenario depends on
the Luttinger parameter $K_c$. Roughly speaking, for repulsive
interactions, $K_c<1$, the insulating case \eqref{insulator} holds,
while for attractive interactions, $K_c>1$, the system remains a
conductor \eqref{conductor}. In Ref.~\cite{us2}, we have shown that
the resulting conductance of a spinless ladder depends not only on the
interactions but also on the local structure of the impurity
potential. For repulsive interactions when the system is in the
CDW$^-$ phase, this results in the curious phenomenon: a single
impurity on one of the legs of the ladder drives the system to an
insulating regime with the low-temperature conductance given by
\eqref{insulator}, whereas a symmetric impurity on both legs keeps the
ladder transparent for transport.

The explanation for this result was quite simple \cite{us2}: in the
former case, the local perturbation couples directly to the local
order parameter operator that describes dominant correlations of the
bulk system (see above Section \ref{subsubhopping} above). In the
latter case, there is no direct coupling -- the impurity looks locally
like the CDW$^+$ while the correlators of these are short ranged in
the bulk CDW$^-$ phase.

The concept of whether the local perturbation couples directly to the
order parameter or not also plays an important role in the present
case of the spinful two-leg ladder. 
There is an important difference however between the present case
and the spinless ladder previously studied.  Within the perturbative RG approach to impurities in the spinful ladder it 
turns out that not only first order but also second order scattering
 terms are relevant when the bulk interactions are 
 repulsive, $K_c^+<1$.
As we will show, this implies that no local potential is
transparent at $K_c^+<1$; the insulating behavior \eqref{insulator}
holds irrespective of the structure of the impurity. The importance of
the local structure of the impurity in the spinful ladder lies in the
fact that the exponent $\gamma'$ does strongly depend on this
structure. In this sense, the metal-insulator transition of the spinless case
transforms to a strong crossover in the present model. 
To be more specific,
in the bulk CDW$^-$ phase and for $K_c^+=1-\delta$, one obtains
$\gamma'\approx 6$ for the case of a single impurity on one chain and
$\gamma'=2\delta$ for the symmetric impurity. A summary of the results
for the exponent $\gamma'$ for both cases as well as the exponent
$\gamma$ for the metallic side (which may occur for attractive
interactions) is given in Table \ref{table-gammas}.

Another difference to the spinless case is that the order parameter
for a large portion of the phase diagram (see
Figs.~\ref{pd1},~\ref{Phase-Diagram} and \ref{hdpt}) is
superconducting.  In particular, for repulsive interactions one
typically has either the CDW$^-$ phase or the $SC^d$ phase, depending
on whether the interchain interaction or the interchain hopping is
dominant. In the latter case, no local density perturbation couples
directly to the order parameter and so the structure of the
perturbation becomes unimportant. On the other hand, this strongly
correlated (quasi-long range order) superconducting state
nevertheless \textit{becomes insulating} (for $K_c^+<1$) when any
impurity is added.

All above considerations apply to energies (temperatures) less than
the dynamically generated gaps in the system, the results being
technically obtained by integrating out the high-energy gapped degrees
of freedom leaving an effective single-channel model for the
current-carrying gapless mode. Close to any of the phase transition
lines (discussed in Section \ref{subsubhopping} above) however, one observes the softening of further modes, leaving a wide temperature range between these new close-to-critical modes and the remaining gapped ones.
The properties of the system then depends crucially on which phase transition we are talking about, and therefore there is a rich variety of
possibilities. Thus transport at or near one of the quantum critical
lines is governed by modified power laws; which are summarized in
Table \ref{table-gamma-transition}. For $K_c^+<1$, we see that the
behavior is again that of an insulator, given by \eqref{insulator}.
Finally, even if the temperature is bigger than all of the gaps, the
system is still an insulator for $K_c^+<1$; in this case, the general
result is given by \eqref{r5}.

\subsubsection{Ionic potential}

As with the local perturbation, we may express an ionic potential
commensurate with the carrier density in terms of the same operators
used to describe the dominant correlations of the unperturbed system.
When the energy scale of the added ionic potential is much larger than
the dynamically generated gaps, the dominant correlations in the
system are clearly going to be of the same type as the potential we
have explicitly added. If the unperturbed ground state of the system
showed different correlations, then the system must evolve from one
sort of correlation to the other as a function of the applied
perturbation $\lambda_b$ in Hamiltonian \eqref{ip0}.

This evolution may happen in one of two ways -- it may be smooth; or
it may undergo certain quantum-phase transitions en route.  Examples
of such phase transitions are known in similar situations in the
literature -- for example, the SU(2)$_1$ transition in dimerized spin
ladders \cite{mss,wan} or zig-zag carbon-nanotubes
\cite{Carr-Gogolin-Nersesyan-2007}; or the \textit{two} transitions
between the Mott insulator and the band insulator in a half-filled
single chain in an ionic potential \cite{fgn}.

Our aim in this work is to categorize the transitions that may or may
not take place when an ionic potential is added to the doped
(incommensurate) two-leg ladder.  Clearly, this depends on both the
ground state of the unperturbed system, as well as the type of ionic
potential added. To be specific, we consider ionic potentials
coupling to the local density (of either CDW$^+$ or CDW$^-$ types)
added to a system with any of the five possible ground states
discussed in this work. To avoid too many cases, we also limit
ourselves to the situation when $K_c^+<2$.

In all cases we consider, we find that the first effect of adding the
ionic potential, $\lambda_b \ne 0$, is that the previously critical
total charge mode $\phi_c^+$ becomes gapped.  In other words, the
system immediately exhibits true long range order and becomes an
insulator.  This doesn't contradict the Hohenberg-Mermin-Wagner
theorem \cite{maw} for there is no spontaneous symmetry breaking --
the external ionic potential explicitly breaks the translational
symmetry.

We then proceed to identify any quantum phase transitions that may
take place as $\lambda_b$ becomes larger, see Table \ref{table-ionic}.
As expected, we find that if the ionic potential has the same
structure as the already existing correlations, then there is no
possibility of a transition. We also find that if the ground state is
in one of the superconducting phases, the evolution is smooth. In all
other cases however, we find there is some critical $\lambda_b^*$ at
which the system becomes critical; the universality class for each
case is listed in Table \ref{table-ionic}.


\section{Capacitively coupled chains}
\label{lowtperp}

Consider a situation where two individual quantum wires are placed
close enough to each other such that the Coulomb interaction between
electrons in different wires is strong enough and, at the same time,
there is no direct contact between the wires preventing the electrons
from tunneling from one wire into another. This system can be
described by the ladder Hamiltonian (\ref{h0}), (\ref{hint}) in the
absence of inter-chain hopping
\[
t_\perp = 0,
\]
and was first studied in Ref.~\cite{lrk}. In this case, the Fermi
momenta for all four channels remain the same, and the single-particle
Hamiltonian (\ref{h0}) is diagonal at the outset. Therefore it is
natural to discuss the case of $t_\perp=0$ using the original chain
basis. We first briefly review the case of two completely decoupled
chains, before adding the inter-chain interaction $V_\perp$ and
analyzing the effect such a term has on the ground state of the
system.

\subsection{Two independent chains}

In the absence of interchain hopping the two chains are connected only
by the interaction $V_\perp$. If one sets $V_{\perp} = 0$, then the
model reduces to two copies of the 1D extended Hubbard model. Its
bosonized form is well-known \cite{gnt} (see \ref{1leg} for details
and notations). The most prominent feature of this model is
spin-charge separation. In particular, the effective low-energy theory
(\ref{bh2}) contains two decoupled sectors describing collective
charge and spin degrees of freedom.

The case of two independent chains is rather trivial -- we now have
two copies of charge and spin fields (\ref{csf}), which we denote as
$\phi_{m,c}$ and $\phi_{m,s}$ ($m=1,2$ is the chain index).  Thus for
$V_\perp=t_\perp=0$ the model Hamiltonian can be written in the form
(\ref{ehboz}) where the charge sector contains two copies of the
Gaussian model
\begin{subequations}
\label{xhm}
\begin{equation}
{\cal H}_c =\frac{v_F}{2} \sum_m
\left[ \Pi_{m,c}(x)^2 + \left(1-\frac{h_c}{\pi v_F}\right) 
         \left[ \partial_x \phi_{m,c}(x) \right]^2 \right],
\end{equation}
and the spin sector -- two copies of the sine-Gordon model
\begin{eqnarray}
&&
{\cal H}_s = \frac{v_F}{2} \sum_m \left[ \Pi_{m,s}(x)^2 + 
\left(1-\frac{h_s}{\pi v_F}\right) 
\left[ \partial_x \phi_{m,s}(x) \right]^2 \right] 
+ \frac{h_s}{2(\pi \alpha)^2} \sum_m
\cos \sqrt{8\pi} \phi_{m,s}(x).
\label{hsh}
\end{eqnarray}
\end{subequations}
As the chains are identical, the interaction parameters are
independent of the chain index $m$. In terms of the original lattice
coupling constants they are given by
\begin{eqnarray}
&&
h_c = - aU - 2aV_\parallel 
\left[ 2 -  \cos (2k_F a) \right], 
\nonumber\\
&&
\nonumber\\
&&
h_{s} = aU + 2aV_\parallel\cos(2k_F a).
\label{eq:sg}
\end{eqnarray}
Notice that the same constant $h_s$ appears in two different terms in
Eq.~(\ref{hsh}). This is a consequence of SU(2) symmetry. In fact,
Eqs.~\eqref{xhm} is more general than the original lattice model, and
one may consider the decoupled legs of the ladder parameterized via
$h_c$ and $h_s$ rather than $U$ and $V_\|$.

\begin{figure}
\begin{center}
\begin{tabular}{c}
\includegraphics[width=8cm]{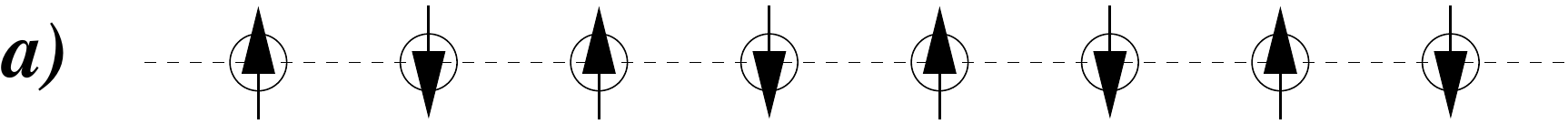} \\
\vspace*{5pt} \\
\includegraphics[width=8cm]{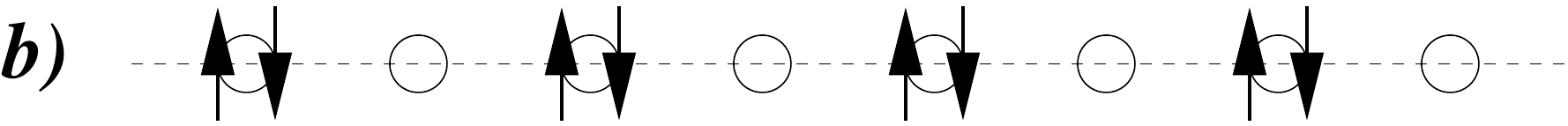}
\end{tabular}
\end{center}
\caption{Cartoon depictions of the two possible phases of the extended
  Hubbard model with repulsive interactions.  In a), the on-site
  interaction is dominant ($h_s>0$), in this case the N\'{e}el order
  is quasi-long range and the spectrum remains gapless.  In b) the
  nearest-neighbor interaction is dominant ($h_s<0)$, in this case the
  singlets don't break any symmetry and the spectrum acquires a finite
  spin gap. In these cartoons (as well as other similar depictions
  below), the circles do not represent lattice sites (we are
  incommensurate), but instead indicate fluctuating charge positions
  an average of $\pi a/2k_F$ apart.}
\label{eh1}
\end{figure}

At $h_s>0$ the perturbation is marginally irrelevant, the model flows
to weak coupling, and the spectrum remains gapless. At $h_s<0$ the
perturbation is relevant, the model flows to strong coupling, and the
spin sector acquires a spectral gap. Note that for fillings between
$1/4<n<1/2$, $\cos(2k_Fa)<0$, so both cases can occur even for purely
repulsive bare interactions (in fact the upper limit of fillings is
$3/4$ but for the sake of concreteness, we will always deal with the
case of less than half filling), depending on whether $U$ or $V$ is
dominant. These two different phases are schematically illustrated in
Fig.~\ref{eh1}.

It is easy to understand within a strong-coupling picture (see
e.g. Fig.~\ref{eh1}) why the interactions $U$ and $V$ are in
competition with each other for fillings between $1/4$ and $1/2$. The
on-site interaction $U$ makes double occupancy energetically
unfavourable -- which for this range of fillings means particles must
be placed in nearest neighbor positions. Such neighboring particles
feel an exchange interaction; which means that in this phase the
dominant correlations are those of the spin-density wave type.  The
interaction $V$ favours opposite configurations, where particles don't
sit in neighboring positions; and therefore there must be some double
occupancy.  In this case, the dominant correlations are of the
charge-density wave type, while the doubly occupied singlets lead to
the presence of a spin-gap.  The transition at the point
$U=-2V\cos(2k_Fa)$ (i.e. $h_s=0$) is exactly of the BKT type.

\subsection{Inter-chain interaction}

Consider now the interchain interaction $V_\perp$ described by the
last term in Eq.~(\ref{hint}). Its bosonized version can be obtained
using representation (\ref{cdd}) for the fermion density (with all
bosonic fields supplied with the additional chain index). The result
is a sum of two terms
\[
\sum_{\sigma\sigma'} n_{i,1,\sigma}n_{i,2,\sigma'} \rightarrow
S_{\rm fs} + S_{\rm bs}.
\]
The first term is the interchain forward scattering
\[
S_{\rm fs} = \frac{1}{\pi} \partial_x \phi_{1,c} \partial_x \phi_{2,c}.
\]
The second term is the interchain backscattering similar to
Eq.~(\ref{bs}): 
\begin{eqnarray*}
&&
S_{\rm bs} = \cos\sqrt{2\pi} \left[ \phi_{1,c} - \phi_{2,c} \right] 
\cos\sqrt{2\pi} \left[ \phi_{1,s} + \phi_{2,s} \right]
\\
&&
\\
&&
\quad\quad\quad
+
\cos\sqrt{2\pi} \left[ \phi_{1,c} - \phi_{2,c} \right] 
\cos\sqrt{2\pi} \left[ \phi_{1,s} - \phi_{2,s} \right].
\end{eqnarray*}
Now we rearrange the bosonic fields into four linear combinations that
correspond to the total and relative charge, as well as total and
relative spin:
\begin{equation}
\label{rcs}
\phi_{c(s)}^{\pm} = \frac{\phi_{1,c(s)} \pm \phi_{2,c(s)}}{\sqrt{2}}.
\end{equation}
This transformation simplifies the interchain backscattering terms and
at the same time diagonalizes the quadratic part of the Hamiltonian,
which now contains four copies of the Gaussian model:
\begin{subequations}
\label{ccc}
\begin{equation}
\label{gm}
{\cal H}_0 = \frac{v_F}{2}
\sum_{\substack{\mu=c,s\\\nu=\pm}}  
\left[ \Pi_\mu^\nu(x)^2 + 
\left(1-\frac{g_\mu^\nu}{\pi v_F}\right) 
\left[ \partial_x \phi_\mu^\nu(x) \right]^2 \right].
\end{equation}
The remaining part of the bosonized Hamiltonian contains the inter-
and intra-chain backscattering terms
\begin{align}
&
{\cal H}_{\rm bs} =
\frac{g_1}{(\pi\alpha)^2} \cos\sqrt{4\pi}\phi_s^+\,\cos\sqrt{4\pi}\phi_s^- 
\nonumber \\
&
\nonumber\\
&
\quad\quad
+
\frac{g_2}{(\pi\alpha)^2} \cos\sqrt{4\pi}\phi_c^- 
\left[\cos\sqrt{4\pi}\phi_s^+ + \cos\sqrt{4\pi}\phi_s^-\right].
\label{hamBS}
\end{align}
\end{subequations}

Backscattering terms (\ref{hamBS}) couple the relative charge mode
($\phi_c^-$) to the total ($\phi_s^+$) and relative ($\phi_s^-$) spin
modes. The corresponding coupling constants in the Hamiltonian
(\ref{ccc}) may be parameterized as
\begin{subequations}
\label{bare1}
\begin{equation}
  g_c^{-} = h_c + h_\perp, \quad g_s^{\pm} = g_1 = h_s, \quad g_2 = h_\perp,
\label{gs}
\end{equation}
where 
\begin{equation}
h_\perp = aV_\perp \label{hs}
\end{equation}
while $h_c$ and $h_s$ are defined in Eqs.(\ref{eq:sg}).
\end{subequations}
The first two parameters in Eq.~(\ref{hs}) coincide with
Eq.~(\ref{eq:sg}) reflecting the fact, that for $V_\perp=0$ the
Hamiltonian~(\ref{ccc}) describes two uncoupled chains
[c.f. Eqs.~(\ref{xhm})]. In fact, this parameterization is valid for
any two spinful TL-liquids coupled via a density-density interaction,
where $h_c$ and $h_s$ parametrize the chains, and $h_\perp$ the
interchain coupling. We will therefore retain this trio
$h_c,h_s,h_\perp$ throughout the paper as a convenient way to
parameterise the phase diagram of the model.

The total charge mode ($\phi_c^+$) is decoupled from the rest of the
spectrum and hence describes a TL liquid with the Luttinger parameter
\begin{equation}
\label{kcplus}
K_c^+ \approx 1 + \frac{g_c^+}{2\pi v_F}.
\end{equation}
As this algebraic relation only holds for weak interactions $|g_c| \ll
1$, while the TL liquid concept is much more general, it is therefore
appropriate to think of the Luttinger parameter $K_c^+$ as the fourth
independent parameter of the ladder\footnote{For the lattice model
  (\ref{h}) the Luttinger parameter can be related to the $h$'s in
  Eq.~(\ref{bare1}) by means of the equality $g_c^+ =
  h_c-h_\perp$. However, this equality is not general and may be
  changed by adding further interactions; for example in
  carbon-nanotubes where the Coulomb interaction is poorly screened,
  $g_c^+$ may become strongly modified by the long range tail of the
  interaction, while the coupling constants in the other sectors of
  the theory remain largely unaffected \cite{nanotube_theory1}.}.
There is another reason to consider $K_c^+$ independently. Much of
our analysis of the Hamiltonian \eqref{ccc} will be based on the
renormalization group, where the parameters $h$ will be allowed to
flow. Clearly however, no renormalization occurs in the decoupled
$\phi_c^+$ sector which is described by the Gaussian model.

\subsection{Refermionized form of the spin sector and symmetries of the model}
\label{refss}

While the bosonized representation \eqref{ccc} is extremely convenient
for a renormalization group analysis, the representation in terms of
four scalar bosonic fields obscures the symmetries of the model,
particularly in the spin sector where we would expect at least that
the SU(2) symmetry is still present, which is not at all apparent in
\eqref{ccc}. For this reason, we give another representation of the
Hamiltonian here, in which the spin sector is \textit{re-fermionized}.
The physical idea behind this technique ultimately goes back to the
seminal works by S.~Coleman \cite{coleman} and Itzykson and Zuber
\cite{iz} in which the equivalence between the sine-Gordon model at
the decoupling point ($\beta^2 = 4\pi$), a free massive Dirac fermion,
and a pair of noncritical two-dimensional Ising models was
established. According to this correspondence the spin sector of the
ladder model (\ref{h}) is equivalent to four weakly coupled quantum
Ising chains \cite{iz}. The latter can be represented in terms of
Majorana fermions \cite{snt,gnt,cft}. While ultimately, this is just
another way of writing the same Hamiltonian, we will see that in the
Majorana representation the symmetries of the Hamiltonian are made
explicit.

We now introduce four Majorana fermion fields $\xi^i_{R(L)}$ to
replace the two bosonic spin fields $\phi_s^\pm$ according to the
re-fermionization rules in (\ref{rf}). The Hamiltonian is split up
into a kinetic (Gaussian) part and interaction (non-linear) terms,
\begin{subequations}
\label{cbrf}
\begin{equation}
{\cal H}={\cal H}_k + {\cal H}_\mathrm{int},
\end{equation}
which are grouped slightly differently as compared to the fully
bosonized case. The kinetic part still includes two Gaussian models
for the bosonized charge modes $\phi^{\pm}_c$ with the forward
scattering amplitudes $g^{\pm}_c$ explicitly included (see
Eq.~(\ref{gm})), and four massless Majorana fields for the spin modes
\begin{equation}
\label{hrf-k}
{\cal H}_k = {\cal H}_0 \left[ \phi_c^+ \right] 
+ {\cal H}_0 \left[ \phi_c^- \right] +
\frac{i v_s }{2} \sum_{a=1,2,3,0}
\left( \xi^a _L \partial_x \xi^a _L -  \xi^a _R \partial_x \xi^a _R \right),
\end{equation}
Marginal interaction in the spin sector, parametrized by $g_{1,2}$, is
included in $ {\cal H}_\mathrm{int}$:
\begin{equation}
\label{hrf-int}
{\cal H}_\mathrm{int} = 
-g_1 \left(\sum_{a=0}^3 \xi^a _R \xi^a _L\right)^2 + \frac{i g_2}{\pi \alpha} 
\cos \sqrt{4\pi} \phi_c^- \sum_{a=0}^3 \xi^a _R \xi^a _L,
\end{equation}
\end{subequations}

There are four independent coupling constants: two of them describe
forward scattering in the charge sectors ($g_c^+$ and $g_c^-$), and
two more describe the spin sectors ($g_1$ and $g_2$), the bare values
being given by Eq.~\eqref{bare1} as before. In the absence of the
interchain hopping the spin symmetry of the ladder model is that of
two decoupled chains: SU(2) $\times$ SU(2) $\approx$ O(4); this is
unbroken by an interchain density-density interaction. This is why the
Hamiltonian contains only O(4) invariants: $(\sum_{a=0}^3 \xi^a _R
\xi^a _L)^2$ which originates from the intra-chain interaction, as
well as the mixed term $\cos \sqrt{4\pi} \phi_c^- \sum_{a=0}^3 \xi^a
_R \xi^a _L$. When interchain hopping processes are allowed, we will
see that they generate interchain exchange terms which reduce the spin
symmetry from SU(2)$\times$ SU(2) to a single SU(2). As we will see
in section \ref{sec:tperp_chain}, the O(4) symmetry present in the
four Majorana fields will split into a spin-triplet mode described by
$\xi^i$ with $i=1,2,3$ and a spin-singlet degree of freedom described
by the remaining $i=0$ fermion.

\subsection{Phase diagram}

We now proceed to construct the phase diagram of the model of
capacitively coupled chains described by Hamiltonian \eqref{h} with
$t_\perp=0$. We will do this carefully using renormalization group
methods in the following, but to begin with let us consider a simple
strong coupling picture in the atomic limit. If the interchain
interaction is weak, then a good reference point is the state of each
individual chain. The single-chain model is described in detail in
\ref{1leg} and illustrated in Fig.~\ref{eh1}.

\begin{figure}
\centering
\subfigure[]{\includegraphics[width=7cm]{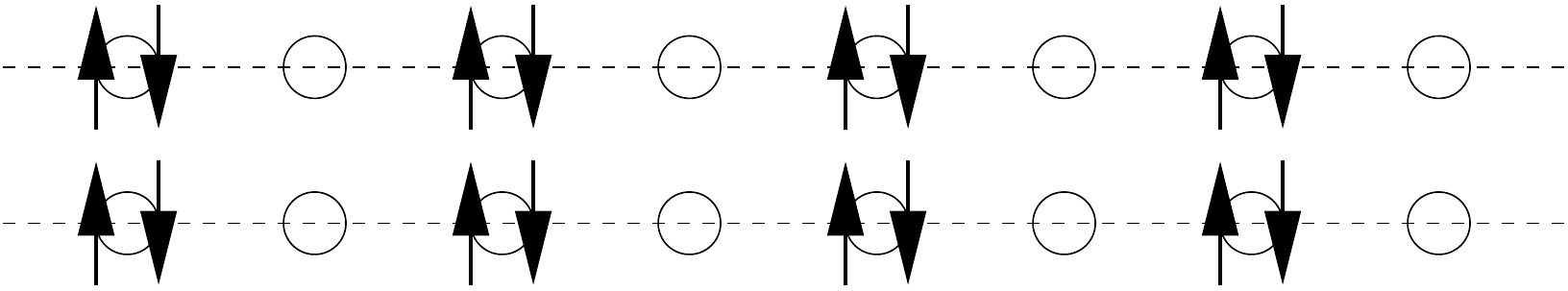}}\hspace{2cm}
\subfigure[]{\includegraphics[width=7cm]{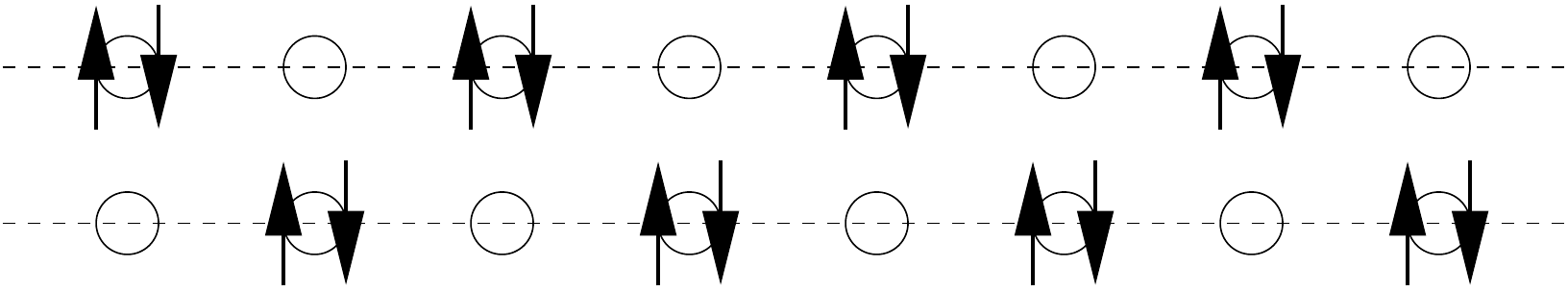}}
\caption{Cartoon depiction of the (a) CDW$^+$ phase and (b) CDW$^-$
  phase of the ladder model. The circles represent fluctuating charge
  positions an average of $k_F^{-1}$ apart.}
\label{cdwp}
\end{figure}

Let us first suppose that $V_\| > U$ (or more precisely, $h_s<0$).
Then each chain exhibits quasi-long-range order of the CDW type as
illustrated in Fig.~\ref{eh1}b. When such chains are coupled by
interchain interaction $V_\perp$, it is clear that all potential
energy terms may be minimized by forming density waves on both changes
with a relative phase of $0$ for attractive interchain interaction and
$\pi$ for the repulsive case. Such states are known as CDW$^+$ and
CDW$^-$ respectively, and are illustrated in Fig.~\ref{cdwp}.
 
In the opposite case when $V_\|< U$ ($h_s>0$) and the dominant
correlations in the uncoupled chains are SDW rather than CDW, the
situation is less obvious as there is now competition between the
different interaction terms.  In fact, the SDW state in the single
chain is quite delicate -- as the SU(2) symmetry prevents the
appearance of a spin-gap in this state.  It will therefore turn out
that the SDW is unstable to arbitrarily small interchain interaction
-- the resulting state being either of the CDW$^\pm$ states mentioned
previously, or remaining a TL liquid with no dynamically acquired
gaps. In order to analyze this, we will now study more closely the
structure of the order parameters.

\subsubsection{Structure of order parameters and semi-classical analysis}
\label{qa}

The general form of the local density operators for the spinful ladder
is given in \ref{op}. At present, we are interested in the charge density
wave (CDW$^\pm$) phases. The corresponding order parameters describe
the staggered (oscillating) components of the total and relative charge
density operators
\[
CDW^+ \leftarrow \sum_{\sigma} \left( c^\dagger_{1\sigma} c_{1\sigma} + c^\dagger_{2\sigma} c_{2\sigma} \right),
\quad
CDW^- \leftarrow \sum_{\sigma} \left( c^\dagger_{1\sigma} c_{1\sigma} - c^\dagger_{2\sigma} c_{2\sigma} \right),
\]
and in the bosonic representation are given by
\begin{align}
{\cal O}_{CDW^+} 
& \sim
e^{-i\sqrt{\pi}\phi_c^+}
\left[\sin \sqrt{\pi} \phi_c^- 
\sin \sqrt{\pi} \phi_s^+ \sin \sqrt{\pi} \phi_s^-
-i
\cos \sqrt{\pi} \phi_c^- 
\cos \sqrt{\pi} \phi_s^+ \cos \sqrt{\pi} \phi_s^-\right],
\nonumber\\
&
\nonumber\\
{\cal O}_{CDW^-} 
& \sim
-e^{-i\sqrt{\pi}\phi_c^+}
\left[
\sin \sqrt{\pi} \phi_c^- 
\cos \sqrt{\pi} \phi_s^+ \cos \sqrt{\pi} \phi_s^-
-i
\cos \sqrt{\pi} \phi_c^- 
\sin \sqrt{\pi} \phi_s^+ \sin \sqrt{\pi} \phi_s^-
\right].
\label{opsboz1}
\end{align}
Notice that the bosonized form of each local operator has a
multiplicative structure involving the fields in each of the four
sectors of the theory
\[
{\cal O} = \sum {\cal O}_{c^+}{\cal O}_{c^-}{\cal O}_{s^+}{\cal O}_{s^-},
\]
and as such, $\langle {\cal O}\rangle = 0$ because the gapless and
decoupled total charge sector always has $ \langle{\cal
  O}_{c^+}\rangle\ne 0$. In other words there is no true long-range
order in the system, which is a direct consequence of the
Hohenberg-Mermin-Wagner theorem \cite{maw}. On the other hand, the
notion of quasi-long-range order of the type ${\cal O}$ assumes a
situation when the correlator $\langle {\cal O} (r) {\cal O}
(0)\rangle$ displays the slowest (among all others) power-law
decay. This is what is also called \textit{algebraic order}. Then we
say that the dominant correlations are those of the operator ${\cal
  O}$. There may be other order parameters with power-low asymptotics
but they decay faster; the remaining order parameters decay
exponentially with a finite correlation length.

We also notice that each operator ${\cal O}$ has a structure of a
product of cosines and sines of the fields added to another product
with the cosines and sines reversed. The origin of this construct is
consistent with the expression Eq.~(\ref{hamBS}) for the
backscattering Hamiltonian. While the latter is invariant under the
transformation $\sqrt{4\pi} \phi_a \rightarrow \sqrt{4\pi} \phi_a +
\pi$ , the operators $\cos\sqrt{\pi} \phi_a$ and $\sin\sqrt{\pi}
\phi_a$ get interchanged in {\em all sectors simultaneously}.  In
order to demonstrate this point, we perform a semi-classical analysis
of the Hamiltonian \eqref{hamBS} by evaluating the field
configurations that minimize the energy. Let us first suppose that
$g_1<0$ (i.e. $h_s<0$ so that each chain forms a CDW). Then to
minimize the potential energy Eq.~\eqref{hamBS}, one locks the fields
$\phi_s^\pm$ either at the set of points $\phi_s^+ = n\sqrt{\pi}$,
$\phi_s^-=m\sqrt{\pi}$; or the set of points $\phi_s^+ = (n+1/2)
\sqrt{\pi}$, $\phi_s^+ = (m+1/2) \sqrt{\pi}$ for integers $n,m$. In
view of the symmetry above, let us pick the first set of points, for
which $\cos\sqrt{4\pi}\phi_s^+=\cos\sqrt{4\pi}\phi_s^-=1$. 
\footnote{We note here that, due to quantum fluctuations, the actual
  expectation values of the cosines of bosonic fields will be less
  than 1.  However, as long as the cosine term is relevant and we
  remain in the strong coupling regime, such semiclassical estimate
  gives a qualitatively correct answer.}  
Now consider the role of the inter-layer interaction, i.e. the $g_2$
term in Eq.~(\ref{hamBS}). If $g_2<0$, then $\cos\sqrt{4\pi}\phi_c^-$
also wants to take the value $+1$. If, on the other hand, $g_2>0$ then
the opposite value $\cos\sqrt{4\pi}\phi_c^-=-1$ is preferred. Thus for
$g_1<0$ both back-scattering terms in Eq.~(\ref{hamBS}) may be
minimized simultaneously.

To understand the resulting phases of the system we turn to the order
parameters.  If both back-scattering coupling terms are negative, then
the three fields $\phi_c^-$, $\phi_s^+$, and $\phi_s^-$ are locked at
e.g. $0$.  Consider now the operator ${\cal O}_0$. The product
$\cos\sqrt{\pi}\phi_c^-\cos\sqrt{\pi}\phi_s^+\cos\sqrt{\pi}\phi_s^-$
now acquires an expectation value and the correlation function
\[
\langle {\cal O}_0(x){\cal O}_0(0)\rangle \sim  
\left[ \langle \cos\sqrt{\pi}\phi_c^-\cos\sqrt{\pi}\phi_s^+\cos\sqrt{\pi}\phi_s^- \rangle \right]^2 
\langle  e^{-i\sqrt{\pi}\phi_c^+(x)}  e^{-i\sqrt{\pi}\phi_c^+(0)} \rangle
\sim  \frac{1}{|x|^\gamma},
\]
will exhibit an algebraic decay with the critical exponent
$\gamma=K_c^+/4$ contributed by the gapless modes of the total charge
sector. All other operators, for example ${\cal O}_z$ involve either
the sines of the above three fields or their dual fields. Therefore
the correlation functions of those operators will decay exponentially
with a finite correlation length
\[
\langle {\cal O}_z(x){\cal O}_z(0)\rangle \sim e^{-|x|/\xi}.
\]
Thus for $g_1, g_2<0$ the dominant correlation is of the CDW$^+$ type
and hence we refer to this phase as CDW$^+$.  If $g_1<0, g_2>0$, then
the situation is just the opposite, and the phase is CDW$^-$, in
agreement with the strong coupling atomic limit previously expounded.
This situation is similar to the spinless case \cite{us1}: if the
dominant order within each chain is CDW-like, then when they are
brought together a repulsive inter-chain interaction will lock the
CDWs with a relative phase $\pi$ while an attractive inter-chain
interaction will lock them with relative phase $0$.

Now we turn to the case $g_1>0$, where the dominant correlations on
the individual chains (in the absence of inter-chain coupling) are of
the SDW type. In this case, it is clear that one cannot
simultaneously minimize both the $g_1$ and the $g_2$ terms in the
back-scattering Hamiltonian, \eqref{hamBS}. Na{\"\i}vely one might
expect a transition between the SDW state (minimizing $g_1$) at
$g_1>2|g_2|$ to the CDW state which minimizes the $g_2$ term at
$g_1<2|g_2|$. However, such an analysis is too simplistic. In fact,
the SDW that would semi-classically minimize the $g_1$ term at $g_1>0$
is critical (unlike the case $g_1<0$) and is thus very sensitive to
perturbations.  Therefore for $g_1>0$ a more rigorous analysis is
required.

\subsubsection{RG equations in the chain basis and phase diagram}
\label{sec:RGwithout_tperp}

To obtain the phase diagram of the model of two capacitively coupled
chains (\ref{ccc}), we now employ the renormalization group technique.
Apart from the total charge sector which is completely decoupled from
the rest of the spectrum, the model contains three independent
coupling constants (\ref{bare1}). It turns out to be instructive to
express the RG equations in terms of the parameters $h_j$ given in
\eqref{hs}, remembering that $h_c$ and $h_s$ parameterise the
intra-chain interactions, while $h_\perp$ is the strength of
inter-chain interaction. The RG equations then take the form
\begin{align}
\frac{\partial h_c}{\partial l} &= \frac{h_\perp}{2} \left( h_c + 3 h_s -3 h_\perp \right), 
\nonumber\\
&
\nonumber\\
\frac{\partial h_s}{\partial l} &= - h_s^2-h_\perp^2 ,
\nonumber\\
&
\nonumber\\
\frac{\partial h_\perp}{\partial l} &= - \frac{h_\perp}{2}  \left( h_c + 3 h_s + h_\perp \right) ,
\label{RG1}
\end{align}
with the initial values given in Eq.~(\ref{hs}).

It is immediately clear that the plane $h_\perp=0$ is invariant,
i.e. if there is no initial inter-chain coupling, none is generated
under the RG flow, and the model is equivalent to two decoupled copies
of the one-dimensional chain.  On this plane the equations (\ref{RG1})
reduce to the SU(2) BKT equation $\frac{\partial h_s}{\partial l} =
-h_s^2$, which describes the physics of the spin sector of the
extended Hubbard model (\ref{sg}), discussed in Appendix~\ref{1leg}.

It is also clear that there is a line of weak-coupling fixed points
\[
h_s=0, \quad h_\perp=0,
\]
where $h_c$ may remain arbitrary. At these fixed points all sectors of
the theory are Luttinger liquids. The RG possesses also strong
coupling fixed points
\[
h_s \to -\infty, \quad h_\perp\to \pm\infty.
\]
These strong coupling fixed points correspond exactly to the scenario
where the bosonic fields are locked at the minima of the cosine
potentials as analysed above.  Therefore this is a phase where all
sectors except total charge are gapped with the dominant order being
of the CDW type. For $h_\perp>0$ this is CDW$^-$, while for
$h_\perp<0$ it is CDW$^+$. By following the RG flow as a function of
bare parameters, we can therefore determine the phase diagram of the
model.
\begin{figure}
\begin{center}
\includegraphics[width=6cm]{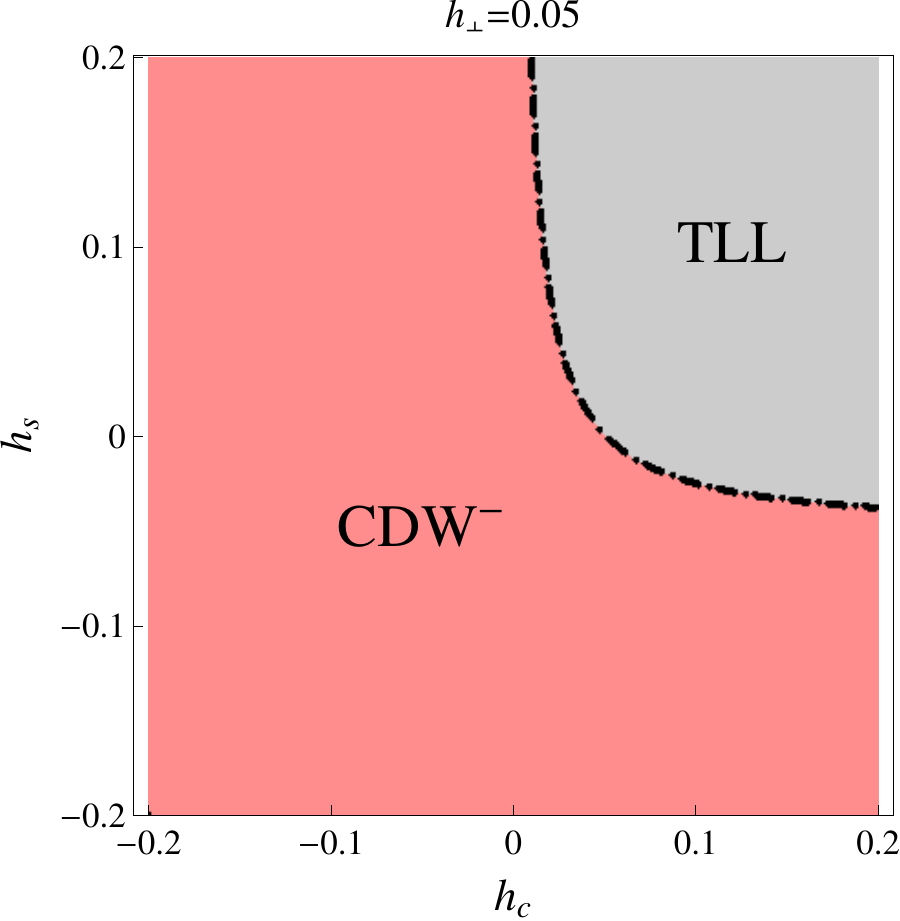} 
\hspace{1cm}
\includegraphics[width=6cm]{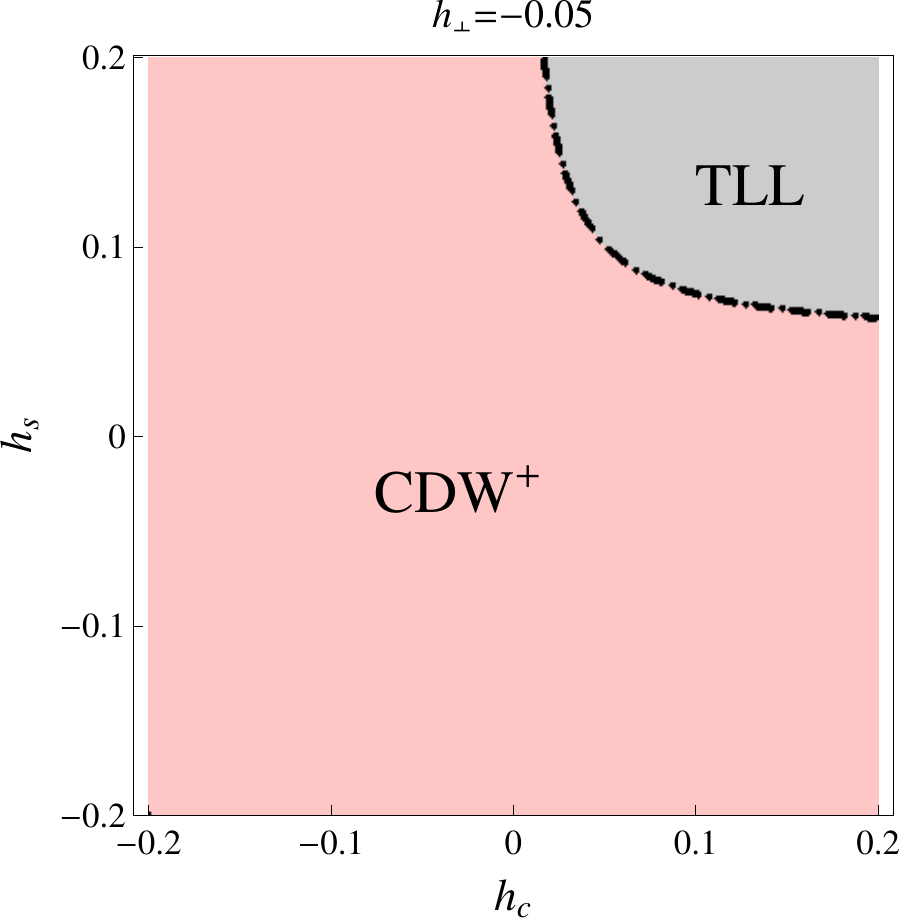} 
\end{center}
\caption{The phase diagram of the system of two capacitively coupled
  chains for fixed inter-chain interaction (left panel corresponds to
  $h_\perp=0.05$, right panel to $h_\perp=-0.05$). Here CDW$^+$ and
  CDW$^-$ are the charge-density waves illustrated in Figs.~\ref{cdwp}
  and TLL stands for the Tomonaga-Luttinger liquid. }
\label{pd1}
\end{figure}

In fact, the boundary between the weak- and strong-coupling phases may be
determined analytically.  As demonstrated in \ref{sec:RGappendix}, 
 the RG possesses an invariant plane
\begin{equation}
\label{ip}
 \quad h_s(h_\perp+h_c)=h_\perp^2.
\end{equation}
which is in fact exactly the phase boundary between the basin of
attraction of the strong and weak coupling fixed points.  The phase
diagram can thus be conveniently parameterised as
\begin{eqnarray}
 \quad h_s(h_\perp+h_c) > h_\perp^2 & : & \mathrm{Weak\ coupling - TLL} \nonumber\\
h_\perp>0, \quad h_s(h_\perp+h_c) < h_\perp^2 & : & \mathrm{Strong\ coupling - CDW}^- \nonumber\\
h_\perp<0, \quad h_s(h_\perp+h_c) < h_\perp^2 & : & \mathrm{Strong\ coupling - CDW}^+
\end{eqnarray}
This resulting phase diagram of the system of two capacitively coupled
chains is illustrated in Fig.~\ref{pd1}.

We see therefore that even in the case that $h_s>0$ and the dominant
correlations on the individual chains before they are coupled was SDW
and not CDW, the interchain coupling may lead to a CDW state being
formed.  We note however that in this case, the resulting dynamically
acquired gap is much smaller than in the case when the bare $h_s<0$
and there was no competition between the different interaction terms.

\subsection{Interchain hopping term in the chain basis}
\label{sec:tperp_chain}

Practical calculations are often performed using a particular set of
basis operators. In principle, physical results have to be independent
of the choice of the basis. On the other hand, this choice does affect
technical complexity of the calculations, which can be greatly reduced
by choosing an appropriate basis.

In the problem of capacitively coupled chains, the choice of the chain
basis (i.e. the set of fermionic operators $c_{i,m,\sigma}$) appears
to be most physical. This way one can trace how the Luttinger
correlations on each chain interplay with each other when the
inter-chain coupling is added to the problem. But once the inter-chain
hopping is taken into account (see e.g. \cite{kar}), the situation
becomes less transparent. Indeed, after bosonization in the chain
basis, the inter-chain hopping operator becomes non-local. The origin
of this issue can be traced back to the single-particle spectrum of
the model, shown in Fig.~\ref{fig:Dispersion}: now that the spectrum
contains two individual branches, it is not immediately clear why the
bosonization procedure in the vicinity of the single-chain Fermi
points is still justified. Instead, it seems natural to rotate the
basis from the original fermionic operators $c_{i,m,\sigma}$ to their
bonding and anti-bonding combinations that diagonalize the
single-particle Hamiltonian (\ref{h0}). This will be done in the next
Section.

Moreover, one can use the above rotated (or {\it band}) basis also in
the absence of $t_\perp$ without incurring significant technical
difficulties (albeit at the expence of physical clarity), since in
this case the spectrum remains invariant under any unitary rotation of
the fermionic basis. But while it is possible to study the whole range
of values of $t_\perp$ using only the band basis (see discussion of
the two-cutoff RG scheme in Section~\ref{PhaseDiagram_tperp}), it
seems reasonable to review several aspects of the role of the
inter-chain hopping in the chain basis, where it can be treated
perturbatively.

In the second order in small $t_\perp$, one finds (in the presence of
interactions) coherent processes \cite{braz,schulz,bac} that foretell much
of the physics that will be derived more formally in the next Section.
Firstly, there are pair hopping terms, which -- if they are the most
relevant operators -- give rise to dominant superconducting
fluctuations. For example in the regime where the un-coupled chains
had a spin gap (thus suppressing single-electron tunneling), singlet
pairs may hop between the chains, leading to strong s-wave
superconducting correlations.

Secondly, inter-chain hopping affects the renornalization of the
interaction coupling constants in the model, which results in
dynamical generation of the inter-chain exchange (see Footnote
\ref{foot-exch}). Without going into details, the main result of such
a term on the low energy effective Hamiltonian is best seen in the
refermionized form, where the marginal interaction in the spin-sector
previously written in \eqref{hrf-int} is modified to become
\begin{equation}
\label{hrf-int-mod}
{\cal H}_\mathrm{int} = 
-g_{1,t} \left(\sum_{a=1}^3 \xi^a _R \xi^a _L\right)^2 
- g_{1,s} \xi^0 _R \xi^0 _L\sum_{a=1}^3 \xi^a _R \xi^a _L + 
\frac{i }{\pi \alpha} 
\cos \sqrt{4\pi} \phi_c^- \left(  g_{2,s} \xi^0 _R \xi^0 _L 
+ g_{2,t} \sum_{a=0}^3 \xi^a _R \xi^a _L\right).
\end{equation}
The interchain exchange splits the previously seen O(4) symmetry down
to O(3)$\times$Z$_2$, corresponding to a spin singlet mode (labelled by
$a=0$), and a spin-triplet mode (labelled by $a=1,2,3$). Physically,
the most important part of ${\cal H}_{\rm int}$ to focus on is the
$g_2$ terms, which -- if relevant -- roughly speaking provide the
Majorana Fermions with mass (see Section~\ref{idpt} for details of the
mechanism of Majorana mass generation). The single parameter $g_2$ has
now split into two parts $g_{2,s}\sim a(V_\perp+J_\perp)$ and
$g_{2,t}\sim a(V_\perp-J_\perp)$. If $g_{2,s}$ and $g_{2,t}$ are the
same sign, then the singlet and triplet modes are split, but other
than this the CDW correlations are fundamentally no different than in
the case $J_\perp=0$. On the other hand, if the (dynamically
generated) inter-chain exchange $J_\perp$ is larger than the
inter-chain interaction $V_\perp$, then $g_{1,s}$ and $g_{1,t}$ will
have opposite signs. In this case, the CDW correlations will decay
exponentially, and the system will be in a fundamentally different
phase. In the case of half filling when the system is a Mott
insulator, this would be a topological (Haldane) phase \cite{snt}; in
the doped (incommensurate) case this will be either an orbital
antiferromagnetic or a superconducting phase, as we will see in the
next Section.


\section{Low-energy effective theory for the two-leg ladder}
\label{hightperp}

The full ladder model (\ref{h}) differs from the model of two
capacitively coupled chains by the presence of the inter-chain hopping
term
\[
-t_\perp \sum_{i\sigma} \left( c_{i,1,\sigma}^\dagger 
c_{i,2,\sigma} + c_{i,2,\sigma}^\dagger c_{i,1,\sigma} \right).
\]
Now, in the original chain basis used in the previous Section, the
single-particle problem is no longer diagonal. A common approach to
the problem is to perform a unitary transformation (or basis rotation)
that diagonalizes ${\cal H}_0$.

If no external magnetic field affecting orbital motion of the
electrons in the ladder is present, then the single-particle problem
is diagonalized by the transformation
\begin{equation}
\label{bands}
c_{\alpha(\beta),i,\sigma} = \frac{c_{i,1,\sigma}\pm c_{i,2,\sigma}}{\sqrt{2}},
\end{equation}
where the new operators $c_{\alpha(\beta)}$ refer to particles
belonging to the two bands (hereafter this set of operators will be
referred to as the {\it band basis}). The single-particle energies of
the two bands are split by $t_\perp$: 
\begin{equation}
\epsilon_{\alpha(\beta)}(k)
= - t_\| \cos (ka) \mp t_\perp.
\label{band_split}
\end{equation}
In contrast to the case of the capacitively coupled chains, now there
are four distinct Fermi points, provided that both bands are partially
filled. The spectrum may be linearized around the four Fermi
points and the resulting chiral fields may be bosonized. 

At the same time, if the inter-chain tunneling is weak, $t_{\perp} \ll
t_{\parallel}$, an alternative approach is possible. One can start
from the limit $t_\perp=0$ and then bosonize the capacitively coupled
chains in the rotated basis. The backscattering part of the
interaction becomes more complicated because the interchain
interaction conserves the particle number in each chain but not in
each band. In other words, there are processes that describe
interband transitions of pairs of fermions
\footnote{One must be very careful, however, not to immediately
  associate such terms with superconducting fluctuations. The physical
  interpretation of terms in this rotated basis is not particularly
  transparent, and one must instead refer to the definitions of the
  local operators in the rotated basis.}.
However, the inter-chain hopping term acquires a simple diagonal form
of the topological charge density:
\begin{equation}
-\frac{t_\perp }{\sqrt{\pi}} \partial_x \phi_c^-.
\label{tperpbos}
\end{equation}
When $t_{\perp}$ is sufficiently large (but still smaller than the
"ultraviolet" cutoff $\sim t_{\parallel}$), the effect of the gradient
term (\ref{tperpbos}) is basically the same as that caused by band
splitting (\ref{band_split}). This follows from the fact that the
emergence of four Fermi points in the single-particle spectrum leads
to the suppression of backscattering processes that involve the field
$\phi_c^-$ and no longer conserve momentum. However, the convenience
of the representation (\ref{tperpbos}) is that it also allows one to
study the small-$t_{\perp}$ limit (where interaction effects may lead
to suppression of the Fermi-point splitting in the excitation spectrum
of the fully interacting problem \cite{and-confinement}), as well as
the nonperturbative regime of a C-IC transition taking place on
increasing $t_{\perp}$, specifically at $t_{\perp}$ comparable with
the mass gap in the relative charge sector. The existence of such
transition follows from the comparison of the phase diagrams of the
model at $t_{\perp} = 0$ and in the limit of large $t_{\perp}$. The
resulting phase diagram for arbitrary $t_\perp$ will be presented in
Sec.~\ref{pts}.

\subsection{Interaction Hamiltonian in the band basis}

Similarly to Eq.~(\ref{ccc}), we represent the bosonized Hamiltonian
in the band basis as a sum of the Gaussian and backscattering parts.
The Gaussian part is formally identical to Eq.~(\ref{gm}), where the
bosonic fields now refer to the band basis, see Eq.~(\ref{bfbb}). The
backscattering part of the Hamiltonian in the band basis has been
derived in many previous works (see
e.g. \cite{fpt,fab,fal,kar,khv,baf,sch,lbf}). For completeness, we
give a derivation in \ref{bb} in terms of symmetries alone, where we
also present the transformation in the bosonized language between the
chain and the band bases, which to our knowledge has never appeared in
literature. 

Consider first the the band basis in the absence of the inter-chain
hopping (i.e. at $t_\perp=0$). Then the processes that involve the
field $\phi_c^-$ conserve momentum and as a result we find the most
general back-scattering Hamiltonian respecting the SU(2) spin
symmetry in the form
\begin{subequations}
\label{fullham1}
\begin{align}
H_{bs} = \frac{1}{2(\pi\alpha)^2}
& \left\{
\cos\sqrt{4\pi} \phi_c^- \left[ \tilde{g}_T \cos \sqrt{4\pi} \phi_s^+ 
+ \frac{\tilde{g}_T-\tilde{g}_S}{2} \cos \sqrt{4\pi} \phi_s^- 
+ \frac{\tilde{g}_T+\tilde{g}_S}{2} \cos\sqrt{4\pi} \theta_s^- \right]  \right.
\nonumber\\
&
\nonumber\\
&
+\cos\sqrt{4\pi} \theta_c^- \left[ g_T \cos \sqrt{4\pi} \phi_s^+ 
+ \frac{g_T-g_S}{2} \cos \sqrt{4\pi} \phi_s^- 
+ \frac{g_T+g_S}{2} \cos\sqrt{4\pi} \theta_s^- \right] 
\nonumber\\
&
\nonumber\\
&
+ \left. \cos\sqrt{4\pi}\phi_s^+ \left[ 
 \frac{g_s^++g_s^-}{2} \cos \sqrt{4\pi} \phi_s^- 
 +  \frac{g^+_s-g_s^-}{2} \cos \sqrt{4\pi} \theta_s^- \right] \right\}.
\label{fullham1bos}
\end{align}
Similarly to Sec.~\ref{refss}, we can make the SU(2) symmetry
explicit by refermionizing the spin sector of the model. Again, the
kinetic (Gaussian) part of the Hamiltonian has the same form as
Eq.~(\ref{hrf-k}), while the interaction part of the refermionized
Hamiltonian in the band basis is given by
\begin{align}
H_{int} =
&
\frac{i}{2\pi\alpha} \cos\sqrt{4\pi}\phi_c^-
\left[\tilde{g}_T \sum_{i=1}^3 \xi^i_R \xi^i_L
- \tilde{g}_S \xi^0_R \xi^0_L\right]
\nonumber \\
&
\nonumber \\
& 
+ \frac{i}{2\pi\alpha} \cos\sqrt{4\pi}\theta_c^-
\left[g_T \sum_{i=1}^3 \xi^i_R \xi^i_L
- g_S \xi^0_R \xi^0_L\right]
\nonumber \\
&
\nonumber\\
&
- g_s^- \xi^0_R \xi^0_L \sum_{i=1}^3 \xi^i_R \xi^i_L
- g_s^+ \left(\sum_{i=1}^3 \xi^i_R \xi^i_L\right)^2.
\label{fullham1referm}
\end{align}
\end{subequations}
The Majorana fermions representing the spin degrees of freedom clearly
split into a triplet mode $\xi^i$ with $i=1,2,3$ and a singlet
$\xi^0$.

We now consider what happens to the above Hamiltonian when the
inter-chain hopping term \eqref{tperpbos} is added. By making the
change of variables
\[
\phi_c^- (x) \rightarrow \phi_c^- (x) - \frac{K_{c-}^2t_\perp}{\sqrt{\pi}v_F} x,
\]
where $K_{c-}=1+g_c^-/(4\pi v_F)$, we see that the term
\eqref{tperpbos} is absorbed into the kinetic part of the Hamiltonian,
while the cosine term is transformed as follows:
\[
\cos \sqrt{4\pi} \phi_c^- \rightarrow 
\cos \left( \sqrt{4\pi} \phi_c^- + \frac{4K_{c-}^2 t_\perp}{v_F} x \right).
\]
The condition of sufficiently large $t_\perp$ is then equivalent to
averaging the cosine over many oscillations which yields zero. More
accurately, when at $t_{\perp} = 0$ a spectral gap $M$ is dynamically
generated in the relative charge sector, the large-$t_{\perp}$
condition actually means that 
\begin{equation}
t_{\perp} \xi/v_F \sim t_{\perp}/|M|
\gg 1, 
\label{large-t-perp-cond} 
\end{equation} 
where $\xi = v_F/|M|$ is the correlation length. Then this cosine
drops out from the interaction, and effectively this is equivalent to
setting $\tilde{g}_T=\tilde{g}_S=0$ in (\ref{fullham1}). Same
conclusion can be reached by noting that for large enough $t_\perp$
there are four Fermi points in the single-particle spectrum and
therefore the processes involving $\phi_c^-$ do not conserve momentum
and should be dropped from the Hamiltonian. At the same time, the
cosine of the dual field remains unchanged by the above
transformation.

The initial (bare) values of the coupling constants can still be
expressed in terms of the four combinations (\ref{bare1}) and
(\ref{kcplus}) of the parameters of the microscopic Hamiltonian
(\ref{h}). In fact, as shown in \ref{bb}, the Luttinger liquid
parameters for the total charge and total spin are the same in both
chain and band bases because the change-of-basis transformation only
involves relative degrees of freedom. The total charge sector remains
decoupled [see the footnote after Eq.~(\ref{kcplus})], while
\begin{subequations}
\begin{equation}
g_s^+ = h_s.
\end{equation}
The remaining coupling constants can be expressed in terms of $h_i$ as
\begin{equation}
g_c^-=g_s^-=h_\perp, \quad g_T = \frac{h_\perp-h_s}{2}, \quad g_S = - \frac{h_c}{2}.
\end{equation}
\label{bbgs}
\end{subequations}
The fact that the initial values of five coupling constants $g_c^-$,
$g_s^\pm$, $g_T$, and $g_S$ can be parametrized by three quantities
$h_i$ reflects the higher symmetry of the \textit{bare} microscopic
interaction Hamiltonian (\ref{hint}), namely that a density-density
interaction always conserves individually the particle number on each
chain.  However, under renormalization, further interactions will be
generated when $t_\perp$ is present that do not respect this symmetry.
These generated interactions manifest themselves in the RG equations
as independent flow of these five parameters.

\subsection{Phase diagram for large $t_\perp$}
\label{rg2}

The strategy here is the same as for the case $t_\perp=0$. We will
perform the semi-classical analysis of the backscattering Hamiltonian
Eq.~\eqref{fullham1bos} (with the first line removed). Defining the
bosonized form of the order parameters in the band representation, we
then associate each of the semi-classical solutions with one of the
phases of the model. We then derive the RG equations for the above
model, identify the strong-coupling fixed points of the RG flow,
relate them to the semi-classical solutions, and thus construct the
phase diagram of the model. As each of these steps is analogous to
those performed in Section \ref{lowtperp}, we will not present full
details here and resort only to an outline and the results.

In addition to the two CDW operators previously discussed, we will
find an operator describing an orbital antiferromagnetic phase, which
is characterized by the staggered part of the local current 
(see \ref{opsbandapp} for details)
\begin{subequations}
\label{otherops}
\begin{equation}
OAF \leftarrow i \left( c_{1\sigma}^\dagger c_{2\sigma} - c_{2\sigma}^\dagger c_{1\sigma} \right),
\end{equation}
as well as two superconducting operators
\begin{align}
{\cal O}_{SCs} &=
\left( c_{1\uparrow} c_{1\downarrow} -  c_{1\downarrow} c_{1\uparrow} \right)
+
\left( c_{2\uparrow} c_{2\downarrow} -  c_{2\downarrow} c_{2\uparrow} \right),
\nonumber \\
&
\nonumber \\
{\cal O}_{SCd} &=
2 \left( c_{1\uparrow} c_{2\downarrow} -  c_{1\downarrow} c_{2\uparrow} \right).
\end{align}
\end{subequations}
The bosonized and refermionized forms of these operators are given in
\ref{opsbandapp}.

As indicated previously, in the large-$t_\perp$ regime, the RG flow is
parametrized by five independent coupling constants (see \ref{AllRG}):
\begin{align}
&
\frac{\partial g_T}{\partial l} = 
\frac{1}{2}\left[ \left(g_c^--2g_s^+\right)g_T 
+ g_s^- g_S \right] ,
\nonumber\\
&
\nonumber\\
&
\frac{\partial g_S}{\partial l} = 
\frac{1}{2}\left[ g_c^- g_S 
+ 3 g_s^- g_T \right] ,
\nonumber\\
&
\nonumber\\
&
\frac{\partial g_c^-}{\partial l} =
\frac{1}{2}\left[ 3g_T^2+g_S^2 \right] ,
\nonumber\\
&
\nonumber\\
&
\frac{\partial g_s^+}{\partial l} =
-\frac{1}{2}\left[ 2g_T^2+\left(g_s^+\right)^2 + \left(g_s^-\right)^2 \right] ,
\nonumber\\
&
\nonumber\\
&
\frac{\partial g_s^-}{\partial l} =
g_T g_S - g_s^+ g_s^-.
\label{RGfive}
\end{align}
The only weak-coupling fixed point of these equations is $g_i=0$ for
all $i$, which is the case of a noninteracting system. This is a
repulsive fixed point. On the other hand, the equations (\ref{RGfive})
allow for the following four strong coupling fixed points:
\begin{align}
&
g_s^-\rightarrow +\infty, \;
g_T\rightarrow + \infty, \;
g_S \rightarrow +\infty \quad \Rightarrow \quad CDW^-,
\nonumber\\
&
\nonumber\\
&
g_s^-\rightarrow +\infty, \;
g_T\rightarrow - \infty, \;
g_S \rightarrow -\infty \quad \Rightarrow \quad OAF,
\nonumber\\
&
\nonumber\\
&
g_s^-\rightarrow -\infty, \;
g_T\rightarrow + \infty, \;
g_S \rightarrow -\infty \quad \Rightarrow \quad SC^s,
\nonumber\\
&
\nonumber\\
&
g_s^-\rightarrow -\infty, \;
g_T\rightarrow - \infty, \;
g_S \rightarrow +\infty \quad \Rightarrow \quad SC^d.
\label{scfps}
\end{align}
The flow of the two remaining parameters in Eq.~(\ref{RGfive}) is the
same in all of the above cases:
\begin{equation}
\label{sc2}
g_c^-\rightarrow\infty, \quad
g_s^+\rightarrow -\infty.
\end{equation}

The identification of the strong-coupling phases follows along the
lines of the analysis of Section~\ref{qa}. For example, when $g^- _s,
g_T, g_S \to +\infty$ then the three terms in the back-scattering
Hamiltonian (\ref{fullham1}) that do not involve the relative spin
field $\phi_s^-$ flow to strong coupling, such that the fields
$\theta_c^-$, $\theta_s^-$, and $\phi_s^+$ get ``locked'' at some
minima of the cosine potentials. The choice of the minima is
degenerate (see the previous discussion on the structure of order
parameters), but for $g_T\rightarrow\infty$ let us choose the minima
such that $\cos\sqrt{4\pi}\theta_c^-\rightarrow 1$, while
$\cos\sqrt{4\pi}\phi_s^+\rightarrow -1$. Then from the conditions
$g_T+g_S\rightarrow\infty$, $g_s^+-g_s^-\rightarrow -\infty$ it
immediately follows that the third field should be locked such that
$\cos\sqrt{4\pi}\theta_s^-\rightarrow -1$. Comparing this result with
the structure of the order parameters (\ref{opsboz}), we conclude that
in this case the correlations of the the operator ${\cal O}_{CDW^-}$
are dominant, i.e. decaying as a power law (in contrast to all other
correlation functions which decay exponentially), similarly to the
discussion in Section~\ref{qa}.

The analysis of the remaining three strong-coupling phases is similar.
It is worth stressing, that both the relative charge density wave
(CDW$^-$) and the orbital antiferromanet (OAF) correspond to the
locking of the $\theta_s^-$ field, while in the ``superconducting''
phases the field $\phi_s^-$ is locked. These two possibilities are
distinguished by the relative sign of the two coupling constants $g_T$
and $g_S$. Let us stress it again (see Section~\ref{qa}) that in any
of the phases discussed here the total charge sector remains gapless
and thus there is never a long-range order in the system.

\begin{figure}
\begin{center}
\includegraphics[width=16cm]{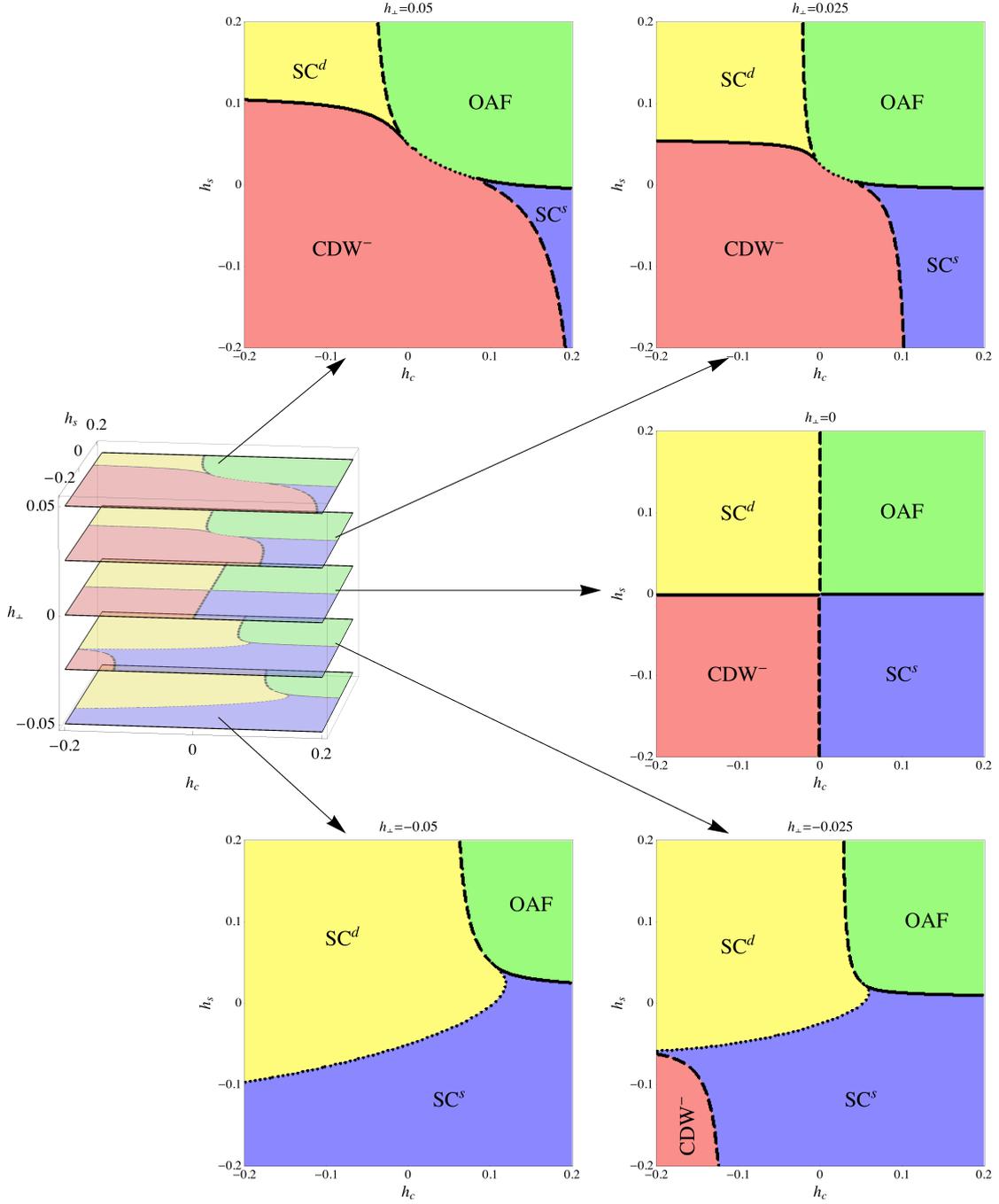}
\end{center}
\caption{The phase diagram for large $t_\perp$, as a function of
  $h_c$, $h_s$ and $h_\perp$, as derived in Sec.~\ref{rg2}. The shape
  of the three-dimensional phase diagram is indicated, as are
  cross-sections for different values of $h_\perp$. The universality
  class of each of the phase transitions, as derived in Sec.~\ref{pts}
  is also indicated. A dashed line between phases indicates a Z$_2$
  (Ising) phase transition, a dotted line a U(1) (Gaussian) phase
  transition, and the solid lines represent phase transitions that are
  weakly first order.}
\label{Phase-Diagram}
\end{figure}

The phase diagram is now determined as the basin of attraction of each
of the strong-coupling fixed points as a function of the bare
parameters of the theory, \eqref{bbgs}. Certain regions may be
determined analytically (see \ref{RGappendix2}), including the
existence and attraction of certain rays in phase space with an
enhanced O(6) symmetry. Ultimately however, the phase diagram is
determined by numerically integrating the flow \eqref{RGfive}, the
results are shown in Fig.~\ref{Phase-Diagram}. As before, the phase
diagram is plotted as a function of the three parameters: $h_c$ and
$h_s$ paramaterizing the correlations on the individual chains before
coupling, and $h_\perp$ which is the strength of the inter-chain
interaction.

The middle panel in Fig.~\ref{Phase-Diagram} -- i.e. the phase diagram
in the case $h_\perp=0$ -- is well known (see e.g. \cite{gb}). In
particular, the most important feature of this phase diagram is that
in the region of purely repulsive (intra-chain) interactions, $h_c<0$,
there is a phase SC$^d$ in which the dominant correlations are
superconducting. Cross-sections for non-zero $h_\perp$ show how
repulsive inter-chain interaction enlarges the region of the CDW$^-$
phase at the expense of the superconducting ones; while attractive
inter-chain interaction has the opposite effect. Curiously, the OAF
phase is only weakly affected by the inter-chain interaction.

Figure \ref{Phase-Diagram} also indicates the universality class of
each of the phase transitions, which is something that can not be
determined from the RG alone, and will be derived in Sec.~\ref{pts}
where an effective strong coupling theory is developed in each of the
phases. Before we look at this however, we first turn our attention to
the case when $t_\perp$ is present, but not necessarily large -- in
other words, to interpolation between the phase diagram shown in
Fig.~\ref{pd1} for $t_\perp=0$, and that of Fig.~\ref{Phase-Diagram}
for large $t_\perp$.

\subsection{Phase diagram at intermediate values of $t_\perp$}
\label{PhaseDiagram_tperp}

Comparison of the $t_\perp=0$ and large-$t_{\perp}$ limits,
Figs.~\ref{pd1} and \ref{Phase-Diagram}, reveals a number of features
caused by inter-chain tunneling. First we note that the TL liquid
phase present at $t_{\perp} = 0$ transforms to an OAF phase when
hopping is present (a similar situation occurs in the spinless chain
\cite{nlk}). As the TLL is critical, we would expect this to happen
for arbitrarily weak inter-chain hopping. The situation is different
if one starts from one of the gapped phases at $t_\perp=0$. Let us
consider first the case of repulsive inter-chain interaction,
$h_\perp>0$. Then we see that depending on the other parameters, the
CDW$^-$ phase at $t_\perp=0$ may become a superconducting phase as
$t_\perp$ is gradually increased. But because of the finite gap
already present at $t_\perp=0$, we would naturally expect such a phase
transition to occur at a small but \textit{non-zero} value of
$t_\perp$.

For attractive inter-chain interaction ($h_\perp<0$), even more
scenarios are possible. In fact, the CDW$^+$ phase which is the ground
state at $t_\perp=0$ does not exist at large $t_\perp$. The reason for
this may be seen by returning to the cartoon depiction of the state in
Fig.~\ref{cdwp} -- where it is clear that the CDW$^+$ state prevents
the electrons from lowering their energy by delocalizing between the
chains, as would be expected at sufficiently large $t_\perp$. A
similar effect is seen in spinless chains \cite{nlk} where this phase
becomes a TL liquid phase at large $t_{\perp}$ via a C-IC
transition. In the present (spinful) case, we see that, depending on
initial parameters, the CDW$^+$ phase at small $t_\perp$ becomes at
larger $t_\perp$ either a superconducting phase or even a CDW$^-$
phase. As we will show in Sec.~\ref{pts}, this occurs again via a C-IC
like transition, but in contrast to the spinless case, this occurs
between \textit{two gapped phases}. The precise meaning of the
nomenclature C-IC in this context will be explained in
Sec.~\ref{cdwpsc}.

Before getting to a description of the phase transitions let us first
ask ourselves: how can we treat the system at \textit{small but
  non-zero} $t_\perp$?  The RG equations (\ref{RGfive}) were derived
under the assumption that $t_\perp$ is sufficiently large [see the
  condition (\ref{large-t-perp-cond})], such that the backscattering
processes involving the relative charge field $\phi_c^-$ can be
neglected as violating momentum conservation. On the contrary, in the
absence of the interchain hopping, one must retain the $\tilde{g}_T$
and $\tilde{g}_S$ terms in Eq.~(\ref{fullham1}). As shown in
\ref{AllRG}, when the RG equations for this full system (which now
involves eight coupling constants) are derived, it is found that the
initial values Eq.~(\ref{bprm}) which are written in terms of three
independent parameters actually flow on a three-dimensional invariant
hyper-surface, with the equations as given in Eq.~(\ref{RG1}). This is
not an accident -- while the backscattering Hamiltonian in the band
basis \eqref{fullham1} in the absence of $t_\perp$ may look
substantially different from that in the chain basis \eqref{hamBS},
they ultimate are two different representations of the same physical
model, and therefore must have the same physical properties.

However, translating this flow from the language of the chain basis to
the band basis has a big advantage -- as one can now utilize a
two-cutoff RG scheme in order to obtain the phase diagram of the model
for arbitrary $t_\perp$.  The main observation is that the first line
of Eq.~\eqref{fullham1bos} involving the field $\phi_c^-$ should only
be present for energies greater than $t_\perp$ -- while the term does
not contribute at energies lower than $t_\perp$ as momentum
conservation can not be satisfied at low energies.  One therefore
defines the following procedure

\begin{enumerate}
\item Starting from the initial bare coupling constants $h_c$, $h_s$,
  and $h_\perp$, one flows from an energy scale $\Lambda$ ($l=0$) down
  to $t_\perp$ ($l=\ln \Lambda/t_\perp$) using the flow equations
  which \textit{include the $\phi_c^-$ field}, namely Eqs.~\ref{RG1}.
\item At energies below $t_\perp$, one continues the flow equations in
  the \textit{absence of the $\phi_c^-$ field}.  Namely, one defines
  the set of parameters $g_S$, $g_T$, $g_s^+$, $g_s^-$ and $g_c^-$ via
  Eq.~\eqref{bbgs} using the renormalized values of $h$ obtained in
  step 1.  The flow then continues with Eq.~\eqref{RGfive}.
\end{enumerate}
In this flow procedure, if any of the coupling constants becomes
of order unity (which may happen during either step 1 or step 2),
strong coupling is reached and the correct strong coupling phase is
identified
\footnote{The exact value $\sim 1$ is arbitrary, and therefore this
  weak coupling RG procedure will not give the exact position of the
  phase boundaries, but is expected to correctly indicate the topology
  of the phase diagram.}.

We comment that in the case when the flow at $t_\perp=0$ already flows
to strong coupling, the energy scale at which the coupling constants
become of order unity is exactly an estimate of the gap, $|M|$,
present in such a ground state.  Under the procedure above, the
presence of $t_\perp$ will therefore not be felt until $t_\perp \sim
|M|$, i.e. there is the possibility of a phase transition at some
finite non-zero value of $t_\perp$.  This regime of maximal
competition can not be correctly described using the two-cutoff RG
approach; in order to study such phase transitions we will develop a
strong coupling approach in the next section.

\begin{figure}
\begin{center}
\includegraphics[width=6cm]{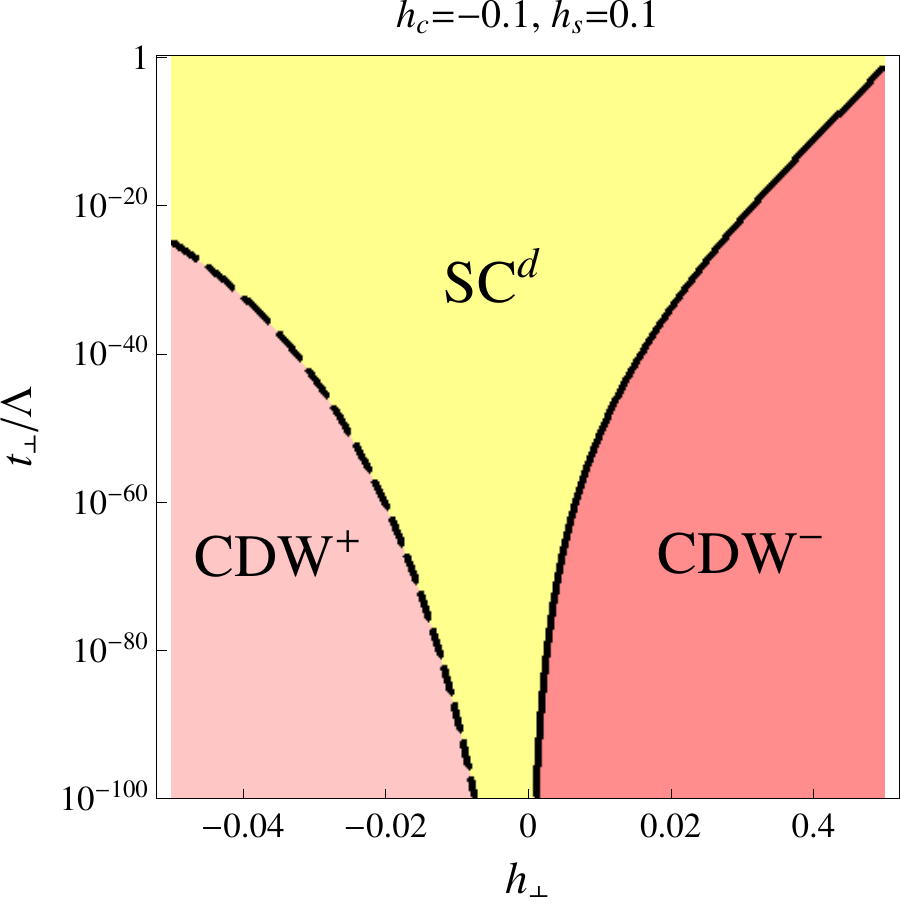} 
\hspace{1cm}
\includegraphics[width=6cm]{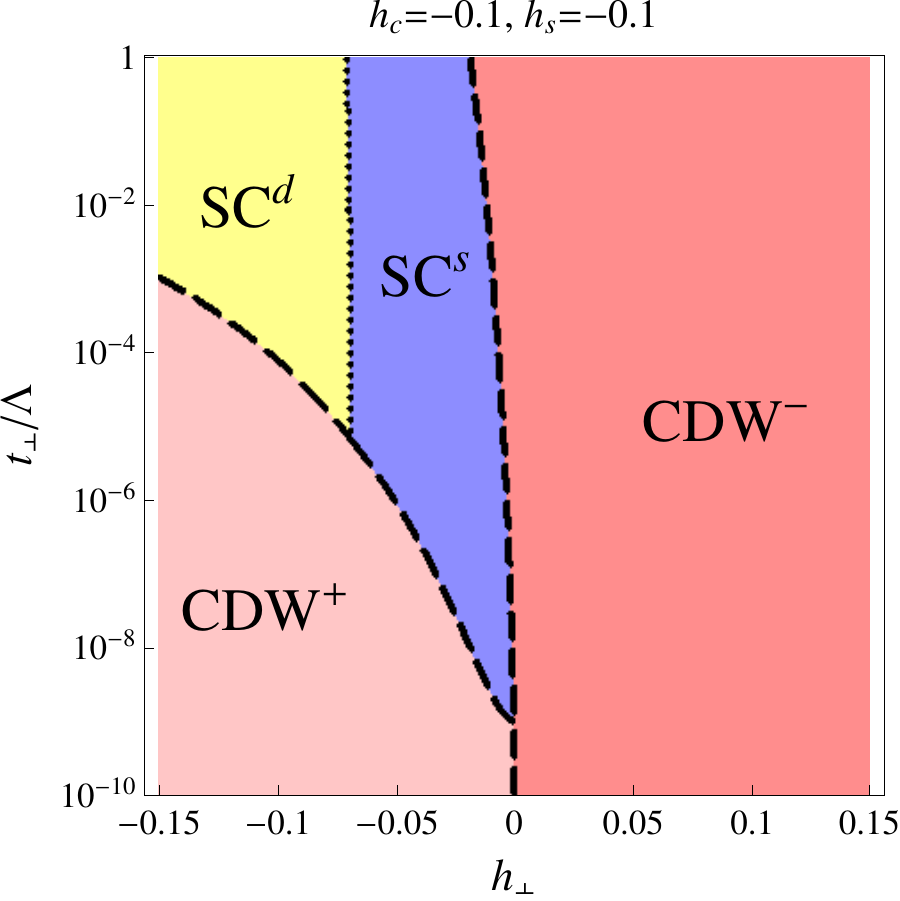}
\end{center}
\caption{Phase diagram of the two-leg ladder model plotted in the
  $t_\perp-h_\perp$ plane with all other parameters fixed. Both panels
  correspond to repulsive intra-chain interaction characterized by
  $h_c=-0.1$. The left panel describes the most physical situation
  where the on-site interaction dominates ($h_s=0.1$), while the right
  panel corresponds to the opposite situation ($h_s=-0.1$).  The solid
  line indicates a (weakly) first order phase transition, the dotted
  line a U(1) transition, while the dashed line between the CDW$^+$
  and SC$^{s/d}$ phases indicates a Z$_2$ commensurate-incommensurate
  transition (see Section \ref{pts}.) While the inter-chain hopping
  $t_\perp$ is plotted on a logarithmic scale, we point out that this
  parameter is typically exponentially small in the separation of the
  two chains, hence this axis may also loosely by thought of as the
  inverse distance between the two nanowires.}
\label{hdpt}
\end{figure}

We may numerically integrate the RG equations using the above RG
procedure to obtain the phase diagram at arbitrary $t_\perp$.  The
phase diagram is now four-dimensional -- a function of the in-chain
parameters $h_c$, $h_s$, and the inter-chain parameters $t_\perp$ and
$h_\perp$ -- and as such is difficult to plot.  We therefore limit
ourselves to the most interesting cases when the interpolation between
the $t_\perp=0$ phase diagram in Fig.~\ref{pd1} and large $t_\perp$ in
Fig.~\ref{Phase-Diagram} is not trivial.  Two such cross sections are
shown in Fig.~\ref{hdpt}.

The left panel corresponds to what might arguably be the most physical
situation, when the in-chain interactions are repulsive $h_c<0$, and
are dominated by on-site interactions $h_s>0$.  For the case when
interchain interaction is also repulsive $h_\perp>0$, we see that the
CDW$^-$ phase is stable for small $t_\perp$ (capacitively coupled
nanowires), while the SC$^d$ phase takes over for larger values of the
inter-chain hopping.  We emphasize however that the transition between
these phases does not occur when the bare parameters are similar
orders of magnitude, but instead when $t_\perp \sim |M|$, the
dynamically generated gap generated by the interactions.  As the
interactions for the doped ladder that we describe are always
marginal, in general the gap is much smaller than the bare coupling
constants $|M| \ll h_i$.  However, in a realistic double-wire
nano-structure, the hopping integral $t_\perp$ is also expected to be
exponentially small in the distance between the wires; hence for
experimentally reasonable parameters a true competition between these
two phases may reasonably be expected.  For this reason, we will pay
particular attention to these two phases when we later discuss
impurity effects.

The right panel is a cross section through the alternative case when
$h_s<0$, when most of the hopping induced phase transitions are seen.
For $h_\perp$ negative but close to zero, we see that the situation
previously mentioned is indeed possible, whereby adding interchain
hopping converts a CDW$^+$ ground state to CDW$^-$.  However, we see
that this is not a direct phase transition, but instead goes through
an intermediate SC$^d$ phase.  We will explain why this is the case in
the next section, when we analyze these phase transitions more closely
using a strong coupling approach.

\section{Phase transitions}
\label{pts}

While the RG approach of the previous two sections has given us the
overall shape of the phase diagram, as a weak-coupling, perturbative
method it cannot provide a reliable way to study the strong-coupling
phases. In particular, it tells us nothing about the nature of the
phase transitions between the different possible ground states. There
are two types of phase transitions in the problem: interaction-driven
transitions and transitions driven by the inter-chain hopping.

The former occur either at $t_\perp=0$ or $t_\perp$ sufficiently large
on varying interaction parameters.  Such transitions have been studied
before (see e.g. \cite{cts}); we include them here to keep the paper
self-contained. By the latter, we mean phase transitions that occur at
fixed value of interaction parameters as $t_\perp$ is varied between
the two limits. In fact, we will show that some phase transitions of
this type may be adiabatically connected the phase boundaries of
transitions we refer to as interaction-driven. On the other hand, we
will also show the presence of a new sort of
commensurate-incommensurate phase transition between two gapped
phases, which rely crucially on the presence of a small but non-zero
$t_\perp$.

\subsection{Interaction-driven phase transitions}
\label{idpt}

\subsubsection{Capacitively coupled chains, $t_\perp=0$}

We begin our discussion of phase transitions with the ladder model in
the absence of the inter-chain hopping. The phase diagram was
discussed in Section~\ref{lowtperp} and is illustrated in
Fig.~\ref{pd1}. There are only two types of phase transitions in the
model: (i) the Berezinskii-Kosterlitz-Thouless (BKT) transition
between the Luttinger liquid and charge-density wave phases, and (ii)
the direct transition between the CDW$^+$ and CDW$^-$ phases at
$h_\perp=0$. The latter transition occurs at the point of complete
decoupling between the two chains ($h_\perp=0$).

\subsubsection{Two-leg ladder in the large $t_\perp$ limit}
\label{ltpt}

In the presence of the sufficiently large inter-chain hopping (see
Section~\ref{hightperp}), the phase diagram (see
Fig.~\ref{Phase-Diagram}) shows four different phases and indicates
possible phase transitions between any two of them. These phase
transitions have been analyzed in Ref.~\cite{cts}, for the sake of
completeness we repeat this analysis here.

The starting point is the refermionized Hamiltonian
\eqref{fullham1referm} which, after dropping the terms involving
$\phi_c^-$, takes the form
\begin{equation}
{\cal H}_{int} =
\frac{i}{2\pi\alpha}
 \cos\sqrt{4\pi}\theta_c^-
\left[g_T\, \sum_{i=1}^3 \xi^i_R \xi^i_L
- g_S\, \xi^0_R \xi^0_L\right]
- g_s^-\, \xi^0_R \xi^0_L \sum_{i=1}^3 \xi^i_R \xi^i_L
- g_s^+\, \left(\sum_{i=1}^3 \xi^i_R \xi^i_L\right)^2.
\label{hpt1}
\end{equation}
This Hamiltonian exhibits a U(1) $\times$ SU(2) $\times$ Z$_2$
symmetry and involves three coupled modes -- the relative charge, the
spin-triplet and spin-singlet modes. Understanding that in the
strong-coupling regime the operators $\cos\sqrt{4\pi}\theta_c^-$,
$\sum_{i=1}^3 \xi^i_R \xi^i_L$ and $\xi^0_R \xi^0_L$ each acquire
nonzero expectation values, we adopt here a \emph{symmetry preserving}
mean-field decoupling of these three modes to obtain
\begin{subequations}
\begin{equation}
{\cal H}_{int} = 
\frac{m_F}{2\pi\alpha}  \cos\sqrt{4\pi}\theta_c^-
+
i m_S \xi^0_R \xi^0_L
+
 i m_T \sum_{i=1}^3 \xi^i_R \xi^i_L
- g_s^+ \left(\sum_{i=1}^3 \xi^i_R \xi^i_L\right)^2,
\label{hpmf1}
\end{equation}
where
\begin{align}
m_F &=  g_T \langle i \sum_{i=1}^3 \xi^i_R \xi^i_L \rangle 
-  g_S \langle i \xi^0_R \xi^0_L \rangle,
\nonumber \\
m_S &= g_S \langle \cos\sqrt{4\pi}\theta_c^-\rangle +
g_s^- \langle i\sum_{i=1}^3 \xi^i_R \xi^i_L\rangle.
\nonumber \\
m_T &= g_T \langle \cos\sqrt{4\pi}\theta_c^-\rangle +
g_s^- \langle i\xi^0_R \xi^0_L\rangle.
\end{align}
\label{hpmf}
\end{subequations}
\noindent
The interaction Hamiltonian (\ref{hpmf1}) consists of mass terms for
each of the three modes along with a marginal interaction in the
spin-triplet sector.
\footnote{To be complete, there is also the forward scattering term
  $g_c^-$ for the relative charge mode. This term however is exactly
  marginal and so (if small) plays very little role. This is in
  contrast to the $g_s^+$ term which may be marginally relevant or
  marginally irrelevant depending on the sign, and so may still gap
  the triplet mode even in the absence of $m_T$.}

\begin{table}[tbp]
\begin{center}
\begin{tabular}{ c c ccc}
Phase && $m_F$ & $m_S$ & $m_T$ \\
\hline
\multirow{2}{*}{CDW$^-$}
&& $+$ & $+$ & $-$ \\
&& $-$ & $-$ & $+$ \\ \hline
\multirow{2}{*}{OAF}
&& $+$ & $-$ & $+$ \\
&& $-$ & $+$ & $-$ \\
\end{tabular}
\hspace{2cm}
\begin{tabular}{ c c ccc}
Phase &&$m_F$ & $m_S$ & $m_T$ \\
\hline
\multirow{2}{*}{SC$^d$}
&& $+$ & $+$ & $+$ \\
&& $-$ & $-$ & $-$ \\ \hline
\multirow{2}{*}{SC$^s$}
&& $+$ & $-$ & $-$ \\
&& $-$ & $+$ & $+$ \\ 
\end{tabular}
\end{center}
\caption{The phase of the model as determined by the sign of the mass
  gap in each of the three modes.  Note that as discussed when the
  order parameters were first introduced, a symmetry of the model
  means that flipping the sign of all three masses simultaneously does
  not affect the phase.}
\label{masstable}
\end{table}

Treating the masses as parameters, the description given by the
mean-field Hamiltonian (\ref{hpmf1}) is expected to lead to a
qualitatively correct phase diagram for the original model
(\ref{hpt1}). It is easy to see by evaluating the expectation values
of the order parameters (as given in \ref{opsbandapp}) that the phase
of the model is now simply determined by the signs of the three
masses; these are shown in Table \ref{masstable}. The phase boundaries
are therefore located exactly at the places where one of these masses
changes sign.  In fact, the region where one of the masses is much
smaller than the other two is exactly where the decoupling
\eqref{hpt1} works best, at least for the soft mode, as at energies
lower than the other two gaps, fluctuations in these fields are
suppressed and the mean-field approximation is well justified. We can
therefore categorize three possible phase transitions:

(i) if the mass $m_S$ of the spin-singlet field $\xi^0$ changes sign,
then the system undergoes an Ising (Z$_2$) transition with central
charge \cite{cft} $c=1/2$;

(ii) the change of sign of the mass of the relative charge sector
corresponds to the U(1) Gaussian phase transition with central
charge $c=1$;

(iii) finally, if the mass $m_T$ of the spin-triplet sector changes
sign, then, depending on the sign of $g_s^+$, the system undergoes
either a first order or an SU(2)$_2$ (central charge $c=3/2$)
transition.

The latter statement can be illustrated further as follows. Suppose
that $m_T$ is the smallest mass in the problem. Then, in the spirit of
the mean-field decoupling above, the effective theory in the
spin-triplet sector at energies lower than $m_F,m_S$ can be described
by the interaction Hamiltonian:
\[
{\cal H}' \sim i m_T \sum_{i=1}^3 \xi^i_R \xi^i_L
- g_s^+ \left(\sum_{i=1}^3 \xi^i_R \xi^i_L\right)^2,
\]
We note that fluctuations of the linear term around the mean-field
solution will contribute to unimportant renormalization of $g_s^+$.

If $m_T=0$, the effective Hamiltonian is that of the O(3)
Gross-Neveu model. For $g_s^+<0$ the interaction is marginally
relevant and generates an exponentially small gap. The resulting state
is degenerate with respect of the sign of the dynamically generated
mass. Adding the explicit mass term in the Hamiltonian lifts this
degeneracy. In other words, the system will exhibit infinite response
to the mass field (similar to the response of a ferromagnet to the
external magnetic field). The transition between the states with
positive and negative $m_T$ is then a first-order transition. If, on
the other hand, $g_s^+>0$, then the interaction is marginally
irrelevant and as $m_T$ changes sign the system undergoes a continuous
transition of the SU(2)$_2$ universality class.

Using Table \ref{masstable} now we comment on the phase diagram shown
in Fig.~\ref{Phase-Diagram}.  Firstly, there are two Ising (Z$_2$)
transitions -- between (i) the CDW$^-$ and SC$^s$ phases and (ii) the
OAF and SC$^d$ phases. In the absence of the inter-chain interaction
($h_\perp=0$) both transition occur as $h_c$ changes sign, the former
transition happening for $h_s<0$ and the latter for $h_s>0$ (in
complete agreement with Ref.~\cite{gb}). In the presence of non-zero
$h_\perp$ the locations of the phase boundaries change, but the
universality class remains the same.

Secondly, there are two U(1) Gaussian transitions -- between (i) the
CDW$^-$ and OAF phases, and (ii) between the two superconducting
phases. The former transition can only happen for repulsive
inter-chain interaction ($h_\perp>0$), while the latter -- for
$h_\perp<0$.

Finally, there are two transitions associated with the change of sign
of the spin-triplet mass $m_T$ -- between (i) the CDW$^-$ and SC$^d$
phases, and (ii) the OAF and SC$^s$ phases. The former transition
occurs for negative $h_c<0$ and gets pushed towards larger negative
values of $h_c$ for larger negative values of $h_\perp$. The opposite
applies to the latter transition. In both cases, the coupling constant
$g_s^+$ flows to large negative values [see Eqs.~(\ref{RGfive}) and
  (\ref{sc2})]. Thus according to the general discussion above, these
are first-order transitions \cite{cts}.  As it is a marginal
interaction that drives it first order, the first-order jump is
expected to be extremely weak, and properties of the hidden SU(2)$_2$
transition are still expected to be seen in a suitable parameter range
close to, but not exactly at, the transition.

The universality class of each phase boundary is indicated in
Fig.~\ref{Phase-Diagram} by the presence of a solid (first-order),
dashed (Z$_2$) or dotted (U(1)) line.

\subsection{``Hopping-driven'' phase transitions}
\label{hdpt-sec}

We now turn to phase transitions that may occur as a nonzero $t_\perp$
is switched on -- the hopping-driven phase transitions oulined in
Sec.~\ref{PhaseDiagram_tperp}. 

Firstly, there is a transition between the TL liquid phase at
$t_\perp=0$ (see Fig.~\ref{pd1}) and the OAF phase appearing for any
$t_\perp \neq 0$ (see Fig.~\ref{Phase-Diagram}). This is a standard
BKT transition that can be described within a perturbative RG
approach: as small $t_\perp$ is introduced to the model, a marginally
relevant perturbation is generated in the second order that drives the
system towards the OAF phase.

Secondly, the CDW$^-$ phase persists in a wide range of values of
$t_\perp$, see Fig.~\ref{hdpt}. For stronger $t_\perp$, there is a
phase transition to one of the superconducting phases. However, these
transitions are of exactly the same universality class as the
equivalent interaction-driven transitions in Fig.~\ref{Phase-Diagram}
-- Z$_2$ for CDW$^-$ to SC$^s$ and first order for CDW$^-$ to
SC$^d$. This is very reasonable -- if one imagines the full four
dimensional phase diagram, the boundaries between the CDW$^-$ phase
and the SC$^{s/d}$ ones are functions of all four parameters, and can
be therefore crossed by varying any one of them, be it an interaction
parameter or the interchain hopping.

This can be seen more formally by looking at the Hamiltonian
\eqref{fullham1} in the band basis. The order parameter corresponding
to the CDW$^-$ involves the field $\theta_c^-$ and not $\phi_c^-$ (see
\ref{opsbandapp}). Therefore, as long as the CDW$^-$ phase is stable
at $t_\perp=0$ (i.e. $\tilde{g}_T$ and $\tilde{g}_S$ flow to weak
coupling) the effect of reducing $t_\perp$ from its ``large'' value
discussed above is completely equivalent to a renormalization of the
interaction parameters. Hence by the same logic used above in
Sec.~\ref{ltpt}, we find no difference in the universality class for
such transitions, regardless of whether they were induced by varying
interaction or by varying $t_\perp$. This argument remains valid
regardless of whether or not the two-cutoff procedure were applied to
$\tilde{g}_T$ and $\tilde{g}_S$.

Finally, the CDW$^+$ phase that exists at $t_\perp=0$ for attractive
interchain interaction also disappears at large enough $t_\perp$. In
this case, the interchain hopping competes with the dynamically
generated gap and the transition takes place at some non-zero value
of $t_\perp$, see Fig.~\ref{hdpt}. We discuss this transition in more
detail in Sec.~\ref{cdwpsc} below.

\subsubsection{Commensurate-incommensurate transition: CDW$^+$ to SC$^{s/d}$}
\label{cdwpsc}

Consider attractive interactions, where in the absence of inter-chain
hopping $t_\perp=0$ a large portion of the phase diagram in
Fig.~\ref{pd1} is occupied by the charge density wave CDW$^+$. At
large $t_\perp$ this phase becomes either superconducting or
CDW$^-$. Here we discuss the former transition (for the latter see the
subsequent subsection).

The ``superconducting'' order parameters are given in
Eq.~(\ref{opsboz2}). Comparing them with ${\cal O}_{CDW^+}$ [${\cal
    O}_0$ from Eq.~(\ref{opsboz})], we observe that the spin-sector
fields remain gapped across the transition line and yield the same
nonzero expectation value for
\[
\langle 
\cos \sqrt{\pi} \phi_s^+ \cos \sqrt{\pi} \phi_s^-
\rangle \ne 0,
\]
or for the similar product where the cosines are replaced by the
sines. In other words, the masses $m_S$ and $m_T$ do not change sign
at such transition, and the latter occurs due to a rearrangement of
the relative charge degrees of freedom.

As in Subsection 5.1.2, here too we adopt a mean-field decoupling
procedure and replace the spin fields by their expectation values in
Eq.~(\ref{fullham1bos}), which yields
\begin{equation}
\label{hcic}
H_{bs} + H_\perp = \frac{g_\phi}{2\pi\alpha} \cos \sqrt{4\pi} \phi_c^- 
+ \frac{g_\theta}{2\pi\alpha} \cos \sqrt{4\pi} \theta_c^- 
- \frac{t_\perp}{\sqrt{\pi}}\partial_x \phi_c^-.
\end{equation}
The parameters $g_\phi$ and $g_\theta$ are given by the appropriate
combinations of expectation values of the spin fields.  Note that we
now must include the inter-chain hopping parameter, as well as the
$\phi_c^-$ field explicitly in the Hamiltonian.

Looking at this Hamiltonian, we can now understand the meaning of
commensurate-incommensurate transition in this context. Starting from
the case $g_\theta=t_\perp=0$, we see that the $g_\phi$ term locks the
field $\phi_c^-$ at one of the classical minima of the potential,
meaning that $\langle e^{i\sqrt{\pi}\phi_c^-} \rangle \ne 0$, and we
are in the CDW$^+$ phase. The inter-chain hopping $t_\perp$ then acts
as a chemical potential for this field. This situation is maintained
as long as the particle number is conserved (i.e. there are no terms
with the dual field in the Hamilonian). Otherwise $t_{\perp}$ affects
the spectrum which is generically massive. The existence of the
transition follows from the qualitative difference between the massive
phases in the two limiting cases.

Now suppose that $t_\perp$ becomes larger than the gap, meaning that
$\phi_c^-$ is no longer locked, and $\langle e^{i\sqrt{\pi}\phi_c^-}
\rangle = 0$. In the absence of $g_\theta$, one then sees CDW$^+$ at
two different wavenumbers corresponding to the splitting of the Fermi
points that would be expected in the non-interacting model.  This is
the traditional commensurate-incommensurate transition, which occurs
in the spinless two-leg ladders \cite{nlk}. The current case is
slightly different however, when one restores the $g_\theta$ term
present in \eqref{hcic}. In this case, as soon as the $t_\perp$ term
prevents $\phi_c^-$ from being locked, the $g_\theta$ term would lock
the dual field instead $\theta_c^-$, meaning that $\langle
e^{i\sqrt{\pi}\theta_c^-} \rangle \ne 0$. This means that correlation
functions involving $e^{i\sqrt{\pi}\phi_c^-}$ (for example, the
CDW$^+$ one) decay exponentially with a finite correlation length.
Nevertheless, such correlation functions would show the interchain
incommensurability, coming from an underlying theory where $t_\perp$
has split the Fermi points. This is the meaning of the C-IC
transition in this context.

To study this transition, we now refermionize the relative charge sector of
the model \eqref{hcic}. Using the conventions in \eqref{rfc}, this gives
\footnote{The same Majorana model appears in
  Ref.~\cite{tsv1990} in the theory of the S=1 biquadratic spin chain
  in a magnetic field parallel to the anisotropy axis.}
\begin{align}
&
{\cal H} = \frac{iv}{2} \sum_{a=\alpha,\beta} 
\left( \xi_L^a \partial_x \xi_L^a - \xi_R^a \partial_x \xi_R^a \right)
-i\mu
\left( \xi_R^\alpha \xi_R^\beta + \xi_L^\alpha \xi_L^\beta \right)
-i \sum_{a=\alpha,\beta} m_a \xi_R^a \xi_L^a.
\label{mh}
\end{align}
Here the ``chemical potential'' of the Majorana fermions $\mu \propto
t_\perp$, while the masses are given by $m_{\alpha(\beta)} = (g_\phi
\pm g_\theta)/2$. In momentum representation the Hamiltonian
(\ref{mh}) may be expressed as a $4\times 4$ matrix acting on the
$4$-component field $\Psi=(\xi_R^\alpha, \xi_R^\beta, \xi_L^\alpha,
\xi_L^\beta)^T$:
\begin{equation}
{\cal H} = \sum_{k>0}  \Psi_k^T {\cal H}_k \Psi_k,
\quad
{\cal H}_k = \mu \hat\sigma_2 + vk \hat\tau_3
+ \left(m_+ + m_- \hat\sigma_3\right)\hat\tau_2,
\end{equation}
where
\[
m_\pm = \frac{m_\alpha\pm m_\beta}{2},
\]
and $\hat\sigma_i$ and $\hat\tau_j$ are the Pauli matrices operating
in the two $2\times 2$ sectors: (1,2) and (R,L), respectively.

The spectrum of this model is given by
\[
\epsilon_k^2 = \mu^2 + v^2k^2 + m_+^2 + m_-^2
\pm 2 \sqrt{m_+^2 m_-^2 + \mu^2 v^2 k^2 + \mu^2 m_+^2}
\]
The spectral gap is defined as
\[
\epsilon_0^2 = \left( m_+ \pm \sqrt{\mu^2 + m_-^2} \right)^2.
\]
Thus, there exists an Ising transition occuring at
\[
\mu = \mu_c = \sqrt{|m_\alpha m_\beta|}.
\]
The sign of the masses in the above relation distinguishes between
transitions to the SC$^d$ and SC$^s$ phases.

We note that an analogous transition has recently been discussed in
the context of the superconductor-insulator transition in
superconducting quantum wires \cite{as1,as2}.

\subsubsection{Absence of direct CDW$^+$ to CDW$^-$ transition}

Finally, we turn to the last remaining potential phase transition in
this model.  Comparing the two phase diagrams, Figs.~\ref{pd1} and
\ref{Phase-Diagram} for zero and large $t_\perp$, respectively, we see
that in a relatively narrow range of interaction parameters, the
addition of interchain hopping may drive the system from a CDW$^+$
phase to a CDW$^-$ one.  As seen from numerical integration of the
two-cutoff RG procedure in the right hand panel of Fig.~\ref{hdpt}
however, this transition appears not to be direct, and instead goes
through an intermediate SC$^s$ phase.  We now give an explantation of
this phenomena.

Comparing the order parameters for CDW$^+$ and CDW$^-$ in the band
basis (${\cal O}_0$ and ${\cal O}_x$ respectively in
Eq.~\eqref{opsboz}), we see that in addition to the change in locking
from the $\phi_c^-$ to the $\theta_c^-$ field as before, the mass of
one of the spin degrees of freedom must also change sign.  While at
finely tuned parameters, both of these things could happen
simultaneously, we will show that this is not the generic situation.
Consequently, the path from CDW$^+$ to CDW$^-$ in general involves two
phase transitions, and therefore an intermediate phase.

The important part of the Hamiltonian \eqref{fullham1} to see this is
\begin{eqnarray*}
&&
{\cal H}' \sim
i\cos\sqrt{4\pi}\theta_c^-
\left[g_T \sum_{i=1}^3 \xi^i_R \xi^i_L
- g_S \xi^0_R \xi^0_L\right]
\\
&&
\\
&&
\quad
+
i\cos\sqrt{4\pi}\phi_c^-
\left[
\tilde{g}_T \sum_{i=1}^3\xi^i_R \xi^i_L - \tilde{g}_S \xi^0_R \xi^0_L
\right] 
+ t_\perp \partial_x \phi_c^-,
\end{eqnarray*}
The CDW$^-$ phase at large $t_\perp$ corresponds to the
strong-coupling fixed point of the RG equations (\ref{RGfive}) where
$g_{T}, g_S \rightarrow\infty$, while the CDW$^+$ phase at $t_\perp=0$
corresponds to the strong-coupling fixed point with $\tilde{g}_T
\rightarrow -\infty, \tilde{g}_S \rightarrow \infty$.

The numerical results depicted in Fig.~\ref{hdpt} can now be
interpreted as follows.  Increasing $t_\perp$ suppresses the field
$\phi_c^-$ and allows the dual field $\theta_c^-$ to lock.  This
changes the nature of the order parameter -- the non-zero expectation
value of $\cos \sqrt{\pi} \phi_c^-$ is replaced by a non-zero
expectation value of $\cos \sqrt{\pi} \theta_c^-$ via the Z$_2$
transition described above.  However, the expectation values of the
terms appearing in the Hamiltonian $\langle \cos \sqrt{4\pi}
\phi_c^-\rangle$ and $\langle \cos \sqrt{4\pi} \theta_c^-\rangle$ are
continuous as one passes through the transition.  As a result, the
spin degrees of freedom are not affected by this first transition, and
therefore the transition is to an intermediate SC$^s$ phase.
 
One can now focus on the spin degrees of freedom via our usual
mean-field decoupling, where in particular, the the spin-triplet mass
$m_T$ is given by
\[
m_T \sim g_T \langle\cos\sqrt{4\pi}\theta_c^- \rangle +
 \tilde{g}_T \langle\cos\sqrt{4\pi}\phi_c^- \rangle.
\]
From the previous analysis of the charge degree of freedom, the
expectation values $\langle\cos\sqrt{4\pi}\theta_c^- \rangle$ and
$\langle\cos\sqrt{4\pi}\phi_c^-\rangle $ change with the increase of
$t_\perp$: the former value increases while the latter decreases.
Given that $g_T>0$ and $\tilde{g}_T<0$ the mass crosses zero when
\[
g_T \langle\cos\sqrt{4\pi}\theta_c^- \rangle =
 |\tilde{g}_T| \langle\cos\sqrt{4\pi}\phi_c^- \rangle.
\]
This is the second transition, to the CDW$^-$ phase; due to the
presence of the marginal term discussed in Sec.~\ref{ltpt}, this
transition becomes weakly first order; as is indicated in
Fig.~\ref{hdpt}.


\section{Effect of a local perturbation}
\label{impurity}

In the preceding Section we have identified the phase diagram of the
ladder model (\ref{h}). In particular, for the most physical case of
predominantly repulsive interaction, the system is either in the
CDW$^-$ or SC$^d$ phase, depending on the relative strength of the
on-site ($U$) and nearest-neighbor ($V_\|$) interactions (see
Fig.~\ref{Phase-Diagram}).

Now we turn to the central issue of this paper, i.e. the effect of a
local perturbation. Previously (see Ref.~\cite{us2}), we have
studied the impurity problem in the context of the spinless ladder
model. Focusing on external probes coupled to the {\it local charge
  density}, we expressed the impurity potential in terms of the
same operators that we used as order parameters that define the
structure of the ground state of the system. The response to the
impurity, i.e. the conductance $G$, is then fully determined by
whether the impurity couples to the dominant order parameter or not:
in the former case the impurity pins the incoherent density wave
related to the order parameter leading to the insulating behavior
($G(T=0)=0$, as in the single-channel Kane-Fisher problem \cite{kaf}),
while in the latter case the impurity has a minimal effect on the
system which remains an ideal conductor (at $T=0$). As a result
\cite{us2}, one can ``switch'' between metallic and insulating
behavior by tuning the parameters of the impurity potential to the
point where it is decoupled from the dominant order parameter in the
system.

In the spinful ladder that is the subject of the present paper the
situation is more complex. In contrast to the spinless case, we
typically have not one but three gapped modes. It turns out that in
this situation, it is not possible to achieve the fully metallic
behavior for any reasonable choice of impurity (as well as the
interaction in the bulk). However, as we show below, there can be a
large difference in the scaling dimension of the impurity, depending
on whether or not the local perturbation couples directly to the order
parameter. This leads to a sizeable crossover effect in the
temperature dependence of the conductance $G(T)$. We also consider
the situation where temperature is lower than most of the gaps in the
system, but greater than the gap in one of the collective modes, as
can happen near one of the phase transition lines.

\subsection{Local perturbation coupled to the dominant order parameter}

Suppose the impurity potential contains the operator describing one of
the massive bulk modes in the problem. As a physical example, consider
the CDW$^-$ phase characterized by the non-zero expectation value 
\begin{equation}
\left\langle \mu_1\mu_2\mu_3\sigma_0
\sin\sqrt{\pi} \theta_c^- 
\right\rangle \ne 0.
\label{nzev}
\end{equation}
Now consider a local perturbation of the form
\begin{equation}
\label{lp1}
{\cal H}_{\rm imp} = \lambda {\cal O}_{CDW^-}(0).
\end{equation}
Given the multiplicative form of the operator ${\cal O}_{CDW^-}$ (see
\ref{op}), we can replace the massive fields in Eq.~(\ref{lp1}) by
their expectation value (\ref{nzev}). Technically, this is equivalent
to ``integrating out'' these massive degrees of freedom. Then the
effective Hamiltonian for the total charge takes the form
\begin{equation}
{\cal H}_{\rm eff} = {\cal H}_0[\phi_c^+] + \tilde{\lambda} \cos\sqrt{\pi}\phi_c^+(0),
\label{he1}
\end{equation}
where
\begin{equation}
\tilde{\lambda} = \lambda 
\left\langle \mu_1\mu_2\mu_3\sigma_0
\sin\sqrt{\pi} \theta_c^- \right\rangle .
\end{equation}
The effective Hamiltonian~(\ref{he1}) describes a boundary sine-Gordon
model (which is equivalent to the single-channel Kane-Fisher problem)
with a local perturbation having scaling dimension $d=K_c^+/4$. Hence
for $K_c^+>4$ the operator is irrelevant and one obtains a power-law
correction to perfect conductance \cite{kaf}
\begin{equation}
G = \frac{4e^2}{h} - {\rm const.} \times \left[{\rm Max}(T,V)\right]^{\gamma}, 
\quad \gamma = 2(d-1) = K_c^+/2 - 2 >0.
\end{equation}
 For $K_c^+<4$ the
perturbation is relevant and flows to strong coupling
\footnote{In this paper we assume the base impurity potential to be
  weak enough, allowing us to consider its RG flow in the presence of
  the dynamically generated bulk masses. It is worth noting that for a
  stronger bare impurity, the RG flows of bulk interactions and the
  impurity may be entangled. This issue lies outside of the scope of
  the present paper and will be discussed elsewhere.}.

The strong-coupling physics is determined solely by the total charge
sector. Mathematically, the resulting behavior is similar to that in
the Kane-Fisher problem. There is, however, an important difference.
In the Kane-Fisher problem \cite{kaf}, the strong-coupling regime is
equivalent to tunneling across the weak link connecting two
semi-infinite chains. In our case, such analogy is also useful,
however, it is confined to the total charge sector only, while the
three massive sectors of the model remain largely unaffected by the
impurity. Thus, the resulting strong-coupling picture is by no means
related to the problem of tunneling between two semi-infinite ladders.
In other words, while in the single-chain case \cite{kaf} the impurity
effectively cuts the chain into two semi-infinite parts, the ladder
remains intact, with only one (out of four) sectors of the effective
low-energy theory being affected by the impurity. Within the total
charge sector, the weak-link analogy can still be used, but with the
caveat that single-electron processes are now forbidden (by the gaps)
and so ``tunneling across the weak link'' refers to four-fermion
processes that do not change any quantum numbers related to the gapped
degrees of freedom. Such a process is represented by the operator
$\cos \sqrt{16\pi} \theta_c^+$, with the scaling dimension
$d=4/K_c^+$. Alternatively \cite{sal}, one may realize that this
operator generates a soliton in the $\phi_c^+$ field between
successive minima of the $\cos\sqrt{\pi}\phi_c^+$ potential; hence
purely on field-theoretic grounds it may be considered as the leading
(irrelevant) operator involved in generating a current in the strong
coupling regime. An RG analysis as studied by Kane and Fisher
\cite{kaf} then leads immediately to the result that the conductance
at temperature $T$ and bias voltage $V$ is given by a power law
\begin{equation}
G \propto \left[{\rm Max}(T,V)\right]^{\gamma'}, \quad\quad\quad
\gamma' = 8/K_c^+ - 2 >0.  
\label{r1}
\end{equation}
While we have arrived at Eq.~(\ref{r1}) specifically for the case of
the CDW$^-$ phase, such a behavior always occurs if some component of
the local impurity couples to the bulk ``order parameter" of the
system so long as temperature and bias voltage remain lower than the
bulk gaps.

\subsection{Local perturbation not directly coupled to the dominant order parameter}
\label{so}

Now, let us suppose that no component of the local perturbation is
identical to the bulk order parameter -- for concreteness we again
consider the case of being in the CDW$^-$ phase, but now with a
symmetric impurity \cite{us2}
\begin{equation}
\label{lp2}
{\cal H}_{imp} = \lambda{\cal O}_{CDW^+}(0).
\end{equation}
To obtain the effective theory of the total charge sector for energies
small compared to the dynamically generated gaps, we again integrate
out the gapped degrees of freedom. In contrast to the previous case,
the impurity potential (\ref{lp2}) does not contribute in the first
order, as becomes clear by comparing the non-zero expectation value
(\ref{nzev}) to the multiplicative structure of the order parameter
[see Eq.~(\ref{opsboz})]
\begin{eqnarray*}
&&
\lambda\left[
\cos\sqrt{\pi}\phi_c^+ \sin\sqrt{\pi}\phi_c^-
\sigma_1\sigma_2\sigma_3\sigma_0 \;
+ \;
\sin\sqrt{\pi}\phi_c^+ \cos\sqrt{\pi}\phi_c^-
\mu_1\mu_2\mu_3 \mu_0 
\right].
\end{eqnarray*}
Focusing on the first term, we find that the second-order contribution
of the impurity to the effective action for $\phi_c^+$ has the form
\begin{equation}
\frac{\tilde{\lambda}^2}{2} \int d \tau_1d \tau_2 C(\tau_1,\tau_2) 
\cos\sqrt{\pi}\phi_c^+(\tau_1) \cos\sqrt{\pi}\phi_c^+(\tau_2),
\label{secondorder}
\end{equation}
where $\tau_i$ are the imaginary times, all quantities are taken at
the impurity site ($x_i=0$), the effective coupling constant is
$\tilde{\lambda}=\lambda\langle \sigma_1\sigma_2\sigma_3\rangle$ and
$C(\tau_1,\tau_2)$ is the short-hand notation for the correlation
function
\begin{eqnarray}
\label{corrf}
C(\tau_1, 0; \tau_2, 0) = 
\left\langle \sin\sqrt{\pi}\phi_c^-(\tau_1, 0) 
\sin\sqrt{\pi}\phi_c^-(\tau_2, 0)\right\rangle
\,
\langle \sigma_0(\tau_1, 0) \sigma_0(\tau_2, 0) \rangle.
\end{eqnarray}
The correlation function (\ref{corrf}) is short-ranged due to the
presence of dynamically generated gaps:
\begin{equation}
C(\tau_1,\tau_2)\sim e^{-|\tau_1-\tau_2|\Delta},
\end{equation}
where $\Delta \sim m_S + m_F$ (up to numerical factors; $m_S$ and
$m_F$ are gaps in the spin-singlet and relative charge sectors,
respectively). Then we can neglect the difference between $\tau_1$ and
$\tau_2$ in the operators in Eq.~(\ref{secondorder}), which leads to
the effective model of the total charge sector
\begin{equation}
{\cal H}_{eff} = {\cal H}_0[\phi_c^+] + \lambda' \cos \sqrt{4\pi}\phi_c^+(0).
\label{he2}
\end{equation}
Unlike the spinless case \cite{us2}, the impurity potential in
Eq.~(\ref{he2}), which has scaling dimension $d=K_c^+$, is still
relevant for overall bulk repulsive interactions $K_c^+<1$, although
significantly less relevant than the similar term in Eq.~(\ref{he1}).
The strong coupling arguments used in the previous subsection may
still be applied, but now the soliton in the fields $\phi_c^+$ is
generated by the operator $\cos \sqrt{4\pi} \theta_c^+$, which
physically corresponds to \textit{pair-hopping} through the strong
impurity. As a result, the conductance is described by the weaker
power law
\begin{equation}
G \propto \left[{\rm Max}(T,V)\right]^{\gamma'}, \quad 
\gamma' = 2/K_c^+ - 2 >0.
\label{r2}
\end{equation}

The metallic behavior found in the case of the spinless ladder
\cite{us2} can appear in the spinful ladder only if the bulk
interactions are attractive $K_c^+>1$.  The impurity potential in
Eq.~(\ref{he2}) is now irrelevant (in the RG sense) and the
conductance remains close to it's maximal value
\begin{equation}
G = \frac{4e^2}{h} - {\rm const.} \times \left[{\rm Max}(T,V)\right]^{\gamma}, 
\quad \gamma = 2K_c^+ - 2 >0.
\label{r3}
\end{equation}

\begin{table}
\begin{center}
\begin{tabular}{ccccccc}
perturbation  && $\gamma'$ && $\gamma$ && $K^*$ \\ 
\hline 
first order coupling && $8/K_c^+ - 2$ && $K_c^+/2 -2$ && 4
 \\ 
  second order only && $2/K_c^+ - 2$ && $2K_c^+ -2$ && 1 \\
\end{tabular}
\end{center}
\caption{Summary of the different power laws seen as a response to
  different impurity structures and in different phases of the bulk
  system.  In each case, one finds insulating behavior for $K_c^+<K^*$
  defined as $G\propto \left[{\rm Max}(T,V)\right]^{\gamma'} $; or
  conducting behavior for $K_c^+>K^*$ defined as $G = 4e^2/h -
  {\rm const.} \times \left[{\rm Max}(T,V)\right]^{\gamma}$.  The
  perturbations are divided into two types: those that couple directly
  to the order parameter of the system (such as a single impurity on
  one of the legs in the CDW$^-$ phase of the system, see \cite{us2});
  and those that don't (for example, any local denisty perturbation in
  one of the superconducting phases of the bulk).}
\label{table-gammas}
\end{table}

The qualitative results obtained so far, can be summarized as follows:
\begin{enumerate}
\item For the system in the bulk CDW$^-$ state, the response of the
  system to a local impurity indeed depends on the nature of the
  impurity, as in the spinless model \cite{us2}. However, in contrast
  to the spinless model, changing the local structure of the impurity
  in the case of repulsive interactions, $K_c^+<1$, does not lead to a
  metal-insulator transition, but rather to a strong crossover at any
  non-zero temperature ($T^{6-8\delta}$ behavior as compared to
  $T^{2\delta}$ behavior for $K_c^+=1-\delta$). The potential-sensing
  applications detailed in \cite{us2} are still therefore realistic in
  the spinful case.
\item The critical $K_c^+$ separating metallic and insulating behavior
  is different for the first- and second-order terms. For certain
  attractive interactions with $1<K_c^+<4$, one may therefore still
  see a metal-insulator transition as the component of the local
  impurity that couples to the order parameter is tuned to zero.
\item If the system is in one of the bulk superconducting states (as
  is most realistic for ladders with strong interchain hopping), the
  response to the local impurity is independent of the structure of
  the impurity.
\end{enumerate}
The results for the different power laws are summarized in
Table~\ref{table-gammas}.

Finally, we mention that the qualitative picture of the strong coupling (insulating)
phase of the impurity, i.e. whether four electrons are required to
tunnel (forming a singlet of all the gapped sectors) or whether pair
tunneling is possible may be extended to SU(N) ladders with all
modes gapped except total charge \cite{aap}.

\subsection{Transport near one of the phase transition lines}

Throughout most of the phase diagram, the three gapped collective
modes all have gaps of the same order of magnitude. However, when the
system is sufficiently close to one of the phase transition lines, the
situation is different.  Here, one of the gaps becomes very small
(vanishing at the transition line), allowing for a sizeable regime
where voltage and/or temperature are much larger than the smallest
gap, but still much smaller than the remaining two. The extra soft
degree of freedom leads to modified exponents in the power laws, but,
as it turns out, only if the impurity couples directly to the order
parameter.  For second order processes, this extra soft degree of
freedom doesn't affect the resulting behavior. In this subsection, we
show how this happens, discussing in detail the situation for
transport near a U(1) transition line, then outlining the
differences for the other cases.

\subsubsection{Transport near the U(1) transition}

As discussed in Section~\ref{ltpt}, U(1) phase transitions (to be
concrete, we will consider the CDW$^-$ to OAF transition) in the
two-leg ladder are accociated with the change of sign of the mass
$m_F$ in the relative charge sector. In the vicinity of the U(1)
transition line, $m_F$ is small compared to the gaps in the spin
sector and we may consider an intermediate temperature regime where
\[
m_F \ll T \ll m_T, m_S.
\]
In this case the relative charge sector should be considered on equal
footing with the gapless total charge sector, while the gapped spin
modes may be integrated out.

Consider now an impurity (\ref{lp1}), which couples to the CDW$^-$
order parameter. In this case the spin part of the impurity potential
has a non-zero expectation value across the transition line.
\footnote{This is true in general for any impurity which couples to
  the ground state of the system to first order on \textit{either}
  side of the transition.  For example, one gets the same scenario if
  the impurity is a local flux, coupling to the OAF order parameter}
Integrating out the spin degrees of freedom, we find the effective
impurity potential for the charge sector in the following form
\begin{equation}
\tilde{\lambda} \sin\sqrt{\pi}\phi_c^+ (0)\cos\sqrt{\pi}\theta_c^-(0), \quad \quad
\tilde{\lambda}=\lambda\langle \mu_1\mu_2\mu_3\sigma_0\rangle.
\label{lp3}
\end{equation}
Up to certain duality transformations and rescaling, the problem
appears to be similar to that of the single impurity in a two-channel
(i.e. spinful) Luttinger liquid \cite{kaf}. Our analysis of the
present case will then be guided by this equivalence. The scaling
dimension of the operator (\ref{lp3}) is
\begin{equation}
d=K_c^+/4+1/4K_c^-.
\end{equation}
For $K_c^-\approx 1$, the operator (\ref{lp3}) is relevant if
$K_c^+<3$. As before, we analyze the strong-coupling phase by
identifying leading irrelevant operators that generate current
flow. Due to the multiplicative structure of the potential
(\ref{lp3}), there are two such operators: (i)
$\cos\sqrt{16\pi}\theta_c^+$, and (ii)
$\cos\sqrt{4\pi}\theta_c^+\cos\sqrt{4\pi}\phi_c^-$. The former
operator (with the scaling dimension $d=4/K_c^+$) creates a full
soliton in the $\phi_c^+$ channel, while the latter (with the scaling
dimension $1/K_c^++K_c^-$) creates a half-soliton in both the
$\phi_c^+$ and the $\phi_c^-$ channels.

The conductance is then dominated by the least irrelevant operator,
which gives
\begin{equation}
G \propto \left[{\rm Max}(T,V)\right]^{\gamma'}, \quad
 \gamma' = 
 \begin{cases}
 2/K_c^+  >0, & K_c^+ < 3, \\
 8/K_c^+ -2, & 3<K_c^+<4.
 \end{cases}
\label{r4}
\end{equation}
We should point out that for values of $3<K_c^+<4$, both the
conducting and insulating fixed points are stable, implying that there
must be an unstable intermediate fixed point.  This is typical in
two-channel Kane-Fisher like problems; in this case however, it occurs
far away in phase space from the most realistic repulsive interactions
$K_c^+<1$ in which we are most interested, so we will say no more
about this in the present work.

As before, we may also consider an impurity potential where the spin
part has no expectation value in the ground state of the system.  For
the CDW$^-$ to OAF transition, an example of such a potential is given
by the symmetric impurity on both legs, Eq.~(\ref{lp2}).  In both the
CDW$^-$ and OAF phases,
$\langle\sigma_1\sigma_2\sigma_3\sigma_0\rangle=0$. Similar to the
discussion in Section~\ref{so}, the leading contribution of the
impurity to the effective theory in the charge sector therefore comes
from the second order term, which may be expressed as
\begin{equation}
{\cal H}_{imp} \sim \left[C_1 + \cos \sqrt{4\pi}\phi_c^+(0)\right]
\left[C_2 + \cos\sqrt{4\pi}\theta_c^- (0)\right].
\label{c1c2}
\end{equation}
The constants $C_1$ and $C_2$ appear due to the structure of the
operator product expansion (OPE) \cite{gnt}:
\begin{equation}
\cos \beta\phi \cos\beta\phi \sim \mathbb{I} + \cos 2\beta\phi,
\label{goodOPE}
\end{equation}
which was used to obtain \eqref{c1c2}.

Now, the only term of relevance is $\cos \sqrt{4\pi}\phi_c^+$ (with
scaling dimension $d=K_c^+$). The problem is thus identical to
Eq.~(\ref{he2}) that arises deep in the ordered phases. Furthermore,
we notice that such a term will always appear in second order due to
the presence of the identity in the OPE \eqref{goodOPE} above. In what
follows we therefore focus only on the cases where the impurity
contributes in the first order, where the presence of the extra
gapless mode modifies the observable properties of the system.

\subsubsection{Transport near the Z$_2$ and SU(2)$_2$ transitions}

We now apply the theory developed in the previous subsection to the
other transition lines. As we have already shown that the the
presence of the transition line does not make a difference in the
situation when the impurity does not couple directly to the order
parameter of the bulk system, we present only the calculations for the
first-order case.

Consider now the situation in the vicinity of the Z$_2$ transition
(for example, between the CWD$^-$ and SC$^s$ phases), and focus on the
impurity potential (\ref{lp1}). Similarly to the U(1) case, within
the interval
\[
m_S \ll T \ll m_T, m_F,
\]
we integrate out the two gapped degrees of freedom and find the
effective potential contributing in the first order
\begin{equation}
{\cal H}_{imp} \sim \cos\sqrt{\pi}\phi_c^+ (0) \sigma_0(0).
\label{lp4}
\end{equation}
The scaling dimension of this operator is
\begin{equation}
d=K_c^+/4+1/8.
\end{equation}
For $K_c^+<7/2$ the impurity potential (\ref{lp4}) is relevant and
again we find two operators that generate a current flow: (i)
$\cos\sqrt{16\pi}\theta_c^+$, and (ii)
$i\xi_R^0\xi_L^0\cos\sqrt{4\pi}\theta_c^+$. As before, the first
operator creates a full soliton in the $\phi_c^+$ channel, while the
second creates a half-soliton in the $\phi_c^+$ channel and at the
same time changes the sign of the Ising order parameter in the
spin-singlet sector.

This can be clarified as follows. Consider the bosonic form of the
impurity operator (\ref{lp1}) [given by Eq.~(\ref{opsboz})]
\[
e^{-i\sqrt{\pi}\phi_c^+}
\sin \sqrt{\pi} \theta_c^- 
\cos \sqrt{\pi} \phi_s^+ \cos \sqrt{\pi} \theta_s^- 
\sim e^{-i\sqrt{\pi}\phi_c^+}
\sin\sqrt{\pi}\theta_c^- \,
\mu_1\mu_2\mu_3 \sigma_0.
\]
The Ising
operator $\sigma_0$ is ``contained'' in the relative spin sector
$\cos\sqrt{\pi}\theta_s^-$. A half-soliton of this field can be
created by the operator $\cos\sqrt{4\pi}\phi_s^-$, which under the
re-fermionization rules (\ref{rf}) corresponds to
$i\xi_R^3\xi_L^3+i\xi_R^0\xi_L^0$. The triplet sector is gapped.
Therefore, given the additive structure of the Gaussian part of the
Hamiltonian, the contribution of the triplet component may be
disregarded here, leading to the above form (ii) of the current
generating operator.

The scaling dimensions of the operators (i) and (ii) are $d=4/K_c^+$
and $d=1/K_c^++1$ respectively.  These are identical to those in the
U(1) case, and thus lead to the same conductance behavior
(\ref{r4}).

Finally, going through the same logic in the vicinity of one of the
SU(2)$_2$ transitions when
\[
m_T \ll T \ll m_S, m_F,
\]
one finds after integrating out the gapped degrees of freedom the
operator
\begin{equation}
{\cal H}_{imp} \sim \cos\sqrt{\pi}\phi_c^+ (0) \sigma_1(0) \sigma_2(0) \sigma_2(0),
\label{lp5}
\end{equation}
so long as the impurity couples to first order.  This operator has
scaling dimension $d=K_c^+/4+3/8$ and is therefore relevant for
$K_c^+<5/2$. Again, there are two soliton-creating operators in the
problem with the scaling dimensions that are the same as in the
previous two cases. The resulting conductance is still given by
Eq.~(\ref{r4}).

A summary of transport near one of the transition lines is given in
Table \ref{table-gamma-transition}. Ultimately for the case of
repulsive interactions $K_c^+<1$ [see the upper case of
  Eq.~(\ref{r4})], the impurity couples directly to the bulk
correlations. For the cases when the impurity does not have a direct
coupling on either side of the transition, the transition line is
irrelevant.

\begin{table*}
\begin{center}
\begin{tabular}{cccccc}
 & SU(2)$_2$ &&  $\gamma=K_c^+/2-5/4$ && $K_c^+>5/2$ \\
 Conducting & U(1) &&  $\gamma=K_c^+/2-3/2$ && $K_c^+>3$ \\
    & Z$_2$ &&  $\gamma=K_c^+/2-7/4$ && $K_c^+>7/2$  \\ \hline
 \multirow{2}{*}{Insulating} & \multirow{2}{*}{all} 
       && $\gamma' = 2/K_c^+$ && $K_c^+<3$ \\
        & && $\gamma' = 8/K_c^+ -2$ && $ 3<K_c^+<4$ \\
  \end{tabular}
\end{center}
\caption{Summary of the impurity effects in the vicinity of phase
  transition lines in the case where the impurity couples directly to
  bulk correlations. The notation for the exponents $\gamma'$ near the
  insulating fixed point and $\gamma$ near the conducting fixed point
  are the same as in Table \ref{table-gammas}. If there is no direct
  coupling between the impurity and the bulk orderparameters, then the
  results are identical to those in Table \ref{table-gammas}. For each
  transition line, there is a range of values of $K_c^+$ for which
  both conducting and insulating fixed points are stable -- these are
  separated by an intermediate unstable fixed point which is beyond
  the scope of the present work.}
\label{table-gamma-transition}
\end{table*}

\subsection{High-temperature transport properties}

Having discussed the conductance of the system at low (i.e. lower than
all the dynamically generated gaps) and intermediate temepratures, we
now turn to the discussion of high-temeprature transport, where the
temperature is highest energy scale in the effective theory, in other
words
\[
T\gg m_S, m_T, \Delta_c^-.
\]
In this case the dynamically generated gaps are not important and one
may consider all four modes as critical. Now, the precise form of the
impurity potential is no longer important. To be concrete however,
below we consider the CDW$^-$ potential (\ref{lp1}) as a
representative example.

The bosonic form of the operator ${\cal O}_{CDW^-}$ is given in
Eq.~(\ref{opsboz}). The scaling dimension of this operator is
\[
d \approx K_c^+/4 + 3/4,
\]
where we assume that all Luttinger liquid parameters except for
$K_c^+$ are of order unity. Thus the operator (\ref{lp1}) is relevant
for $K_c^+<1$, which is the case for overall bulk repulsive
interaction. As before, we now need to identify operators, that can
generate the current flow. There are now three possibilities: (i) the
operator $\cos\sqrt{16\pi}\theta^+$ (with the scaling dimension
$d=4/K_c^+$), creating a soliton in the total charge sector; (ii)
operators creating half-solitons in the total charge and
simulataneously in one other sector ($d=1/K_c^++1$); and (iii) an
operator creating ``quarter-solitons'' in all four sectors. This is
possible due to the double degeneracy of all local operators in
Eq.~(\ref{opsboz}). A ``quarter-soliton'' corresponds to the
one-quarter period shift of bosonic fields, that ``switches'' between
the two additive terms in the bosonic form of the local operators. In
the case of ${\cal O}_{CDW^-}$, such an object can be created by the
operator
\footnote{Naively the operator creating ``quarter-solitons'' appears
  to break the SU(2) symmetry. This, however, is not so, since the
  duality transformation used to derive this operator involves a
  particle-hole transformation that changes the SU(2)
  representation. Keeping this in mind it is not difficult to see that
  the operator is in fact SU(2)-invariant.}
\begin{eqnarray*}
&&
\cos\sqrt{\pi}\theta_c^+ \sin\sqrt{\pi}\phi_c^-
\cos\sqrt{\pi}\theta_s^+ \cos\sqrt{\pi}\phi_s^- 
\;
\sim
\; \cos\sqrt{\pi}\theta_c^+ \sin\sqrt{\pi}\phi_c^-
\mu_1\sigma_2\sigma_3\sigma_0,
\end{eqnarray*}
with the scaling dimension
\begin{equation}
\label{dimp-highT}
d = 1/(4K_c^+) + 3/4
\end{equation}
Putting this all together gives us the expression for conductance
\begin{eqnarray}
G \propto \left[{\rm Max}(T,V)\right]^{\gamma'}, &
\gamma' = 1/(2K_c^+) - 1/2  >0, & K_c^+ < 1 \nonumber \\
G=\frac{4e^2}{h} - {\rm const.} \times \left[{\rm Max}(T,V)\right]^{\gamma}, & 
\gamma = K_c^+/2 - 1/2  >0, & K_c^+ > 1.
\label{r5}
\end{eqnarray}

Finally, the case when the temperature is bigger than all but one of
the gaps (as may happen in the regions of the phase diagram when two
phase boundaries meet), the situation is very similar to this
high-temperature phase; particularly in that here the structure of the
local impurity is not important.  Aside from some slight changes in
scaling dimensions from the single ``locked" degree of freedom, and
the absence of the quarter-soliton in all sectors, the analysis of
this case is identical to the present section.  We will however not
delve further into details of this final uncommon case.


\section{The effect of an ionic potential}
\label{ionic}

We now consider the situation where the ladder is subject to a
periodic external potential commensurate with the carrier density in
the system, i.e. the potential is characterized by the wave-vector
$2k_F$.  As discussed in the introduction, this potential may be there
naturally in certain materials \cite{tnc1,tnc2} or artificially added
in other setups \cite{ipo,bichromatic1,bichromatic2}. Such a potential
is similar to the charge-density wave order parameters CDW$^\pm$ and
couples directly to the particle density in a similar manner to the
local perturbations considered in the previous Section. We therefore
express the external potential in terms of the same local operators
(\ref{opsboz}) that we have used as order parameters to describe the
bulk phases of the system, as well as to describe the impurity
potentials (\ref{lp1}) and (\ref{lp2}). In the presence of such a
perturbation, the Hamiltonian of the system is
\begin{equation}
{\cal H} = {\cal H}_{\rm bulk} + \left[ \lambda_i {\cal O}_{CDW^{\pm}} + H.c. \right],
\label{ipot}
\end{equation}
where ${\cal H}_{\rm bulk}$ is the Hamiltonian (\ref{h}).  Without
loss of generality, we may allow $\lambda_i$ to be real.

Let us reiterate, that in contrast to the local impurity potential, we
now subject the system to a {\it bulk} perturbation. The perturbation
therefore will affect the bulk state of the system. Indeed, when the
characteristic energy of the perturbation is the largest energy scale
in the problem, the dominant correlations in the system will be
determined by said perturbation, and not by the interactions which may
favour a different phase. There is therefore a possibility of
exhibiting a quantum phase transition as $\lambda_i$ is increased from
zero. In the equivalent situation in a half-filled single chain, there
may occur \textit{two} quantum phase transitions \cite{fgn} on the
path from Mott Insulator to band insulator. Adding a dimerizing
potential to the spin ladder may lead to an SU(2)$_1$ transition
\cite{mss,wan}, a possibility which also naturally occurs in
\textit{half-filled} zig-zag carbon nanotubes
\cite{Carr-Gogolin-Nersesyan-2007}. The present case of adding such a
potential to a doped ladder has been recently discussed \cite{ret} in
the context of unusual ground states, such as pair-density waves.

As discussed in Section~\ref{hightperp} and illustrated in
Fig.~\ref{Phase-Diagram}, in a typical ground state of the system in
the absence of the external potential the relative charge sector and
the two spin sectors of the theory are gapped. In such a state, one of
the order parameters (\ref{opsboz}) is characterized by power-law
correlations, while all other operators are short-ranged. Therefore we
expect two typical scenarios: (i) the ionic potential may couple to
the dominant order parameter in the system; we expect that in this
case the ionic potential will stablize the existing correlations and
lead to the appearance of the true long-range order; or (ii) the ionic
potential may couple to one of the short-ranged density operators. In
this case the competition between the ionic potential and the
pre-existing dominant correlations may lead to phase transitions at
some critical values of the strength of the potential. Below we
decsribe both situations. The resulting phase transitions are
summarized in Table~\ref{table-ionic}.

\subsection{Ionic potential coupled to the dominant order parameter}

In this Section we consider the situation where the ionic potential
couples directly to the dominant order parameter in the system. This
happens if either the CDW$^-$ potential is applied to the system in
the CDW$^-$ phase, or the CDW$^+$ potential is applied to the CDW$^+$
phase.

\subsubsection{CDW$^-$ ground state}

We start by considering the ladder with repulsive interaction, such
that the system is in the CDW$^-$ phase (see
Fig.~\ref{Phase-Diagram}). This phase is characterized by the non-zero
expectation value (\ref{nzev}), which we repeat for convenience:
\begin{equation}
\left\langle \mu_1\mu_2\mu_3 \sigma_0 \sin\sqrt{\pi}\theta_c^-\right\rangle \ne 0.
\label{ev1}
\end{equation}
Suppose that the external potential is also of the CDW$^-$ type.
Integrating out all gapped degrees of freedom, we find the effective
perturbation in the total charge sector
\begin{equation}
\delta{\cal H}_{\rm eff} = \lambda_i 
\left\langle \mu_1\mu_2\mu_3 \sigma_0 
\sin\sqrt{\pi}\theta_c^-\right\rangle \cos\sqrt{\pi}\phi_c^+  .
\label{ip1}
\end{equation}
This is a relevant perturbation which effectively pins the incoherent
charge-density wave and opens a gap in the total charge sector,
turning the system into an insulator.  In other words, the
quasi-long-range order CDW$^-$ phase exhibits true long-range order as
soon as the external potential is non-zero.  This does not violate the
Hohenberg-Mermin-Wagner theorem \cite{maw} because there is no
spontaneous symmetry breaking -- the symmetry is explicitly broken by
the external potential.

\subsubsection{CDW$^+$ ground state}

An equivalent situation occurs in the case of attractive inter-layer
interaction and very weak (or vanishing) inter-chain hopping, where
the ground state of the system is CDW$^+$ and the system is
characterized by the expectation value
\begin{equation}
\left\langle \sigma_1\sigma_2\sigma_3 \sigma_0 
\sin\sqrt{\pi}\phi_c^-\right\rangle \ne 0.
\label{ev2}
\end{equation}
In full analogy to the previous case of repulsive interaction, we
integrate out the gapped degrees of freedom and find the same
effective perturbation (\ref{ip1}), although with a different
effective coupling constant
\[
\delta{\cal H}_{\rm eff} = \lambda_i 
\left\langle \sigma_1\sigma_2\sigma_3 \sigma_0 
\sin\sqrt{\pi}\phi_c^-\right\rangle
\cos\sqrt{\pi}\phi_c^+  ,
\]
which has the same effect of opening the gap in the total charge
sector and forming a state with true long-range order.

\subsection{Ionic potential coupled to short-ranged correlations}

Now we turn to the case where the ionic potential couples not to the
dominant order parameter in the system, but instead to one of the
short-ranged operators. Since we consider ionic potentials of the
CDW type (\ref{ipot}), here we discuss the situation, where the
CDW$^+$ perturbation is applied to the CDW$^-$ phase of the ladder
(and vice versa), as well as the effect of such potentials on either
the SC$^{d(s)}$ or OAF ground states.

In what follows, we concentrate only on the case where $K_c^+<2$ (thus
includes the physical case of repulsive interactions $K_c^+<1$). For
$K_c^+>2$, the situation is quite different from that presented here,
as the second-order terms are not relevant in this case. However, as
values of $K_c^+$ greater than two correspond to unphysically large
attractive interacitons, we do not discuss this case further in the
present work.

\subsubsection{Ground states with superconducting correlations}

Consider now ground states with superconducting correlations. For the
case of repulsive interaction, this is the state with SC$^d$
correlations, characterized by the expectation value (again, without
loss of generality, we may choose one of the two possibilities)
\begin{equation}
\langle \cos\sqrt{\pi}\theta_c^-
\mu_1\mu_2\mu_3 \mu_0 \rangle \ne 0.
\label{ev3}
\end{equation}

Suppose, that the ionic potential is of the CDW$^+$ type. This
operator has the same structure of the spin degrees of freedom as the
expectation value \eqref{ev3}.  The perturbation will therefore not
affect the spin sector, and we may integrate out the spin degrees of
freedom and focus on the effective theory of the charge sector
\begin{subequations}
\label{ehsp}
\begin{equation}
{\cal H}_{\rm eff} = {\cal H}_0\left[\phi_c^+\right] + {\cal H}_0 \left[\phi_c^-\right] 
- g_\theta \cos\sqrt{4\pi} \theta_c^- + \delta{\cal H}_{\rm eff},
\end{equation}
\begin{equation}
\delta{\cal H}_{\rm eff} = \lambda_i 
\left\langle \mu_1\mu_2\mu_3 \mu_0 \right\rangle 
\sin\sqrt{\pi}\phi_c^+  \cos\sqrt{\pi}\phi_c^-.
\label{ip3}
\end{equation}
\end{subequations}
We now argue, that the effective perturbation (\ref{ip3}) leads to a
gap opening in the total charge sector. As in the local impurity case,
the perturbation $\delta{\cal H}_{\rm eff}$ does not couple to first
order to the correlations of the bare system. However, in the second
order, a term $\cos \sqrt{4\pi} \phi_c^-$ is generated. This is
generally relevant (for $K_c^+<2$), and so will lead to a gap opening
in the $\phi_c^+$ mode, i.e. to a non-zero expectation value
$\langle\sin\sqrt{\pi}\phi_c^+\rangle\ne 0$. Hence, we may integrate
out the total charge and focus on the relative charge sector, which is
now described by the effective Hamiltonian
\begin{equation}
{\cal H}_{\rm eff} = {\cal H}_0 \left[\phi_c^-\right] 
- g_\theta \cos\sqrt{4\pi} \theta_c^- + \tilde{\lambda_i} \cos\sqrt{\pi}\phi_c^-,
\label{ehspr}
\end{equation}
where
\[
\tilde{\lambda_i} = \lambda_i \left\langle \sin\sqrt{\pi}\phi_c^+\right\rangle 
\left\langle \mu_1\mu_2\mu_3 \mu_0 \right\rangle.
\]
Naively, Eq.~(\ref{ehspr}) is the Hamiltonian of the double
sine-Gordon model. However, the appearance of both the field
$\phi_c^-$ and its dual field $\theta_c^-$ means that there is no
phase transition: as soon as the coupling constants in
Eq.~(\ref{ehspr}) are non-zero $g_\theta \ne 0$ and
$\tilde\lambda_i\ne 0$ we find {\it two} non-zero expectation values
\begin{equation}
\left\langle \cos\sqrt{\pi}\theta_c^- \right\rangle \ne 0, 
\quad
\left\langle \cos\sqrt{\pi}\phi_c^- \right\rangle \ne 0.
\end{equation}
Thus we conclude, that the ionic potential immediately drives the
system into a state where the CDW$^+$ correlations are fully
locked. The resulting state is characterized by a non-zero expectation
value of both the CDW$^+$ order parameter, \textit{and} the operator
\begin{equation}
\langle \sin\sqrt{\pi}\phi_c^+ \cos\sqrt{\pi}\theta_c^-
\mu_1\mu_2\mu_3 \mu_0 \rangle \ne 0,
\end{equation}
which corresponds to a spin-liquid type order.

If the ground state is of the SC$^s$ type, the effect is the same,
with the only difference that all occurrences of
$\cos\sqrt{\pi}\theta_c^-$ should be replaced with
$\sin\sqrt{\pi}\theta_c^-$.

Now, if the external potential is of the CDW$^-$ type and the ground
state is SC$^d$, then applying the same line of arguement as above, we
find that one must focus on the triplet sector, and the effective
perturbation (having integrated out the spin-singlet and
relative-charge degrees of freedom) is
\begin{equation}
\sin\sqrt{\pi}\phi_c^+ \langle \cos\sqrt{\pi}\theta_c^- \rangle
\sigma_1\sigma_2\sigma_3 \langle \mu_0\rangle .
\end{equation}
As above, treating this operator in the second order, we find that the
total charge sector acquires a gap as soon as the ionic potential is
applied. The remaining question is the physics in the spin-triplet
sector, where the effective Hamiltonian is
\begin{equation}
{\cal H}_{\rm eff} \sim {\cal H}_0 \left[ \xi_i \right] 
+ im_T \sum_{i=1}^3 \xi^i_R\xi^i_L + \tilde{\lambda}_i \sigma_1\sigma_2\sigma_3.
\end{equation}
The sign of $m_T$ is such that $\langle \mu_1\mu_2\mu_3\rangle\ne 0$
in the absence of the $\tilde{\lambda}_i$ term. However, this term
does not induce a phase transition and similarly to the above we find
the state with both $\langle \mu_1\mu_2\mu_3\rangle\ne 0$ and $\langle
\sigma_1\sigma_2\sigma_3\rangle\ne0$.

Finally, consider the case where the external potential is of the
CDW$^-$ type and the ground state is SC$^s$. As already mentioned
above for the case of the CDW$^+$ perturbation, the analysis remains
the same, but we need to replace $\cos\sqrt{\pi}\theta_c^-$ with
$\sin\sqrt{\pi}\theta_c^-$. In particular, instead of the expectation
value (\ref{ev3}) we now have
\begin{equation}
\langle \sin\sqrt{\pi}\theta_c^-
\mu_1\mu_2\mu_3 \mu_0 \rangle \ne 0
\end{equation}
in the ground state, so that the effective perturbation is
\begin{equation}
\cos\sqrt{\pi}\phi_c^+ \langle \sin\sqrt{\pi}\theta_c^- \rangle
\langle \mu_1 \mu_2 \mu_3 \rangle \sigma_0.
\end{equation}
The subsequent analysis is the same as for the SC$^d$ state, but with
the triplet sector replaced by the singlet one.  

In summary, applying an ionic potential to one of the superconducting
states leads to states with unconventional correlations, but does not
cause any quantum phase transitions.

\subsubsection{CDW$^+$ perturbation in the CDW$^-$ phase}

Consider again the case of repulsive interaction, where the ladder is
in the CDW$^-$ phase. Suppose now that we apply the ionic potential of
the CDW$^+$ type. This is perhaps the most interesting case, where the
dominant CDW$^-$ correlations compete with the external potential that 
tries to force the system to form the CDW$^+$ state.

The CDW$^-$ phase of the ladder model is characterized by the non-zero
expectation value (\ref{ev1}). Within the mean-field decoupling
procedure, we find that now the effective perturbation has the form
\begin{eqnarray}
\label{ip2}
&&
\delta{\cal H}_{\rm eff} = \lambda_i 
\left[ \sin\sqrt{\pi}\phi_c^+ \cos\sqrt{\pi}\phi_c^-
\langle\mu_1\mu_2\mu_3 \rangle \mu_0 
\;+\;
\cos\sqrt{\pi}\phi_c^+ \sin\sqrt{\pi}\phi_c^-
\sigma_1\sigma_2\sigma_3\langle \sigma_0\rangle\right].
\nonumber
\end{eqnarray}
As before, at energies less than the gaps, the operator (\ref{ip2})
contributes to the effective action in the second order yielding the
relevant term
\begin{equation}
\tilde{\lambda}_i \cos\sqrt{4\pi}\phi_c^+.
\end{equation}
Thus the total charge field acquires a gap and true long-range order
is formed in the system.

However, in contrast to the case of superconducting ground states
above, the nature of the resulting state depends on the sign of the
effective coupling constant $\tilde{\lambda}_i$ which is given by
\begin{eqnarray}
&&
\tilde{\lambda}_i \sim -\lambda_i^2 \left[ 
\left\langle \mu_1\mu_2\mu_3\right\rangle^2 
\int dx \left\langle \cos\sqrt{\pi}\phi_c^-(x)\cos\sqrt{\pi}\phi_c^-(0)\right\rangle
\left\langle \mu_0(x)\mu_0(0) \right\rangle\right. 
\nonumber\\
&&
\nonumber\\
&&
\quad\quad\quad\quad\quad\quad\quad\quad\quad
-
\left.
\left\langle \sigma_0 \right\rangle^2  
\int dx \left\langle \sin\sqrt{\pi}\phi_c^-(x) \sin\sqrt{\pi}\phi_c^-(0) \right\rangle
\left\langle \sigma_1(x)\sigma_1(0) \right\rangle^3 \right].
\end{eqnarray} 
Note, that the sign of $\tilde{\lambda}_i$ is not fixed, and depends
on the relative magnitudes of $|m_F|,|m_S|,|m_T|$.

Suppose for the sake of argument that $\tilde{\lambda}_i<0$ such that
$\langle\cos\sqrt{\pi}\phi_c^+\rangle\ne 0$. Then the signs of the
mass terms in the total charge and spin-singlet sectors is fixed and
we may integrate them out. As a result we find the effective model for
the relative charge and spin-singlet sectors
\footnote{As the gapped fields are integrated out, we also generate a
  term proportional to
  $\sum\xi^i_R\xi^i_L\cos\sqrt{4\pi}\theta_c^-$. This term is less
  relevant than the terms already written, and therefore can be
  disregarded.}
\begin{equation}
\label{mostinteresting}
{\cal H}_{\rm eff} \sim {\cal H}_0 \left[ \phi_c^- \right] 
+  {\cal H}_0 \left[ \xi_i \right]  +im_T \sum_{i=1}^3 \xi^i_R\xi^i_L
-A \cos\sqrt{4\pi}\theta_c^- 
+ \tilde{\lambda}_i \sin\sqrt{\pi}\phi_c^-\,\sigma_1\sigma_2\sigma_3.
\end{equation}
The sign of the triplet mass term $m_T$ is such that it alone would
generate a state with $\langle\mu_1\mu_2\mu_3\rangle\ne 0$.

Now there are two limiting cases in the problem: $|m_T|\gg |m_F|$
and $|m_T| \ll |m_F|$. Focusing on the former case, we integrate out
the spin-triplet fields. The last term in Eq.~(\ref{mostinteresting})
contributes in second order and we find
\begin{equation}
{\cal H}_{\rm eff} \sim {\cal H}_0 \left[ \phi_c^- \right]  
-A \cos\sqrt{4\pi}\theta_c^-  + B  \cos\sqrt{4\pi}\phi_c^- .
\end{equation}
As $B$ is increased relative to $A$, there occurs a Z$_2$ transition
in the relative charge field.  As soon as this happens, the
perturbation starts to contribute to the spin-triplet sector in first
order. At this point we find a state with two order parameters
present, similar to the situation with the ground states with
superconducting correlations considered above.

In the opposite limit one applies the same procedure, but integrating
out the relative charge field. We are then left with the effective
theory in the spin-triplet sector -- in the second order, we find a
contribution similar to the mass term, but with the opposite sign.
Thus the resulting effective mass in the spin-triplet channel may
change sign and the system undergoes an SU(2)$_2$ transition (similar
to the transition described in Section~\ref{ltpt}).

Finally, in the case when $\tilde{\lambda}_i>0$ one finds the similar
behavior, but now with the spin-singlet sector instead of the triplet.
The only difference with the above is that the spin-singlet sector
undergoes the Ising (Z$_2$) transition.

We therefore see that adding a non-zero ionic potential $\lambda_i$,
the system immediately becomes fully gapped (and insulating); however
the CDW$^-$ correlations of the $\lambda_i=0$ system persist until
some critical $\lambda_i^*$; whereby one sees a single quantum phase
transition of either the Z$_2$ or SU(2)$_2$ universality class,
depending on the relative magnitudes of the gaps in the system.

\subsubsection{CDW$^-$ perturbation in the CDW$^+$ phase}

Let us now briefly outline the calculation in the converse case where
the CDW$^-$ ionic potential is applied to the ladder in the CDW$^+$
phase, where the system possesses the non-zero expectation value
(\ref{ev2}). The effective perturbation is qualitatively similar to
Eq.~(\ref{ip2})
\begin{eqnarray}
\label{ip4}
&&
\delta{\cal H}_{\rm eff} = \lambda_i 
\left[ \sin\sqrt{\pi}\phi_c^+ \cos\sqrt{\pi}\theta_c^-
\langle\sigma_1\sigma_2\sigma_3 \rangle \mu_0 
\;+\;
\cos\sqrt{\pi}\phi_c^+ \sin\sqrt{\pi}\theta_c^-
\mu_1\mu_2\mu_3\langle \sigma_0\rangle\right].
\nonumber
\end{eqnarray}
Thus the subsequent arguments are practically identical to the those
in the previos case. Treating the perturbation in the second order, we
generate the relevant operator $\pm \cos\sqrt{4\pi}\phi_c^+$ and find
true long-range order. The sign depends on the relative magnitude of
the mass gaps in the spin sector. If $|m_S| \ll |m_T|$, then
$\langle\sin\sqrt{\pi}\phi_c^+\rangle \ne 0$. Integrating out the
total charge and spin-singlet degrees of freedom we arrive at the
effective Hamiltonian
\begin{eqnarray}
&&
{\cal H}_{\rm eff} \sim {\cal H}_0 \left[ \phi_c^- \right] 
+  {\cal H}_0 \left[ \xi_i \right]+ im_T \sum_i\xi^i_R\xi^i_L
 -A \cos\sqrt{4\pi}\phi_c^-  
+ \tilde{\lambda}_i \sin\sqrt{\pi}\theta_c^-\,\sigma_1\sigma_2\sigma_3,
\end{eqnarray}
which is almost identical to Eq.~(\ref{mostinteresting}). The
conclusions are also the same: depending on the relation between
$|m_T|$ and $|m_F|$ one finds either a Z$_2$ transition in the
relative charge sector, or the SU(2)$_2$ spin-triplet transition.

\subsubsection{OAF ground state}

Last, but not least, let us consider the effect of the ionic potential
in the case when the ladder model is in the OAF phase. The only
difference between the OAF order parameter and that of CDW$^-$ is the
substitution
\[
\cos\sqrt{\pi}\theta_c^- \quad\leftrightarrow\quad  
\sin\sqrt{\pi}\theta_c^-.
\]
Therefore, if the CDW$^+$ ionic potential is applied, the behavior is
the same as in the case when the CDW$^+$ ionic potential is applied to
the CDW$^-$ ground state.

If on the other hand the CDW$^-$ ionic potential is applied, then a
slightly different situation arises. Indeed, instead of the
expectation value (\ref{ev1}) we now have
\begin{equation}
\left\langle  
\mu_1\mu_2\mu_3\sigma_0
\cos\sqrt{\pi}\theta_c^-
\right\rangle \ne 0,
\label{ev5}
\end{equation}
so that the effective perturbation is different from (\ref{ip1})
\begin{eqnarray}
\label{ip5}
&&
\delta{\cal H}_{\rm eff} \sim \lambda_i \left[
\sin\sqrt{\pi}\phi_c^+ \left\langle\cos\sqrt{\pi}\theta_c^-\right\rangle
\sigma_1\sigma_2\sigma_3 \mu_0 
\; + \;
\cos\sqrt{\pi}\phi_c^+ \sin\sqrt{\pi} \theta_c^-
\left\langle\mu_1\mu_2\mu_3 \sigma_0\right\rangle\right].
\end{eqnarray}
Which term is dominant depends on the relative magnitude of the gaps
in the relative charge and spin sectors. Let us suppose that the spin
gaps are larger, which allows us to focus on the second term.
Integrating out the gapped degrees of freedom we generate a relevant
term in the total charge sector, such that
$\langle\cos\sqrt{\pi}\phi_c^+ \rangle\ne 0$. Thus we arrive at the
effective model for the relative charge
\begin{equation}
{\cal H}_{\mathrm eff} = {\cal H}_0 \left[ \theta_c^- \right]
-g_\theta \cos\sqrt{4\pi}\theta_c^- - \tilde{\lambda}_i
\cos\sqrt{\pi}\theta_c^-.
\end{equation}
This is the classic double sine-Gordon model which undergoes an Ising
(Z$_2$) transition as the strength of the ionic potential is
increased.


\begin{table*}
\begin{tabular}{cccc}
ground state  &&CDW$^+$ ionic potential & CDW$^-$ ionic potential \\
\hline
CDW$^-$ && Z$_2$ (flavor) / Z$_2$ (spin) / SU(2)$_2$ (spin)
& --- \\
CDW$^+$ && --- & 
Z$_2$ (flavor) / Z$_2$ (spin) / SU(2)$_2$ (spin)\\
OAF && Z$_2$ (flavor) / Z$_2$ (spin) / SU(2)$_2$ (spin) & 
Z$_2$ (flavor) \\
SC$^{d(s)}$ && --- & ---
\end{tabular}
\caption{Summary of the phase transitions in the ladder model
  subjected to the external ionic potential. Where more than one
  transition is indicated, the behavior of the system depends on the
  relative magnitudes of the dynamically generated gaps. Empty entries
  indicate the absence of any phase transitions in the cases where the
  system acquires a gap in the total charge sector as soon as the
  ionic potential is applied and then remains in the ordered state
  independent of the strength of the effective coupling constants. The
  relative charge sector is denoted as ``flavor''.}
\label{table-ionic}
\end{table*}

To summarize the results of this section, we see that the addition of
an ionic potential (of either the CDW$^+$ or CDW$^-$ type) to the
incommensurate two-leg ladder will immediately lead to a fully gapped,
insulating, state. However in some situations, one may also find a
critical strength of ionic potential $\lambda_i^*$ at which the system
undergoes a further phase transition at which correlations of the
original order parameter of the system becomes short range.  The
possible transitions are summaraized in Table \ref{table-ionic}.


\section{Conclusions and outlook}
\label{conclusions}

In this work, we have considered interacting spin-$1/2$ fermions on a
two-leg ladder, as described by Hamiltonian \eqref{h}, focussing on
the case where the lattice is at an incommensurate filling. The ground
state of such a model is typically characterized as having a gapless
charge-carrying mode and three gapped collective modes, conventionally
known as \textit{flavour} or relative charge, \textit{spin singlet}
and \textit{spin triplet}. The critical charge mode means there is no
long-range order in the system; however certain operators show power
law correlations, which may be considered as the order parameters for
quasi-long range order.

Our main goal was to study the response of the system to one of two
perturbations: either a local impurity \eqref{imp}, or an ionic
potential \eqref{ip0}. Such analysis requires the complete knowledge
of the phase diagram of the system in the absence of the
impurity. Although the ground states of the ladder have been
extensively studied in previous literature, we have described all
possible phases and phase transitions to make the paper
self-contained. The contents of this work may thus be summarized as
follows:
\begin{enumerate}
\item We reviewed the phase diagram of the clean system for both zero
  (see Fig.~\ref{pd1}) and large (see Fig.~\ref{Phase-Diagram}
  interchain hopping $t_\perp$.  We also reviewed (see
  Sec.~\ref{idpt}) the classification of the transitions between these
  phases. In doing so, we also presented (\ref{bb}) a comprehensive
  description of the relationship between the chain and band bases
  used to study these two limits.
\item We extended the known phase diagram to arbitrary $t_\perp$,
  shown in Fig.~\ref{hdpt}, highlighting transitions that may occur as
  a function of $t_\perp$ in Sec.~\ref{hdpt-sec}. In particular, we
  identified a new sort of $Z_2$ commensurate-incommensurate
  transitions between the CDW$^+$ phase and a superconducting phase.
  This is a transition between two gapped phases; however, on one side
  of the transition, there are no (interchain) incommensurate
  correlations, while on the other side certain operators
  (characterized by exponentially decaying correlation functions) do
  show such incommensurate correlations.
\item We showed in Sec.~\ref{impurity} that the response of the system
  to a local impurity may depend on the detailed structure of the
  applied potential. For repulsive interactions $K_c^+<1$, this
  manifests itself in strong crossover behavior (and not the
  metal-insulator transition seen in the spinless case \cite{us2}). We
  also analyzed transport properties at the critical lines, where the
  conductance is still characterized by the power-law temperature
  dependence, but with a different set of exponents.
\item Finally, we showed in Sec.~\ref{ionic} that the addition of an
  ionic potential to such a system will always lead to a gap opening
  in the total charge sector, and therefore turn the system into an
  insulator. However, in certain circumstances, further phase
  transitions may be seen which we have classified in Table
  \ref{table-ionic}.
\end{enumerate}

While we have made every effort to be as comprehensive as possible in
this work, let us finish by mentioning a few important points that
have not been discussed. We have limited ourselves to weak
impurities. This means not only that the bare impurity is weak, but
also that the renormalized impurity at the energy scale of the bulk
gaps is also weak. As the impurity in the high-temperature phase is in
general a relevant perturbation (see Eq.~\eqref{dimp-highT}), an
alternative hierarchy of energy scales may be realized, even for weak
bare impurities. In this case, the impurity would indeed split the
ladder into two semi-infinite parts, and one must study tunneling
between them \textit{without} first projecting onto the gapless total
charge sector. Certain aspects of this have been discussed in
Ref.~\cite{smh}; however to our knowledge the crossover between this
behavior and that presented by us has never been discussed. One can
also envisage a situation when the component of the impurity coupled
to the dominant order of the system is small but non-zero, which would
lead to a crossover energy or temperature scale separating the
different exponents. While it is more or less clear how such
crossovers would be, we believe a more serious study of these effects
to be necessary in order to have a complete picture of transport in
the ladder systems.

Studying the local impurity, it was crucial whether the impurity
coupled directly to the dominant bulk order parameter in the
system. This could only happen if the system was in one of the density
wave ground states -- in the superconducting states, the structure of
the local impurity was unimportant. However, one could study the dual
problem of tunneling into the ladder. In this case, such ``matches"
would occur in the superconducting states and not the density wave
ones, again giving rise to strong changes in power-law
exponents. Results for this setup will be presented elsewhere
\cite{sac}.

Finally, we mention the possibility of extending the work to ladders
with a larger number of legs. For example, could the $Z_N$ parafermion
criticality discussed recently in the context of $N$-leg
\textit{bosonic} ladders \cite{lan,tak} also be seen in fermionic
ladders? And if so, what would be the transport properties of such a
system and the response to local impurities? We expect, that systems
with a gapless charge mode but gapped spin excitations are conductors,
however as strongly correlated states they may exhibit unusual
responses to disorder. This is therefore an extremely fertile ground
for investigation, which is becoming technologically more and more
important as well, as quantum wires shrink in size and such
correlations may become important to model their behavior.

The authors would like to thank the Abdus Salam International Center
for Theoretical Physics for its hospitality during part of this work.
They also thank P.~Azaria,  T.~Giamarchi, E.~K\"onig, S.~Ngo Dinh, M.~Sch\"utt and E.~Shimshoni, for useful discussions. A.A.N. is supported
by Grant IZ73Z0-128058/1.


\appendix

\section{Extended Hubbard model}
\label{1leg}

The most representative model of spinful fermions on a single chain
with short-range interaction and unbroken SU(2) spin symmetry is the extended
Hubbard model \cite{gnt,gb}. The Hamiltonian of this model includes the
single-particle hopping
\begin{equation} 
H_0 =
-\frac{t_\|}{2} \sum_{i,\sigma} 
c_{i,\sigma}^\dagger c_{i+1,\sigma} + H.c.,
\label{ham-kin}  
\end{equation}
and the interaction term consisting 
of an on-site interaction $U$ and nearest-neighbor interaction $V$:
\begin{equation} 
H_{\text{int}} = U\sum_{i} n_{i,\uparrow} n_{i,\downarrow} + 
V\sum_{i,\sigma,\sigma'} n_{i,\sigma}n_{i+1,\sigma'}.
\label{ham-int}  
\end{equation}
Bosonization of this model proceeds by linearizing the spectrum around
the Fermi-points. In the continuum limit the lattice fermionic
operators undergo chiral decomposition:
\begin{equation}
c_{i,\sigma} \to \sqrt{a}
\left[e^{ik_F x} R_\sigma (x) + e^{-i k_F x} L_\sigma (x) \right]. 
\label{cd}
\end{equation} 
Here $x=ai$, $a$ being the lattice spacing, $k_F$ is the Fermi
wavevector, and $R_\sigma (x)$ and $L_\sigma (x)$ are slowly varying
fields defined in the vicinity of the right and left Fermi points.
The local density operator can then be written as the sum of a smooth
part and a rapidly oscillating staggered part
\begin{equation}
\label{cdd}
n_{i\sigma} = a\left[ R^\dagger_\sigma R_\sigma + L^\dagger_\sigma L_\sigma 
+ e^{-2ik_F x} R^\dagger_\sigma L_\sigma 
+ e^{2ik_F x} L^\dagger_\sigma R_\sigma \right].
\end{equation}
For the case of incommensurate filling (i.e. $2k_F\ne \pi/a$) one
neglects all rapidly oscillating terms that would appear after
substituting the density (\ref{cdd}) into the interaction Hamiltonian
(\ref{ham-int}). It is then straightforward to express the Hamiltonian
in terms of the chiral fields $R_\sigma$ and $L_\sigma$. Yet, as a
small subtlety, one should be careful not to break Galilean invariance
when dealing with the equal-spin nearest-neighbor term
$n_{i,\sigma}n_{i+1,\sigma}$. This is a well known and unimportant
problem of the regularization appearing in models with the infinite
linear spectrum. The standard approach for the extended Hubbard model
is to replace the product $n_{i,\sigma}n_{i+1,\sigma}$ by the term
\begin{equation}
n_{i,\sigma}n_{i+1,\sigma} \sim a^2\left(1 - \cos(2k_F a) \right) 
 \left(  R_\sigma^\dagger R_\sigma + L_\sigma^\dagger L_\sigma\right)
\left( R_{\sigma}^\dagger R_{\sigma} + L^\dagger_{\sigma} L_{\sigma} \right), 
\nonumber
\end{equation}
which is Galilean-invariant and reproduces correct answers for small
$V$. For larger values of $V$ the parameters appearing in the
bosonized theory must be considered phenomenological anyway, with no
simple relationship to the bare parameters of the lattice model. No
such problem exists when $\sigma\ne\sigma'$.

We now apply the standard bosonization procedure (see
e.g. Refs. \cite{gnt,gb}), with the following conventions
\begin{eqnarray}
&&
R_{\sigma} = \frac{\kappa_{\sigma}}{\sqrt{2\pi\alpha}} \;
e^{i\sqrt{4\pi} \phi^R_{\sigma}},  \quad
L_{\sigma} = \frac{\kappa_{\sigma}}{\sqrt{2\pi\alpha}} \;
e^{-i\sqrt{4\pi} \phi^L_{\sigma}}, 
\nonumber\\
&&
\nonumber\\
&&
\quad \;
\left[ \phi^R_{\sigma}, \phi^L_{\sigma'} \right] = \frac{i}{4} \delta_{\sigma\sigma'}, 
\quad
\left\{ \kappa_{\sigma}, \kappa_{\sigma'} \right\} = 2 \delta_{\sigma\sigma'}, 
\nonumber\\
&&
\nonumber\\
&&
\quad\quad \; \; \;
\phi_{\sigma} = \phi^R_{\sigma} + \phi^L_{\sigma}, \quad
\theta_{\sigma} =  \phi^L_{\sigma} -\phi^R_{\sigma},
\label{bfh}
\end{eqnarray}
where $\alpha$ is the bosonic ultraviolet cut-off. The Klein factors
$\kappa_\sigma$ are introduced to ensure anti-commutation between
operators with different spins. They are not dynamic variables, so we
are free to choose a particular representation:
\begin{equation}
\kappa_\uparrow\kappa_\downarrow = - \kappa_\downarrow\kappa_\uparrow = i, 
\quad \kappa_\sigma^2=1.
\label{kfsc}
\end{equation}
Note that the correct anti-commutation between the two different
chiral branches are fixed by the non-local {\em commutation} relations
between $\phi_R$ and $\phi_L$.

Using the above definitions
we find that
\begin{eqnarray}
&&
R^\dagger_{\sigma} L_{\sigma} = -\frac{i}{2\pi \alpha} \; e^{-i\sqrt{4\pi} \phi_{\sigma} }, 
\nonumber\\
&&
\nonumber\\
&&
R^\dagger_{\sigma} R_{\sigma} + L^\dagger_{\sigma} L_{\sigma}=
\sqrt{\frac{1}{\pi}} \; \partial_x \phi_{\sigma}. \label{bilinears}
\end{eqnarray}

In terms of the bosonic fields (\ref{bfh}) the Hamiltonian of the extended
Hubbard model can be written as a sum of three parts
\begin{subequations}
\label{bh1}
\begin{equation}
{\cal H} = {\cal H}_0 + {\cal H}_f + {\cal H}_b.
\end{equation}
The kinetic part of the Hamiltonian is 
\begin{equation}
{\cal H}_0 = \frac{v_F}{2} \sum_\sigma 
\left( \Pi_\sigma(x)^2 + \left[ \partial_x \phi_\sigma(x) \right]^2 \right),
\end{equation}
where $\Pi_\sigma(x)=\partial_x \theta_\sigma(x)$ is the momentum
conjugate to $\phi_\sigma(x)$ 
\[
[ \Pi_\sigma(x) ,
  \phi_{\sigma'}(y) ] = - i\delta(x-y)\delta_{\sigma\sigma'}. 
\]
The interaction consists of the forward-scattering terms
\begin{equation}
{\cal H}_f = \sum_\sigma \left( 
\frac{f_{2}}{\pi}  \left[ \partial_x \phi_\sigma(x) \right]^2
+\frac{f'_{2}}{2\pi} \partial_x\phi_\sigma(x)\partial_x\phi_{-\sigma}(x) 
\right),
\end{equation}
and the back-scattering 
\begin{equation}
{\cal H}_b = \frac{f_{1}}{2(\pi\alpha)^2} 
\cos\left[ \sqrt{4\pi} \left( \phi_\uparrow(x)-\phi_\downarrow(x) \right)\right].
\label{bs}
\end{equation}
\end{subequations}
The parameters appearing in Eqs.~(\ref{bh1}) can (for weak coupling)
be related to the ``lattice'' interaction constants $U$ and $V$ as
follows
\begin{eqnarray}
&&
f_{2} = aV \left(1-\cos(2k_F a) \right), 
\nonumber\\
&&
\nonumber\\
&&
f'_{2}  = aU + 2aV, 
\nonumber\\
&&
\nonumber\\
&&
f_{1} = aU + 2aV\cos (2k_F a).
\end{eqnarray}
Notice that as a consequence of the SU(2) symmetry these parameters
are not independent but satisfy the relation
\[
f_1 = f_2'-2f_2.
\]
This notation is slightly different from the common $g$-ology
\cite{gnt} insofar that here we keep the Galilean invariance explicit
by not splitting the $g_2$ and $g_4$ terms.

A simple rotation of the bosonic basis
\begin{equation}
\label{csf}
\phi_c = \frac{\phi_\uparrow+\phi_\downarrow}{\sqrt{2}}, \quad 
\phi_s = \frac{\phi_\uparrow-\phi_\downarrow}{\sqrt{2}},
\end{equation}
leads to the well-known charge-spin separated \cite{gnt} structure of 
the Hamiltonian (\ref{bh1}):
\begin{subequations}
\label{bh2}
\begin{equation}
{\cal H} = {\cal H}_c + {\cal H}_s, \quad [{\cal H}_c, {\cal H}_s] = 0
\label{ehboz}
\end{equation}
The charge sector is described by a Gaussian model
\begin{eqnarray}
{\cal H}_c &=&\frac{v_F}{2} \left[ \Pi^2 _c(x) + \left(1-\frac{g_c}{\pi v_F}\right) 
\left[ \partial_x \phi_c(x) \right]^2 \right] \nonumber\\
&=& \frac{v_c}{2} \left[ K_c \Pi^2 _c (x)
+ K^{-1}_c \left( \partial_x \phi_c (x) \right)^2 \right]
\label{charge_TL}
\end{eqnarray}
where $v_c$ is the group velocity of the gapless collective charge excitations,
whereas
$K_c$ is the Luttinger-liquid coupling constant. The two parameters
satisfy the condition following from Galilean invariance: 
\begin{equation}
v_c K_c = v_F,  \label{v_K_charge}
\end{equation}
the parameters being given by
\begin{equation}
K_c \approx 1 + \frac{g_c}{2\pi v_F}
+ O(g^2 _c),\label{charge_param} 
\end{equation}
where in terms of the original lattice parameters
\begin{equation}
g_c = - f'_{2}-2f_{2}  = -aU-2aV \left( 2 -  \cos(2k_F a) \right).
\end{equation}
For repulsive interactions $U,V>0$, we find that
$K_c<1$.
\medskip

The spin sector is represented by a sine-Gordon model
\begin{eqnarray}
&&
{\cal H}_s = \frac{v_F}{2} \left[ \Pi_s(x)^2 + \left(1-\frac{g_s}{\pi v_F}\right) 
\left[ \partial_x \phi_s(x) \right]^2 \right] 
\nonumber\\
&&
\nonumber\\
&&
\quad\quad\quad\quad
+ \frac{g_s}{2(\pi \alpha)^2} \cos \sqrt{8\pi} \phi_s(x).
\label{sg}
\end{eqnarray}
where 
\begin{equation}
g_{s} = f'_{2}-2f_{2}  = f_{1} = aU + 2aV\cos(2k_F a).
\label{gcgs}
\end{equation}
\end{subequations}
Due to the SU(2) symmetry, manifesting in the fact that there is
only one parameter in Eq.~(\ref{sg}), the RG flow of this model lies
on one of the two separatrices of the Kosterlitz-Thouless phase
diagram, and is given by
\begin{equation}
\frac{\partial g_s}{\partial l} = - g_s^2.
\end{equation}
Consequently it can be seen that if the bare $g_s>0$, the flow is to a
weak coupling fixed point where the spin-sector is described by a
Luttinger liquid, while if $g_s<0$ then the flow is to a strong
coupling fixed point characterized by a spin gap.

\section{Refermionization conventions}
\label{referm}

The principle of refermionization comes from the equivalence of the
sine-Gordon model and the massive Thirring model \cite{coleman}.  This
equivalence is based on the correspondence between bosonic vertex
operators with scaling dimension 1 and fermionic bilinears:
\[
e^{i \sqrt{4\pi} \Phi} ~\leftrightarrow~ - 2\pi i \alpha L^{\dagger}R, ~~~
e^{i \sqrt{4\pi} \Theta} ~\leftrightarrow~ 2\pi i \alpha L^{\dagger}R^{\dagger}
\]
If the sine-Gordon Hamiltonian is defined as
\begin{equation}
\label{hsg}
H_{SG} = H_0 + \frac{m}{\pi \alpha} \cos \sqrt{4\pi \Phi},
\end{equation}
then using the above correspondence between the bosonic and fermionic
operators (such that $\cos\sqrt{4\pi\Phi}\to i
\pi\alpha\left(R^{\dagger}L - h.c.\right)$ one finds the refermionized
perturbation to have the form of the fermion mass term:
\[
+ i m (R^{\dagger} L - h.c.).
\]

This becomes particularly useful if one replaces a Dirac fermion by a
pair of two Majorana fermions (for details, see e.g. \cite{gnt})
\[
R^\dagger = \frac{\xi^1_R+i\xi^2_R}{\sqrt{2}}, \quad\quad 
L^\dagger = \frac{\xi^1_L+i\xi^2_L}{\sqrt{2}}, \quad\quad
(\xi^a _{R,L})^{\dagger} = \xi^a _{R,L} .
\]
In terms of the Majorana fields the cosine perturbation becomes
\[
+ i m \left( \xi^1 _R \xi^1 _L + \xi^2 _R \xi^2 _L \right)
\]
So, the original $\beta^2 = 4\pi$ sine-Gordon model now consists of
two massive Majorana modes
\footnote{The refermionized Hamiltonian in general also has marginal
  four Majorana mode interactions, which appear as a result of
  refermionizing the forward-scattering $g_2$ terms.},
each of which is formally equivalent to the \textit{off-critical}
transverse-field (or quantum) Ising model after a Jordan-Wigner
transformation \cite{iz}:
\[
H_{SG} \rightarrow \sum_{j=1,2} 
H_M^{(j)} = \sum_{j=1,2} \left[
\frac{iv}{2}  \left( \xi^j_L \partial_x \xi^j_L 
-  \xi^j_R \partial_x \xi^j_R \right)
+ i m  \xi^j_R \xi^j_L \right].
\]
This allows potential order parameters, such as $\cos \sqrt{\pi} \phi$
or $\sin \sqrt{\pi} \phi$, to be rewritten in terms of the order
($\sigma$) and disorder ($\mu$) parameters of the corresponding
quantum Ising chains.

With the sign convention adopted in Eq.~(\ref{hsg}), the case $m>0$
describes the ordered phase of the quantum Ising model with $\langle
\sigma \rangle \ne 0$, $\langle \mu\rangle =0$, while the case $m<0$
corresponds to the disordered phase with $\langle \sigma \rangle = 0$,
$\langle \mu\rangle \ne 0$. The case $m=0$ corresponds to a
\textit{critical} Ising model, and hence a Z$_2$ quantum criticality.

\subsection{Spin modes}

For the two-leg ladder, it is convenient to refermionize the two
bosonic modes corresponding to the spin degrees of freedom of the
original model \cite{snt}.  There are therefore four Majorana fermions
$\xi_{R(L)}^i$, which are related to the bosonic fields according to
the following convention:
\begin{align}
&
\cos \sqrt{4\pi} \phi_s^+ \rightarrow i\pi\alpha \left(\xi^1_R \xi^1_L+\xi^2_R \xi^2_L\right),
\quad
\partial_x \phi_s^+ \rightarrow i \sqrt{\pi}
\left(\xi^1_R \xi^2_R+\xi^1_L \xi^2_L\right),
\nonumber\\
&
\nonumber\\
&
\cos \sqrt{4\pi} \theta_s^+ \rightarrow i\pi\alpha \left(\xi^1_R \xi^1_L-\xi^2_R \xi^2_L\right),
\nonumber\\
&
\nonumber\\
&
\cos \sqrt{4\pi} \phi_s^- \rightarrow i\pi\alpha \left(\xi^3_R \xi^3_L+\xi^0_R \xi^0_L\right),
\quad
\partial_x \phi_s^- \rightarrow i \sqrt{\pi}
\left(\xi^3_R \xi^0_R+\xi^3_L \xi^0_L\right),
\nonumber\\
&
\nonumber\\
&
\cos \sqrt{4\pi} \theta_s^- \rightarrow i\pi\alpha \left(\xi^3_R \xi^3_L-\xi^0_R \xi^0_L\right),
\label{rf}
\end{align}
Re-fermionizing the order parameters makes use the Ising order and disorder
operators \cite{gnt} as follows:
\begin{align}
&
\cos\sqrt{\pi} \phi_s^+ \rightarrow \mu_1 \mu_2, \quad
\sin\sqrt{\pi} \phi_s^+ \rightarrow \sigma_1 \sigma_2,
\nonumber\\
&
\nonumber\\
&
\cos\sqrt{\pi} \theta_s^+ \rightarrow \mu_1 \sigma_2, \quad
\sin\sqrt{\pi} \theta_s^+ \rightarrow \sigma_1 \mu_2,
\nonumber\\
&
\nonumber\\
&
\cos\sqrt{\pi} \phi_s^- \rightarrow \mu_3 \mu_0, \quad
\sin\sqrt{\pi} \phi_s^- \rightarrow \sigma_3 \sigma_0,
\nonumber\\
&
\nonumber\\
&
\cos\sqrt{\pi} \theta_s^- \rightarrow \mu_3 \sigma_0, \quad
\sin\sqrt{\pi} \theta_s^- \rightarrow \sigma_3 \mu_0.
\label{rfrules}
\end{align}

\subsection{Relative charge mode}

In the study of the hopping-driven phase transitions, we also bosonize
the relative charge sector.  Here, the following convention is used
\begin{align}
&
\cos \sqrt{4\pi} \phi_c^- \rightarrow i\pi\alpha 
\left(\xi^\alpha_R \xi^\alpha_L+\xi^\beta_R \xi^\beta_L\right),
\quad 
\partial_x \phi_c^- \rightarrow i \sqrt{\pi}
\left(\xi^\alpha_R \xi^\beta_R+\xi^\alpha_L \xi^\beta_L\right),
\nonumber\\
&
\nonumber\\
&
\cos \sqrt{4\pi} \theta_c^- \rightarrow i\pi\alpha 
\left(\xi^\alpha_R \xi^\alpha_L-\xi^\beta_R \xi^\beta_L\right).
\label{rfc}
\end{align}
and therefore the order parameters are
\begin{align}
&
\cos\sqrt{\pi} \phi_c^- \rightarrow \mu_\alpha \mu_\beta, \quad
\sin\sqrt{\pi} \phi_c^- \rightarrow \sigma_\alpha \sigma_\beta,
\nonumber\\
&
\nonumber\\
&
\cos\sqrt{\pi} \theta_c^- \rightarrow \mu_\alpha \sigma_\beta, \quad
\sin\sqrt{\pi} \theta_c^- \rightarrow \sigma_\alpha \mu_\beta.
\label{rfrulesc}
\end{align}

\section{Klein factors in $4$-channel case}
\label{kf}

Bosonization of the $4$-channel model proceeds similar to the
$2$-channel case (\ref{bfh}):
\begin{eqnarray}
&&
R_{\mu\sigma} = \frac{\kappa_{\mu\sigma}}{\sqrt{2\pi\alpha}} \;
e^{i\sqrt{4\pi} \phi^R_{\mu\sigma}},  \quad
L_{\mu\sigma} = \frac{\kappa_{\mu\sigma}}{\sqrt{2\pi\alpha}} \;
e^{-i\sqrt{4\pi} \phi^L_{\mu\sigma}}, 
\nonumber\\
&&
\nonumber\\
&&
\quad \;
\left[ \phi^R_{\mu\sigma}, \phi^L_{\mu'\sigma'} \right] = \frac{i}{4} \delta_{\mu\mu'}\delta_{\sigma\sigma'}, 
\quad
\left\{ \kappa_{\mu\sigma}, \kappa_{\mu'\sigma'} \right\} = 2 \delta_{\mu\mu'}\delta_{\sigma\sigma'}, 
\nonumber\\
&&
\nonumber\\
&&
\quad\quad \; \; \;
\phi_{\mu\sigma} = \phi^R_{\mu\sigma} + \phi^L_{\mu\sigma}, \quad
\theta_{\mu\sigma} =  \phi^L_{\mu\sigma} -\phi^R_{\mu\sigma}.
\label{bfh4}
\end{eqnarray}
However, the situation with the Klein factors $\kappa_{\mu\sigma}$,
which ensure anti-commutation between fermion operators corresponding
to different channels
\footnote{Anti-commutation between left and right movers in the same
  channel are ensured by our convention of bosonic commutation
  relations.},
is now more involved.

The anti-commutation relations of the Klein factors (\ref{bfh4}) imply
\[
\kappa_{\mu\sigma}^2=1.
\]
Aside from the identity, the only other term that may appear in our
Hamiltonian (in the Klein factor subspace) is
\begin{equation}
\Gamma = \kappa_{1\uparrow}\kappa_{1\downarrow} \kappa_{2\uparrow}\kappa_{2\downarrow}
\end{equation}
It is easy to see that $\Gamma$ is a Hermitian operator, and satisfies
$\Gamma^2=1$. As $\Gamma$ is not a dynamic variable in this model, we
can choose the sector $\Gamma=1$.

For local operators, we will also need the bi-linear combinations
\begin{equation}
h_\sigma = \kappa_{2\sigma}\kappa_{1\sigma}, \quad
h'_\mu = \kappa_{\mu\uparrow} \kappa_{\mu\downarrow}
\end{equation}
These combinations are anti-Hermitian and satisfy
\[
h_\sigma^2={h'_{\mu}}^2=-1. 
\]
All of these bi-linears commute with $\Gamma$, but they do not commute
with each other and thus cannot be diagonalized simultaneously.
Strictly speaking, this means that one ought to keep the Klein
factors in all operators (or order parameters) which are combinations
of products of two fermionic operators.
However, in this paper we always discuss various order parameters
``one at a time'', i.e. we never fuse different order parameters. Even
in the case of the ionic potential where we may need to consider the
second order in the ionc perturbation, we would fuse two {\it
  identical} order parameters and thus would need only the squared
bi-linears $h$. 

In other words, Klein factors only matter for the resulting structure
of the effective Hamiltonian. Once the vacua for a given
strong-coupling phase are identified, the fate of the ordering does
not depend on what specific Klein prefactor stands in the definition
of that order parameter. Only its bosonic structure is important.

As a result, in all expressions for the local operators (see \ref{op})
we will replace the Klein-factor bi-linears by c-numbers as follows
(here the actual values are chosen to represent a consistent
representation of the Clifford algebra)
\begin{equation}
\Gamma=1, \quad h_\sigma=i, \quad h'_{\mu}=i(-1)^\mu,
\end{equation}
or explicitly
\begin{eqnarray}
&& 
\kappa_{1\uparrow}\kappa_{1\downarrow} \kappa_{2\uparrow}\kappa_{2\downarrow}=1 ,
\nonumber\\
&&
\nonumber\\
&&
\kappa_{2\uparrow}\kappa_{1\uparrow} = i,  \quad \kappa_{1\uparrow}\kappa_{2\uparrow}=-i ,
\nonumber\\
&&
\nonumber\\
&&
\kappa_{2\downarrow}\kappa_{1\downarrow}= i, \quad \kappa_{1\downarrow}\kappa_{2\downarrow}=-i ,
\nonumber\\
&&
\nonumber\\
&&
\kappa_{1\uparrow}\kappa_{1\downarrow}=- i, \quad \kappa_{1\downarrow}\kappa_{1\uparrow}=i, 
\nonumber\\
&&
\nonumber\\
&&
\kappa_{2\uparrow}\kappa_{2\downarrow}= i, \quad \kappa_{2\downarrow}\kappa_{2\uparrow}=-i.
\end{eqnarray}

\section{Bosonized and refermionized form of local operators}
\label{op}

The general form of the of a local density operator for the spinful
ladder is given by
\begin{equation}
\tilde{\cal O}_a^\alpha = \sum_{m, m';\sigma,\sigma'} 
c^\dagger_{m,\sigma} \tau^a_{\mu\mu'} \sigma^\alpha_{\sigma\sigma'} c_{m',\sigma'}, 
\label{ops}
\end{equation}
where $\tau^a$ and $\sigma^\alpha$ are Pauli matrices in the space of
chains and spin respectively (including the identity operators,
$\tau^0 = I$ and $\sigma^0 = I$).

Using the chiral decomposition (\ref{cd}), we can split any operator
(\ref{ops}) into smooth and oscillating (staggered) components:
\[
\tilde{\cal O}_a^\alpha = \left(\tilde{\cal O}_a^\alpha\right)_{\rm smooth}
+ \left(\tilde{\cal O}_a^\alpha\right)_{\rm stag},
\]
where the smooth part of the operator $\tilde{\cal O}_a^\alpha$ is
given by
\[
\left(\tilde{\cal O}_a^\alpha\right)_{\rm smooth} =
\sum_{m, m';\sigma,\sigma'} 
\left[
R^\dagger_{m,\sigma} \tau^a_{\mu\mu'} \sigma^\alpha_{\sigma\sigma'} R_{m',\sigma'}
+
L^\dagger_{m,\sigma} \tau^a_{\mu\mu'} \sigma^\alpha_{\sigma\sigma'} L_{m',\sigma'}
\right],
\]
and the staggered part is
\begin{equation}
\left(\tilde{\cal O}_a^\alpha\right)_{\rm stag} =
e^{-2ik_Fx}{\cal O}_a^\alpha + e^{2ik_Fx}\left({\cal O}_a^\alpha\right)^\dagger,
\quad
{\cal O}_a^\alpha = \sum_{m, m';\sigma,\sigma'} 
R^\dagger_{m,\sigma} \tau^a_{\mu\mu'} \sigma^\alpha_{\sigma\sigma'} L_{m',\sigma'}.
\label{opo}
\end{equation}

The oscillating parts -- the density waves -- are the most
interesting, as these act as local order parameters to distinguish the
different phases of the model~
\footnote{While $4k_F$ density waves are also possible, they only
  become dominant at very strong interaction $K_c^+<1/2$, which we do
  not discuss here.}.
Therefore in this paper we concentrate on the ${\cal O}_a^\alpha$
component of the density operators.

For our purposes, we need only the operators with $\alpha=0$, as the
operators coupling to spin-density can never act as local order
parameters (for their bosonized form, see e.g. \cite{wlf}). With our
convention for Klein factors, their bosonized and refermionized form
is
\begin{align}
{\cal O}_0 & \sim e^{-i\sqrt{\pi}\phi_c^+}
\left[\sin \sqrt{\pi} \phi_c^- 
\sin \sqrt{\pi} \phi_s^+ \sin \sqrt{\pi} \phi_s^-
-i
\cos \sqrt{\pi} \phi_c^- 
\cos \sqrt{\pi} \phi_s^+ \cos \sqrt{\pi} \phi_s^-\right]
\nonumber\\
&
\quad\quad\quad
\sim
e^{-i\sqrt{\pi}\phi_c^+}
\left[
\sin\sqrt{\pi}\phi_c^- \,
\sigma_1\sigma_2\sigma_3\sigma_0
-i
\cos\sqrt{\pi}\phi_c^- \,
\mu_1\mu_2\mu_3 \mu_0
\right],
\nonumber\\
&
\nonumber\\
{\cal O}_x & \sim -e^{-i\sqrt{\pi}\phi_c^+}
\left[
\sin \sqrt{\pi} \theta_c^- 
\cos \sqrt{\pi} \phi_s^+ \cos \sqrt{\pi} \theta_s^- 
-i 
\cos \sqrt{\pi} \theta_c^- 
\sin \sqrt{\pi} \phi_s^+ \sin \sqrt{\pi} \theta_s^- 
\right]
\nonumber\\
&
\quad\quad\quad
\sim -e^{-i\sqrt{\pi}\phi_c^+}
\left[
\sin\sqrt{\pi}\theta_c^- \,
\mu_1\mu_2\mu_3 \sigma_0
-i 
\cos\sqrt{\pi}\theta_c^- \,
\sigma_1\sigma_2\sigma_3 \mu_0
\right],
\nonumber\\
&
\nonumber\\
{\cal O}_y & \sim -e^{-i\sqrt{\pi}\phi_c^+}
\left[
\cos \sqrt{\pi} \theta_c^- 
\cos \sqrt{\pi} \phi_s^+ \cos \sqrt{\pi} \theta_s^- 
+i
\sin \sqrt{\pi} \theta_c^-
\sin \sqrt{\pi} \phi_s^+ \sin \sqrt{\pi} \theta_s^- 
\right]
\nonumber\\
&
\quad\quad\quad
-e^{-i\sqrt{\pi}\phi_c^+}
\left[
\cos\sqrt{\pi}\theta_c^- \,
\mu_1\mu_2\mu_3 \sigma_0
+i
\sin\sqrt{\pi}\theta_c^- \,
\sigma_1\sigma_2\sigma_3 \mu_0
\right],
\nonumber\\
&
\nonumber\\
{\cal O}_z & \sim -e^{-i\sqrt{\pi}\phi_c^+}
\left[
\sin \sqrt{\pi} \phi_c^- 
\cos \sqrt{\pi} \phi_s^+ \cos \sqrt{\pi} \phi_s^-
-i
\cos \sqrt{\pi} \phi_c^- 
\sin \sqrt{\pi} \phi_s^+ \sin \sqrt{\pi} \phi_s^-
\right]
\nonumber\\
&
\quad\quad\quad
\sim -e^{-i\sqrt{\pi}\phi_c^+}
\left[
\sin\sqrt{\pi}\phi_c^- \,
\mu_1\mu_2\mu_3 \mu_0
-i
\cos\sqrt{\pi}\phi_c^- \,
\sigma_1\sigma_2\sigma_3 \sigma_0
\right].
\label{opsboz}
\end{align}

\subsection{Chain basis}

In the chain basis, the association of the operators \eqref{opsboz}
above with their physical meaning is trivial, as \eqref{ops} is
naturally written in the chain basis. The operators with $a=0$ and
$a=z$ correspond to charge density waves (CDWs) with a relative phase
of $0$ (CDW$^+$) and $\pi$ (CDW$^-$), respectively (as depicted in
Fig.~\ref{eh1}). The operator with $a=x$ is a bond density wave
(BDW), while the operator with $a=y$ corresponds to the orbital
anti-ferromagnet (OAF) \cite{nlk}; which is also known in the
literature as the staggered-flux phase or $d$-density wave.

The BDW order parameter turns out to be not terribly important in the
physics of the two leg ladder, unless a magnetic field affecting the
orbital motion of the electrons is added \cite{us1}; a situation that
we do not consider in the present work.

\subsection{Band basis}
\label{opsbandapp}

As \eqref{opsboz} is a complete set of local density operators, the
rotation to the band basis does not modify the set, but merely changes
the physical association of each operator. Ultimately, the rotation
to the band basis is a rotation of $\pi/2$ around the $y$ axis (in
chain/band space); hence the only difference with the chain-basis is
the interchanging of the roles of ${\cal O}_x$ and ${\cal O}_y$.  For
reference the association is now:
\begin{center}
$a=0$ : CDW$^+$, $\quad a=x$ : CDW$^-$, $\quad a=y$ : OAF, $\quad a=z$ : BDW.
\end{center}

It turns out that in addition to the local density operators, there
are two further superconducting operators that are important for the
two-leg ladder \cite{fpt,fal,fab}.  These are known as $s$-wave and
$d$-wave superconductivity (SC$^s$ and SC$^d$) as they are truncations
of the order parameters of these symmetries on the two-dimensional
lattice, and are defined as (as usual $1,2$ refers to the chains,
while $\alpha,\beta$ refers to the bands)
\begin{align}
{\cal O}_{SCs} &=
\left( c_{1\uparrow} c_{1\downarrow} -  c_{1\downarrow} c_{1\uparrow} \right)
+
\left( c_{2\uparrow} c_{2\downarrow} -  c_{2\downarrow} c_{2\uparrow} \right)
\nonumber \\
&
= 
\left( c_{\alpha\uparrow} c_{\alpha\downarrow} -  c_{\alpha\downarrow} c_{\alpha\uparrow} \right)
+
\left( c_{\beta\uparrow} c_{\beta\downarrow} -  c_{\beta\downarrow} c_{\beta\uparrow} \right),
\nonumber \\
&
\nonumber \\
{\cal O}_{SCd} &=
2 \left( c_{1\uparrow} c_{2\downarrow} -  c_{1\downarrow} c_{2\uparrow} \right)
\nonumber \\
&
= 
\left( c_{\alpha\uparrow} c_{\alpha\downarrow} -  c_{\alpha\downarrow} c_{\alpha\uparrow} \right)
-
\left( c_{\beta\uparrow} c_{\beta\downarrow} -  c_{\beta\downarrow} c_{\beta\uparrow} \right).
\label{opsc}
\end{align}
We give the bosonized and refermionized forms of these operators only
in the band representation:
\begin{align}
{\cal O}_{SCs} &=
\exp\left(i\sqrt{\pi}\theta_c^+\right) 
\Big[
\sin\sqrt{\pi}\theta_c^-
\cos\sqrt{\pi}\phi_s^+\cos\sqrt{\pi}\phi_s^-
-i
\cos\sqrt{\pi}\theta_c^-
\sin \sqrt{\pi}\phi_s^+\sin \sqrt{\pi}\phi_s^- \Big]
\nonumber\\
&\quad =
\exp\left(i\sqrt{\pi}\theta_c^+\right) 
\Big[
\sin\sqrt{\pi}\theta_c^- \,
\mu_1\mu_2\mu_3 \mu_0
-i
\cos\sqrt{\pi}\theta_c^- \,
\sigma_1\sigma_2\sigma_3\sigma_0 \Big].
\nonumber\\
&
\nonumber\\
{\cal O}_{SCd} &=
\exp\left(i\sqrt{\pi}\theta_c^+\right) 
\Big[
\cos\sqrt{\pi}\theta_c^-
\cos\sqrt{\pi}\phi_s^+\cos\sqrt{\pi}\phi_s^-
-i
\sin\sqrt{\pi}\theta_c^-
\sin \sqrt{\pi}\phi_s^+\sin \sqrt{\pi}\phi_s^- \Big]
\nonumber\\
&\quad =
\exp\left(i\sqrt{\pi}\theta_c^+\right) 
\Big[
\cos\sqrt{\pi}\theta_c^- \,
\mu_1\mu_2\mu_3 \mu_0
-i
\sin\sqrt{\pi}\theta_c^- \,
\sigma_1\sigma_2\sigma_3\sigma_0 \Big].
\label{opsboz2}
\end{align}

\section{Analysis of RG equations for capacitively coupled chains}
\label{sec:RGappendix}

The phase diagram (see Fig.~\ref{pd1}) of the system consisting of two
capacitively coupled chains was obtained by solving the RG flow
equations \eqref{RG1}. These RG equations allow for analytical
analysis, which can be simplified by re-writing the equations
\eqref{RG1} by means of simple linear transformations of the coupling
constants.

Firstly, we can formulate the RG equations in terms of the coupling
constants $g_j$ appearing in the Hamiltonian (\ref{ccc}), rather than
in terms of the parameters $h_j$ as in Eqs.~\eqref{RG1}:
\begin{align}
\frac{\partial g_1}{\partial l} &= - g_1^2-g_2^2 ,
\nonumber\\
&
\nonumber\\
\frac{\partial g_2}{\partial l} &= - \frac{g_2}{2}  \left( g_{c}^- + 3 g_1  \right) ,
\nonumber\\
&
\nonumber\\
\frac{\partial g_{c}^-}{\partial l} &= -2g_2^2, 
\label{RG1-g}
\end{align}
with the initial values of $g_i$ are still given by Eqs.~(\ref{bare1}).

The RG flow (\ref{RG1-g}) possesses a higher-symmetry line. Indeed,
the model (\ref{ccc}) becomes O(6)-symmetric if
\[
g_1=\pm g_2=g_c^-=g, \quad\quad\quad ( h_s = \pm h_\perp = h_c + h_\perp = g)
\]
with the corresponding RG equation being
\[
\frac{\partial g}{\partial l} = -2g^2.
\]
Clearly, this model flows to strong coupling if the bare $g<0$, and to
weak coupling otherwise.

Secondly, consider a linear combination of the three fields (\ref{hs})
\[
h_w = h_c + 3 h_s + h_\perp.
\]
Using this new field $h_w$ in place of $h_c$, the RG equations (\ref{RG1}) may be 
re-written as
\begin{eqnarray}
&&
\frac{\partial h_s}{\partial l} = - h_s^2-h_\perp^2 ,
\nonumber\\
&&
\nonumber\\
&&
\frac{\partial h_\perp}{\partial l} = - \frac{h_\perp h_w}{2}  ,
\nonumber\\
&&
\nonumber\\
&&
\frac{\partial h_w}{\partial l} = - 3 h_s^2 - 5 h_\perp^2 . 
\label{RGw}
\end{eqnarray}
These equations are invariant under the sign change $h_\perp \rightarrow
-h_\perp$.  Thus we will only analyze the case where the bare $h_\perp>0$, the
opposite case follows from this symmetry.

There is a line of weak-coupling fixed points
\[
h_s=0, \quad h_\perp=0,
\]
and a strong coupling fixed point 
\[
h_s\to -\infty, \quad h_\perp\to\infty, \quad h_w\to -\infty,
\]

To determine the boundary between the basins of attraction of the weak
and strong coupling fixed points, let us combine the flow equations
(\ref{RGw}) as follows
\begin{equation}
\label{fb}
\frac{\partial}{\partial l}
\left(h_s h_w - 3 h_s^2 - h_\perp^2\right) = 
- h_s \left(h_s h_w - 3 h_s^2 - h_\perp^2\right).
\end{equation}
It is now clear that the RG possesses an invariant plane
\begin{equation}
\label{ipa}
h_s h_w = 3 h_s^2 + h_\perp^2 \quad \Rightarrow \quad h_s(h_\perp+h_c)=h_\perp^2.
\end{equation}
Moreover, the combination $h_s(h_\perp + h_c) - h_\perp^2$ can never change
sign. In particular, if the initial values of the interaction
parameters are such that $h_s(h_\perp + h_c) > h_\perp^2$ with $h_s>0$, then
both parameters $h_s$ and $h_w$ can never become negative. The RG then
flows to weak coupling.

On the other hand, if the initial values of $h_s$ and $h_\perp + h_c$
are negative, then they will always remain negative under the RG and
flow to strong coupling. Same result will be achieved for any initial
values such that $h_s(h_\perp + h_c) < h_\perp^2$. In that case
nothing prevents either of $h_s$ and $h_\perp + h_c$ to become
negative followed by the flow to $-\infty$. Thus we conclude that the
surface (\ref{ipa}) with positive $h_s$ and $h_w$ defines the phase
boundary in the model (\ref{ccc}), as discussed and plotted in
Sec.~\ref{sec:RGwithout_tperp} of the main text.

\section{Relationship between the chain and band bases}
\label{bb}

The unitary transformation Eq.~(\ref{bands}) diagonalizes the
single-particle Hamiltonian (\ref{h0}). For large values of $t_\perp$
the eigen-basis of ${\cal H}_0$ is the band basis. However, for
$t_\perp=0$ the single-particle Hamiltonian is diagonal already in the
chain basis. In this case, the two bases are completely equivalent. At
the same time, the precise form of the interaction Hamiltonian
(\ref{hint}) in terms of the chiral fermionic fields as well as their
bosonic counterparts strongly depends on the choice of the basis. Here
we discuss the connection between the two and derive the most general
form of ${\cal H}_{int}$ in the bosonic representation.

\subsection{Rotation of chiral fields between the chain and band bases}

Consider the the unitary transformation (\ref{bands}) between the
chain and the band bases in the case where $t_\perp=0$. Then the Fermi
momenta are the same in both bases and a similar transformation
relates the chiral fields as well
\begin{eqnarray}
&&
R_{\alpha(\beta),\sigma} = \frac{R_{1,\sigma}\pm R_{2,\sigma}}{\sqrt{2}}, 
\nonumber\\
&&
\nonumber\\
&&
L_{\alpha(\beta),\sigma} = \frac{L_{1,\sigma}\pm L_{2,\sigma}}{\sqrt{2}}.
\label{cfbb}
\end{eqnarray}
Now we apply this transformation to all the terms in the interaction
Hamiltonian and observe how the coupling constants in the two bases
are related to each other. In order to do this, we write down all
possible two-particle interaction terms disregarding all constraints
due to symmetries and conservation laws (these will be enforced later
by requiring certain relations between the coupling constants). All
such terms can be arranged into three distinct categories: (here
$\mu=\pm$ denotes the two fermionic flavors; these should be
understood as ``chains'' -- that we usually denote by $m=1,2$ -- in
the chain basis and ``bands'' -- usually denoted as
$\nu=\alpha(\beta)$ in the band basis).

\noindent
(A) interactions between particles of the same
spin
\begin{subequations}
\begin{eqnarray}
\star && f_1^A \sum_{\mu\sigma} 
\left( R_{\mu,\sigma}^\dagger R_{\mu,\sigma} + L_{\mu,\sigma}^\dagger L_{\mu,\sigma} \right)
\left( R_{\mu,\sigma}^\dagger R_{\mu,\sigma} +  L_{\mu,\sigma}^\dagger L_{\mu,\sigma} \right) ,
\nonumber\\
&&
\nonumber\\
\star && f_2^A \sum_{\mu\sigma} 
\left( R_{\mu,\sigma}^\dagger R_{\mu,\sigma}  + L_{\mu,\sigma}^\dagger L_{\mu,\sigma} \right)
\left( R_{-\mu,\sigma}^\dagger R_{-\mu,\sigma} + L_{-\mu,\sigma}^\dagger L_{-\mu,\sigma}  \right) ,
\nonumber\\
&&
\nonumber\\
\star && f_3^A \sum_{\mu\sigma} 
\left( R_{\mu,\sigma}^\dagger L_{\mu,\sigma} L_{-\mu,\sigma}^\dagger R_{-\mu,\sigma} 
+  L_{\mu,\sigma}^\dagger R_{\mu,\sigma} R_{-\mu,\sigma}^\dagger L_{-\mu,\sigma} \right) ,
\nonumber\\
&&
\nonumber\\
&& f_4^A \sum_{\mu\sigma} 
\left( R_{\mu,\sigma}^\dagger L^\dagger_{\mu,\sigma} L_{-\mu,\sigma}R_{-\mu,\sigma} 
+  L_{\mu,\sigma}^\dagger R_{\mu,\sigma}^\dagger R_{-\mu,\sigma} L_{-\mu,\sigma} \right);
\end{eqnarray}
(B) interaction between particles of opposite spin, but without spin
tranfer between left and right Fermi points
\begin{eqnarray}
\star && f_1^B \sum_{\mu\sigma}
\left( R_{\mu,\sigma}^\dagger R_{\mu,\sigma} + L_{\mu,\sigma}^\dagger L_{\mu,\sigma} \right)
\left( R_{\mu,-\sigma}^\dagger R_{\mu,-\sigma} +  L_{\mu,-\sigma}^\dagger L_{\mu,-\sigma} \right) 
\nonumber\\
&&
\nonumber\\
\star && f_2^B \sum_{\mu\sigma} 
\left( R_{\mu,\sigma}^\dagger R_{\mu,\sigma}  + L_{\mu,\sigma}^\dagger L_{\mu,\sigma} \right)  
\left( R_{-\mu,-\sigma}^\dagger R_{-\mu,-\sigma}  + L_{-\mu,-\sigma}^\dagger L_{-\mu,-\sigma}  \right) 
\nonumber\\
&&
\nonumber\\
&& f_3^B \sum_{\mu\sigma} 
\left( R_{\mu,\sigma}^\dagger R_{-\mu,\sigma} + L_{\mu,\sigma}^\dagger L_{-\mu,\sigma} \right) 
\left( R_{\mu,-\sigma}^\dagger R_{-\mu,-\sigma} +  L_{\mu,-\sigma}^\dagger L_{-\mu,-\sigma} \right) 
\nonumber\\
&&
\nonumber\\
&& f_4^B \sum_{\mu\sigma} 
\left( R_{\mu,\sigma}^\dagger R_{-\mu,\sigma} +   L_{\mu,\sigma}^\dagger L_{-\mu,\sigma} \right)
\left( R_{-\mu,-\sigma}^\dagger R_{\mu,-\sigma} +  L_{-\mu,-\sigma}^\dagger L_{\mu,-\sigma}  \right)
\end{eqnarray}
(C) backscattering with respect to spin degrees of freedom:
\begin{eqnarray}
\star && f_1^C \sum_{\mu\sigma} 
\left( R_{\mu,\sigma}^\dagger L_{\mu,\sigma} L_{\mu,-\sigma}^\dagger R_{\mu,-\sigma} 
+  L_{\mu,\sigma}^\dagger R_{\mu,\sigma} R_{\mu,-\sigma}^\dagger L_{\mu,-\sigma} \right) 
\nonumber\\
&&
\nonumber\\
\star && f_2^C \sum_{\mu\sigma} 
\left( R_{\mu,\sigma}^\dagger L^\dagger_{\mu,\sigma} L_{-\mu,-\sigma}R_{-\mu,-\sigma} 
+  L_{\mu,\sigma}^\dagger R_{\mu,\sigma}^\dagger R_{-\mu,-\sigma} L_{-\mu,-\sigma} \right) 
\nonumber\\
&&
\nonumber\\
&& f_3^C \sum_{\mu\sigma} 
\left( R_{\mu,\sigma}^\dagger L_{-\mu,\sigma} L_{\mu,-\sigma}^\dagger R_{-\mu,-\sigma} 
+  L_{-\mu,\sigma}^\dagger R_{\mu,\sigma} R_{-\mu,-\sigma}^\dagger L_{\mu,-\sigma} \right) 
\nonumber\\
&&
\nonumber\\
&& f_4^C \sum_{\mu\sigma} 
\left( R_{\mu,\sigma}^\dagger L^\dagger_{-\mu,\sigma} L_{-\mu,-\sigma}R_{\mu,-\sigma} 
+  L_{-\mu,\sigma}^\dagger R_{\mu,\sigma}^\dagger R_{\mu,-\sigma} L_{-\mu,-\sigma} \right)
\end{eqnarray}
\label{it}
\end{subequations}

\noindent
Terms, that respect the particle-number conservation on each
individual chain are marked with a star (these are the terms that are
present in the chain basis). A technical reason to arrange the
interaction terms (\ref{it}) into the three categories is that these
categories remain ``invariant'' under the basis rotation (\ref{cfbb}).
The coupling constants corresponding to these interaciton terms
transform within their categories. The transformation rules are:
\begin{subequations}
\label{tr}
\begin{equation}
\begin{pmatrix} 
\tilde{f}_1^A  \cr \tilde{f}_2^A \cr \tilde{f}_3^A \cr \tilde{f}_4^A 
\end{pmatrix}
= \frac{1}{2} 
\begin{pmatrix}
1 & 1 & -1 & 1 \cr 
1 & 1 & 1 & -1 \cr 
-1 & 1 & 1 & 1 \cr 
1 & -1 & 1 & 1 
\end{pmatrix}
\begin{pmatrix}
f_1^A  \cr f_2^A \cr f_3^A \cr f_4^A 
\end{pmatrix}
\label{tr1}
\end{equation}
\begin{equation}
\begin{pmatrix} 
\tilde{f}_1^B  \cr \tilde{f}_2^B \cr \tilde{f}_3^B \cr \tilde{f}_4^B 
\end{pmatrix}
= \frac{1}{2} 
\begin{pmatrix}
1 & 1 & 1 & 1 \cr 
1 & 1 & -1 & -1 \cr 
1 & -1 & 1 & -1 \cr 
1 & -1 & -1 & 1 
\end{pmatrix}
\begin{pmatrix}
f_1^B  \cr f_2^B \cr f_3^B \cr f_4^B 
\end{pmatrix}
\end{equation}
\begin{equation}
\begin{pmatrix} 
\tilde{f}_1^C  \cr \tilde{f}_2^C \cr \tilde{f}_3^C \cr \tilde{f}_4^C 
\end{pmatrix}
= \frac{1}{2} 
\begin{pmatrix}
1 & 1 & 1 & 1 \cr 
1 & 1 & -1 & -1 \cr 
1 & -1 & 1 & -1 \cr 
1 & -1 & -1 & 1 
\end{pmatrix}
\begin{pmatrix}
f_1^C  \cr f_2^C \cr f_3^C \cr f_4^C 
\end{pmatrix}
\end{equation}
\end{subequations}
The transformation matrices satisfy $\widehat R ^2 = 1$ as they must.

\subsection{Interaction Hamiltonian in the fermionic representation}

Unlike Eqs.~(\ref{it}) which simply list all possible interaction
terms, the interaction Hamiltonian of a physical model is resitricted
by symmetries and conservation laws.

\subsubsection{Chain basis}

Consider first the ladder model (\ref{h}) in the chain basis.  Then
the interaction Hamiltonian(\ref{hint}) respects the conservation of
the number of particles on each individual chain. As a result
$f_4^A=f_3^B=f_4^B=f_3^C=f_4^C=0$. The remaining seven terms are
marked in Eqs.~(\ref{it}) with a star. These terms are further
restricted by the SU(2) symmetry, as discussed in
Sec.~\ref{lowtperp}. The ensuing condition
\begin{equation}
\label{cbsu2}
f_1^A+f_1^C = f_1^B, \quad f_2^A=f_2^B, \quad f_3^A = f_2^C,
\end{equation}
reduces the number of free parameters to 4: three that enter the RG
equations (\ref{RG1}), and the Luttinger-liquid parameter in the total
charge sector. The parameters in Eqs.~(\ref{it}) are related to those
in Eqs.~(\ref{RG1}) by 
\begin{eqnarray}
&&
f_1^{A(B)} = -\frac{g_c^+ + h_c + h_2 \pm 2 h_1}{4}, \quad
f_3^A=f_2^C = h_2,
\nonumber\\
&&
\nonumber\\
&&
f_2^{A(B)} = - \frac{g_c^+ - h_c - h_2}{4}, \quad
f_1^C=h_1.
\label{cbh}
\end{eqnarray}

\subsubsection{Band basis}

The band basis looks very similar to the chain basis, with the
exception of one important difference: there is no particle-number
conservation for each individual band. In fact, if $t_\perp=0$, one
still has to require particle-number conservation for each of the
chains, since this is a physical conservation law independent of the
choice of the basis. This constrains the coupling constants $f_i^a$ as
\begin{equation}
\label{bbcc}
f_1^B = f_2^B, \quad f_1^C=f_2^C, \quad 
f_3^B = f_4^B, \quad f_3^C = f_4^C.
\end{equation}
Furthermore, one has to respect the global SU(2) symmetry. This
yields the following additional restrictions
\begin{eqnarray}
&&
f_1^A+f_1^C = f_1^B, \quad f_3^C = f_3^B - f_4^A 
\nonumber\\
&&
\nonumber\\
&&
f_2^A = f_2^B - f_4^C, \quad f_3^A = f_2^C - f_4^B.
\label{bbsu2}
\end{eqnarray}
The SU(2) conditions (\ref{bbsu2}) are valid also in the
chain basis, but there they are combined with vanishing of the five
coupling constants due to the particle-number conservation leading to
Eq.~(\ref{cbsu2}). In the band basis we need to use both relations
(\ref{bbcc}) and (\ref{bbsu2}) to reduce the number of independent
parameters. Since the total charge sector (described by $g_c^+$)
always decouples we are left with 7 parameters describing 
the remaining sectors of the theory.

\subsection{Interaction Hamiltonian in the bosonic representation}
\label{gfih}

As usual, the low-energy effective Hamiltonian corresponding to the
twelve interaction terms (\ref{it}) can be represented as a sum of 
forward- and backscattering terms.

\subsubsection{Forward scattering}

Four out of the twelve interaction terms, $f_1^A, f_2^A, f_1^B$ and
$f_2^B$ are regular current-current forward scattering terms.
They can be bosonized to a quadratic form
\begin{eqnarray}
&&
\frac{1}{2\pi} \sum_{\mu\sigma} 
\left[f_1^A \partial_x \phi_{\mu\sigma} \partial_x \phi_{\mu\sigma} + 
f_2^A \phi_{\mu\sigma} \partial_x \phi_{-\mu,\sigma} \right. 
\nonumber\\
&& 
\nonumber\\
&& 
\quad\quad
\left. 
+ f_1^B \partial_x \phi_{\mu\sigma} \partial_x \phi_{\mu,-\sigma} 
+ f_2^B \phi_{\mu\sigma} \partial_x \phi_{-\mu,-\sigma} \right].
\label{fsff}
\end{eqnarray}
Introducing the usual total and relative charge and spin
\begin{eqnarray}
&&
\phi_c^{\pm} = \frac{\phi_{1\uparrow}+\phi_{1\downarrow} \pm 
( \phi_{1\uparrow} + \phi_{2\downarrow} )}{2} ,
\nonumber\\
&& 
\nonumber\\
&& 
\phi_s^{\pm} = \frac{\phi_{1\uparrow}-\phi_{1\downarrow} \pm 
( \phi_{1\uparrow} - \phi_{2\downarrow} )}{2},
\label{bfbb}
\end{eqnarray}
we find that each mode is the Luttinger liquid with the corresponding
Luttinger parameter $K_\mu^\nu=1+\gamma_\mu^\nu/4\pi v_F$, where the 
effective coupling constants are
\begin{eqnarray}
&&
\gamma_c^{+} = -\left( f_1^A + f_2^A + f_1^B + f_2^B \right) ,
\nonumber\\
&& 
\nonumber\\
&& 
\gamma_c^{-} = -\left( f_1^A - f_2^A + f_1^B - f_2^B \right) ,
\nonumber\\
&& 
\nonumber\\
&& 
\gamma_s^{+} = -\left( f_1^A + f_2^A - f_1^B - f_2^B \right), 
\nonumber\\
&& 
\nonumber\\
&& 
\gamma_s^{-} = -\left( f_1^A - f_2^A - f_1^B + f_2^B \right).
\end{eqnarray}
Note that when combined with the transformation (\ref{tr}), we see
that $\gamma_c^{+}$ and $\gamma_s^{+}$ are invariant, as they must be.

\subsubsection{Backscattering}

The eight remaining interaction terms in Eq.~(\ref{it}) yield the
backscattering part of the effective Hamiltonian. The bosonization of
these terms is straightforward (please note our choice of the Klein
factors, described in Appendix~\ref{kf}). As a result we find eight
regular backscattering terms (there are also two extra terms with
non-zero conformal spin; such terms do not contribute to the one-loop
RG and therefore we disregard them here)
\begin{align}
&
{\cal H}_{BS} = \frac{1}{2(\pi\alpha)^2} \left\{
\cos\sqrt{4\pi} \phi_c^- \left[
\gamma_1 \cos \sqrt{4\pi} \phi_s^+ + \gamma_2 \cos \sqrt{4\pi} \phi_s^- 
+ \gamma_3 \cos\sqrt{4\pi} \theta_s^- \right] \right.
\nonumber\\
&&
\nonumber\\
&
\quad\quad\quad\quad\quad\quad\quad
+  \cos\sqrt{4\pi} \theta_c^- \left[
\gamma_4 \cos \sqrt{4\pi} \phi_s^+ + \gamma_5 \cos \sqrt{4\pi} \phi_s^-
 + \gamma_6 \cos\sqrt{4\pi} \theta_s^- \right] 
\nonumber\\
&
\nonumber\\
&
\quad\quad\quad\quad\quad\quad\quad
\left. + \cos\sqrt{4\pi}\phi_s^+ \left[ \gamma_7 \cos \sqrt{4\pi} \phi_s^- 
+ \gamma_8 \cos \sqrt{4\pi} \theta_s^- \right] \right\}.
\label{intbos}
\end{align}
Here we have re-labeled the coupling constants as follows:
\begin{align}
&
\gamma_1 = f_2^C, \quad \gamma_2 = f_3^A, \quad \gamma_3 = f_4^B,
\nonumber\\
&
\nonumber\\
& 
\gamma_4 = -f_3^C, \quad \gamma_5 = -f_3^B, \quad \gamma_6 = f_4^A
\nonumber\\
&
\nonumber\\
&
\gamma_7 = f_1^C, \quad \gamma_8 = f_4^C.
\label{icb}
\end{align}

\subsubsection{Refermionized representation and SU(2) spin symmetry}

By applying the refermionization rules in \ref{referm}, we obtain an
alternative representation (terms involving the decoupled $\phi_c^+$
field are dropped for convenience)
\begin{align}
&
{\cal H}_{FS} + {\cal H}_{BS} = 
\frac{i}{2\pi\alpha} 
\cos\sqrt{4\pi} \phi_c^- \left[
\gamma_1 (\xi_R^1\xi_L^1 + \xi_R^2 \xi_L^2 )
+ \gamma_2  (\xi_R^3\xi_L^3 + \xi_R^0 \xi_L^0 ) 
+ \gamma_3 (\xi_R^3\xi_L^3 - \xi_R^0 \xi_L^0 ) \right] 
\nonumber\\
&
\nonumber\\
&
\quad\quad\quad\quad\quad\quad\quad
+  \frac{i}{2\pi\alpha}  \cos\sqrt{4\pi} \theta_c^- \left[
\gamma_4 (\xi_R^1\xi_L^1 + \xi_R^2 \xi_L^2 )
+ \gamma_5  (\xi_R^3\xi_L^3 + \xi_R^0 \xi_L^0 ) 
+ \gamma_6 (\xi_R^3\xi_L^3 - \xi_R^0 \xi_L^0 ) \right]
\nonumber\\
&
\nonumber\\
&
\quad\quad\quad\quad\quad\quad\quad
+ \frac{\gamma_7}{2}  (\xi_R^1\xi_L^1 + \xi_R^2 \xi_L^2 )(\xi_R^3\xi_L^3 + \xi_R^0 \xi_L^0 )
+ \frac{\gamma_8}{2} (\xi_R^1\xi_L^1 + \xi_R^2 \xi_L^2 )(\xi_R^3\xi_L^3 - \xi_R^0 \xi_L^0 )
\nonumber\\
&
\nonumber\\
&
\quad\quad\quad\quad\quad\quad\quad
+ \frac{\gamma_s^+}{2} \xi_R^1\xi_L^1 \xi_R^2 \xi_L^2  
+ \frac{\gamma_s^-}{2} \xi_R^3\xi_L^3 \xi_R^0 \xi_L^0  
- \frac{\gamma_c^-}{\pi} (\partial_x \phi_c^-)^2.
\label{icbrf}
\end{align}
By the rules governing the addition of two spin-half objects, we know
that there should be one triplet spin-mode (here represented by
$\xi^i$ with $i=1,2,3$) and one singlet mode ($\xi^0$). We therefore
see that the SU(2) spin symmetry is satisfied so long as the
following relations between the parameters hold
\begin{align}
\gamma_1 = \gamma_2 + \gamma_3, &\quad \gamma_s^{+} = \gamma_7+\gamma_8, \nonumber \\
\gamma_4=\gamma_5+\gamma_6, & \gamma_s^{-} = \gamma_7-\gamma_8.
\label{su2general}
\end{align}
The above set of conditions reduce by four the number of free
parameters in the most general interaction Hamiltonian written above;
and is furthermore preserved by the rotation between chain and band
basis, Eq.~\eqref{tr1}, as it must be.  Before analysing further
symmetries of the model in the chain and band basis however, it is
convenient to derive the renormalization group flow for the above
model.

\subsection{RG equations}
\label{AllRG}

The easiest way to derive the RG flow equations for the most general
form of the coupled sine-Gorden model, Eq.~\eqref{icb} along with the
relevant forwards scattering terms, Eq.~\eqref{fsff}, is using the
operator product expansion method, as described in
e.g. Ref.~\cite{senechal}. This gives to second order the set of
equations
\begin{alignat}{3}
\frac{\partial \gamma_1}{\partial l} &= -\gamma_1 \left( \frac{\gamma_c^{-} 
+ \gamma_s^{+}}{2} \right) - \gamma_2 \gamma_7 - \gamma_3 \gamma g_8,
& \quad\quad
\frac{\partial \gamma_7}{\partial l} &= -\gamma_7 \left( \frac{\gamma_s^{+} 
+ \gamma_s^{-}}{2} \right) - \gamma_1 \gamma_2 - \gamma_4 \gamma_5,
\nonumber \\ 
 \frac{\partial \gamma_2}{\partial l} &= -\gamma_2 \left( \frac{\gamma_c^{-} 
+ \gamma_s^{-}}{2} \right) - \gamma_1 \gamma_7,
& \quad\quad
\frac{\partial \gamma_8}{\partial l} &= -\gamma_8 \left( \frac{\gamma_s^{+} 
- \gamma_s^{-}}{2} \right) - \gamma_1 \gamma_3 - \gamma_4 \gamma_6,
\nonumber \\ 
 \frac{\partial \gamma_3}{\partial l}  &= -\gamma_3 \left( \frac{\gamma_c^{-} 
- \gamma_s^{-}}{2} \right) - \gamma_1 \gamma_8,
 & \quad\quad
  & 
\nonumber  \\
\frac{\partial \gamma_4}{\partial l} &= -\gamma_4 \left( \frac{-\gamma_c^{-} 
+ \gamma_s^{+}}{2} \right) - \gamma_5 \gamma_7 - \gamma_6 \gamma_8,
& \quad \quad
\frac{\partial \gamma_c^{-}}{\partial l} &= -\gamma_1^2 -\gamma_2^2 - \gamma_3^2 
+ \gamma_4^2 + \gamma_5^2 + \gamma_6^2,
\nonumber \\
\frac{\partial \gamma_5}{\partial l} &= -\gamma_5 \left( \frac{-\gamma_c^{-} 
+ \gamma_s^{-}}{2} \right) - \gamma_4 \gamma_7,
& \quad\quad
\frac{\partial \gamma_s^{+}}{\partial l} &= - \gamma_1^2 - \gamma_4^2 
- \gamma_7^2 - \gamma_8^2, 
\nonumber \\
\frac{\partial \gamma_6}{\partial l} &= -\gamma_6 \left( \frac{-\gamma_c^{-} 
- \gamma_s^{-}}{2} \right) - \gamma_4 \gamma_8,
& \quad\quad
\frac{\partial \gamma_s^{-}}{\partial l} &=  - \gamma_2^2 - \gamma_5^2 
- \gamma_7^2 + \gamma_3^2 + \gamma_6^2 + \gamma_8^2.
\label{longRG}
\end{alignat}
It is easy to check that the hyperplane described by conditions
\eqref{su2general} is invariant under the RG flow. We now reduce
these general equations to the specific cases of the chain and the
band bases.

\subsubsection{Chain basis}

In the chain basis only three terms in Eq.~(\ref{intbos}) respect the
particle-number conservation so that in the chain basis
\begin{equation}
\label{cbpc}
\gamma_3=\gamma_4=\gamma_5=\gamma_6=\gamma_8=0.
\end{equation}
Adding the SU(2) symmetry, which leads to the condition
\begin{equation}
\label{cbsu}
\gamma_1=\gamma_2, \quad g_s^+ = g_s^- = \gamma_7
\end{equation}
we see that the backscattering Hamiltonian takes the form
(\ref{hamBS}), reparameterised as
\begin{equation}
\gamma_7=g_1, \quad\quad \gamma_1=\gamma_2=g_2, \quad\quad  \gamma_c^- = g_c^-.
\label{cprm}
\end{equation}
 By parameterizing the RG flow in the same way, the general RG
 equations above \eqref{longRG} reduce to those given ipreviously,
 Eq.~\eqref{RG1-g}.

\subsubsection{Band basis}
\label{AppBand}

The backscattering interaction, Eq.~\eqref{fullham1} presented in the
main text may be obtained from \eqref{intbos} by a reparameterization
of the SU(2) conditions, \eqref{su2general}, specifically
\begin{align}
& \gamma_s^+ = g_s^+, \quad \gamma_s^- = g_s^-, \quad \gamma_7
=\frac{g_s^++g_s^-}{2}, \quad \gamma_8=\frac{g_s^+-g_s^-}{2}, 
\nonumber \\
& \gamma_1 = \tilde{g}_T, \quad \gamma_2 
= \frac{\tilde{g}_T-\tilde{g}_S}{2}, \quad \gamma_3=\frac{\tilde{g}_T+\tilde{g}_S}{2}, 
\nonumber \\
& \gamma_4 = {g}_T, \quad \gamma_5 
= \frac{{g}_T-{g}_S}{2}, \quad \gamma_6=\frac{{g}_T+{g}_S}{2}.
\end{align} 
The bare values of these parameters are conveniently obtained from
applying the transformations \eqref{tr1} to the values parameterizing
the chain basis. The rotation of the condition \eqref{cbpc} into the
band basis gives
\begin{equation}
\gamma_c^-=\gamma_s^-, \quad\quad
\gamma_1=\gamma_7, \quad\quad
\gamma_3=-\gamma_5, \quad\quad
\gamma_4=-\gamma_8.
\label{bbpc}
\end{equation}
Combining this with the SU(2) symmetry gives eight conditions on the
eleven parameters of Hamiltonian \eqref{intbos} which again leads to a
parameterization in terms of three free parameters:
\begin{alignat}{3}
\gamma_c^- = \gamma_s^- &= g_2,
& \quad\quad
\gamma_s^+ &= g_1,
 \nonumber \\
 \gamma_1=\gamma_7 &= \frac{1}{2}{g_1+g_2}
 & \quad\quad
 \gamma_2 &= \frac{1}{4}(g_1+2g_2+g_c^-),
 \nonumber \\
 \gamma_3=-\gamma_5 &=\frac{1}{4} (g_1-g_c^-),
 &\quad\quad
  \gamma_6 &= \frac{1}{4} (2g_2-g_1-g_c^-),
 \nonumber \\
 \gamma_4=-\gamma_8 &=\frac{1}{2}(g_2-g_1).
 &&
 \label{bprm}
\end{alignat}
The equivalent parameterization in terms of the set of $h_i$
(Luttinger liquid parameters for the decoupled chains) is given in the
main text in Eq.~\eqref{bbgs}.

Substituting this parameterization \eqref{bprm} into the RG flow
\eqref{longRG}, one finds that it is indeed an invariant hyperplane of
the flow, and furthermore leads to the same RG equations \eqref{RG1-g}
for $g_1$, $g_2$, and $g_c^-$ as found in the chain basis.  This is
not surprising -- the change of basis in which the Hamiltonian is
written should not change any physical observable.

At energy scales less than $t_\perp$ however, the flow is altered, as
processes involving $\cos \sqrt{4\pi}\phi_c^-$ should be dropped due
to the rapid oscillations of this term which average out to zero.  In
other words, one sets $\gamma_4=\gamma_5=\gamma_6=0$ in the above flow
and sees what happens.  In this case, although the initial values of
$\gamma_i$ may be parameterized as in \eqref{bprm} above, the RG flow
will now leave this three parameter hyperplane.  Physically, the
renormalization in the presence of $t_\perp$ generates interaction
terms that do not conserve particle number individually on each chain.
The SU(2) symmetry remains unbroken though, so out of the initially
eleven parameters, one drops three of them relating to $\phi_c^-$ and
SU(2) symmetry fixes another three of them, leaving five free
parameters.  Substituting these conditions into the general RG flow
\eqref{longRG} above gives the set of five RG equations \eqref{RGfive}
in the main text.

\section{Analysis of RG equations in large $t_\perp$ limit}
\label{RGappendix2}

Here, we summarize a few important properties of the RG flow equations
for the model in the large $t_\perp$ limit, given by
Eq.~\eqref{RGfive}.

\subsection{Duality and O(6) symmetry}

First, we point out that there are two transformations between the
parameters that leave the RG flow invariant:

(i) $g_T\rightarrow -g_T$ and $g_s^- \rightarrow -g_s^-$ simultaneously; and

(ii) $g_S\rightarrow -g_S$ and $g_s^- \rightarrow -g_s^-$ simultaneously.

\noindent 
These are manifestations of duality transformations, discussed in
detail in Ref.~\cite{bal}, that effectively map the flow to one of the
ordered phases to one of the others.

This duality becomes particularly useful when combined with another
property of the equations -- namely, the existence of high-symmetry
rays.  Explicitly, we see that if one sets
\begin{equation}
g_T = g_S = g_s^+ = -g_s^- = -g_c^- = g
\label{o6tune}
\end{equation}
then the flow remains along this single-parameter line
\begin{equation}
\frac{\partial g}{\partial l} = - 2 g^2,
\label{o6flow}
\end{equation}
and furthermore, the Hamiltonian on this ray in phase space has an
explicit O(6) symmetry. By also applying the transformations above,
we see that in fact there are a total of four high-symmetry rays,
related by duality transformations.

The existence of such high-symmetry rays was first discussed in
\cite{lbf} for the half-filled case, where the enhanced symmetry turns
out to be SO(8). The importance of this is the theory of two
leg-ladders stems from two important facts -- firstly, these
high-symmetry rays are attractive (in the RG sense), a phenomena known
as dynamical symmetry enhancement. In other words, small deviations
from this high symmetry flow are irrelevant, meaning that the analysis
of such rays is more general than the fine-tuned parameters
\ref{o6tune} may suggest. Secondly, the high-symmetry strong-coupling
fixed point turns out to be described by an integrable model -- in
this case the Gross-Neveu model -- meaning that non-perturbative
properties may be calculated.

A full analysis of the consequences of this for the doped case with
which we are currently dealing was performed in Ref.~\cite{cts}. 
Indeed, with the following modification in notation,
\begin{equation}
g_s^+ = \frac{g_{\sigma+}}{2}, \quad
g_s^- = - \frac{g_{\sigma-}}{2}, \quad
g_c^- = - \frac{g_{\rho-}}{2},  \quad
g_{T(S)} = \frac{g_{c, st(ss)}}{2},
\end{equation}
our RG equations \eqref{RGfive} become equivalent to those in Equation
5 of \cite{cts}.  In particular, the stability of these rays against
small perturbations breaking the O(6) symmetry was explicitly shown
in this paper; along with an analysis of the correlation functions and
elementary excitations in the strong coupling phases.

\subsection{Qualitative analysis of the RG flow}

Unlike the flow equations \ref{RG1} for the case $t_\perp=0$, we were
unable to analytically determine the phase boundaries from the RG
equations \eqref{RGfive} for large $t_\perp$. However, we can still
make a few observations about the equations which allow us to
determine the phase for a wide variety of parameters, and aids the
numerical determination of the fixed point of the flow for other
initial parameters.

Consider the flow of the product of the two coupling constants $g_T$
and $g_S$. By combining the equations in Eq.~\ref{RGfive}, we see that
\begin{equation}
\frac{\partial (g_T g_S)}{\partial l} =  
\left[ g_c^- - g_s^+ \right] g_T g_S 
+ \frac{g_s^-}{2}\left[ 3g_T^2 + g_S^2\right],
\label{gtgs}
\end{equation}
and at the same time
\begin{equation}
\frac{\partial (g_s^- g_T g_S )}{\partial l} 
=
\left[ g_c^- - 2 g_s^+ \right] 
g_T g_S g_s^- 
+ \frac{1}{2}\left[ 3g_T^2 \left(g_s^-\right)^2 
+ g_S^2\left(g_s^-\right)^2
+ 2 g_T^2 g_S^2  \right].
\label{gsgtgs}
\end{equation}
From these combinations, we can therefore state the following:

(i) The second term in \eqref{gsgtgs} is always positive. Therefore
if the product $g_s^- g_T g_S>0$ is positive, then it can never change
sign. As a consequence, when combined with \eqref{gtgs},

(ii) if at any point during the RG flow both $g_s^->0$ {\it and} $g_T
g_S>0$ then the field $\theta_s^-$ gets locked and the RG equations
flow to either the CDW$^-$ or OAF fixed point, depending on the sign
of $g_T$ [as indicated in Eq.~(\ref{scfps})]. This situation is
realized for repulsive interactions if the intra-chain
nearest-neighbor coupling dominates over the on-cite Hubbard
interaction (see Fig.~\ref{eh1}). Then $h_c<0$, $h_s<0$, and
$h_\perp>0$. The bare values of all three coupling constants $g_T$,
$g_S$, and $g_s^-$ are positive and therefore the system flows to the
CDW$^-$ phase.

(iii) alternatively, if at any point during the RG flow both $g_s^-<0$
{\it and} $g_T g_S<0$, then the resulting phase is superconducting.
This situation is realized in the particular case, where each chain is
described by a Hubbard model, i.e. for $V_\perp=V_\|=0$, the bare
values of the coupling constants are [according to Eqs.~(\ref{bbgs})
  and (\ref{hs})]:
\begin{equation}
g_s^+=2g_S=aU, \quad g_T=-aU/2,  \quad g_c^-=g_s^-=0,
\label{gh}
\end{equation}
and the system flows to the superconducting state in
full agreement with known results (see e.g. Ref.~\cite{gb}).

In the case when $g_s^- g_T g_S<0$, such that $g_s^- $ and $g_T g_S$
have opposite signs, no general statement similar to (i) above can be
made and one needs to integrate the RG equations (\ref{RGfive})
numerically.  However, the above analysis allows one to stop the
numerical integration and determine the $l\rightarrow \infty$ fixed
point when any of the previous conditions is met.  This enables a fast
and efficient numerical determination of the phase diagram, as given
by the RG equations \eqref{RGfive}.


\end{document}